\documentclass[12pt,psfig]{article}
\usepackage{graphicx}
\usepackage[abs]{overpic}
\usepackage{subfigure}
\usepackage[centertags]{amsmath}
\usepackage{amssymb}
\usepackage{dsfont}
\usepackage{cite}
\numberwithin{equation}{section}
\providecommand{\openone}{\leavevmode\hbox{\small1\kern-3.8pt\normalsize1}}

\newcommand{\tfkto}[2]{\vartheta\bigg[ {\begin{array}{*{20}c}#1  \\#2  \\\end{array}}\bigg]}
\newcommand{\tfktob}[2]{\vartheta \Big[{#1 \atop #2}\Big]}
\newcommand{\tfktsq}[2]{\vartheta^2\bigg[ {\begin{array}{*{20}c}#1  \\#2  \\\end{array}}\bigg]}
\newcommand{\tfktsqb}[2]{\vartheta^2 \Big[{#1 \atop #2}\Big]}
\newcommand{\tfktvier}[3]{\vartheta^4\bigg[ {\begin{array}{*{20}c}#1  \\#2  \\\end{array}}\bigg]\left( #3 \right)}
\newcommand{\barre}[1]{%
        \setbox1=\hbox{$#1$} \dimen2=\ht1 \dimen3=\dp1 \dimen4=\wd1
        \setbox2=\hbox{\sl /}
        \dimen1=\wd1 \advance\dimen1 by -\wd2 \divide\dimen1 by 2
        \advance\dimen1 by \wd2 \advance\dimen1 by 0.4pt
        \setbox3=\hbox to \wd1{\hss \box1 \kern -\dimen1 \box2\hss}
        \ht3=\dimen2 \dp3=\dimen3 \wd3=\dimen4
        \box3
        }

\begin{document}
\begin{titlepage}
\begin{flushright}
KUL-TF-05/22\\
hep-th/0509108
\end{flushright}
\vspace{.5cm}
\begin{center}
\baselineskip=20pt {\LARGE    A new picture for 4-dimensional `spacetime'\\
\vskip 0.2cm
from intersecting D-branes on the $T^9$
}\\
\vfill
{\Large Tassilo Ott$^\dagger$} \\
\vfill
{Instituut voor Theoretische Fysica, Katholieke Universiteit
Leuven,\\
       Celestijnenlaan 200D B-3001 Leuven, Belgium.
 }
\end{center}
\vfill
\begin{center}
{\bf Abstract}
\end{center}
{\small A factorization of spacetime of the form
${time}\times\mathcal{M}^3\times\mathcal{M}^3\times\mathcal{M}^3$
is considered in this paper as the closed string background in
type IIA. The idea behind this construction is that each
$\mathcal{M}^3$ might give rise to one large spatial dimension of
4-dimensional spacetime in the closed string sector. In the open
string sector, intersecting D6-branes can be constructed for the
simple choice of an orientifolded $\mathcal{M}^3=T^3$ in a similar
way as on the prominent $T^6=T^2\times T^2\times T^2$ using exact
CFT. The D6-branes then are allowed to span general 2-cycles on
each $T^3_i$. The intersection 1-cycles between two stacks of
branes on one $T^3_i$ can be understood as one spatial dimension
of the effective 4-dimensional `spacetime' for the massless chiral
fermions charged under these two stacks. Additionally to the known
solutions to the R-R tadpole equations conserving
(3+1)-dimensional Poincare invariance, this allows for solutions
with globally just (2+1)- or (1+1)-Poincare invariance. For
non-supersymmetric solutions, a string tree-level and one-loop
potential for the scalar moduli (including the spacetime radii) is
generated in the NS-NS sector. This potential here is interpreted
dynamically for radii and dilaton in order to describe the global
evolution of the universe. In the late time picture,
(3+1)-dimensional global Poincare invariance can be restored well
within experimental bounds. This approach links particle
properties (the chiral massless fermion spectrum) directly to the
global evolution of the universe by the scalar potential, both
depending on the same topological wrapping numbers. In the future,
this might lead to much better falsifiable phenomenological
models.
 }\vspace{2mm} \vfill
 \hrule width 3.cm
 {\footnotesize \noindent
$^\dagger$  tassilo.ott@fys.kuleuven.be }
\end{titlepage}
\newpage


\setcounter{page}{1} \pagestyle{plain}

\section{Introduction} \label{intro}
All matter of the standard model is chiral and appears in
bifundamental representations of unitary and special unitary gauge
factors.

In 1996, Berkooz, Douglas and Leigh discovered that intersecting
D-branes in general lead to chiral fermions in the massless open
string sector when embedded in a closed string type IIA string
theory\cite{Berkooz:1996aa}. Bifundamental massless fermions in
this picture correspond to strings, stretching between two stacks
of D-branes and then shrinking to zero size\footnote{which is
possible if the two stacks are at a non-vanishing angle}. Every
stack of D-branes (with a stacksize $N$) gives rise to a unitary
gauge factor U($N$). The bifundamental representations of chiral
fermions then are of the type $(N_a,\bar N_b)$ or $(N_a,N_b)$,
where $a$ is the stack where the one endpoint and $b$ the
second stack where the other endpoint of the open string lies.

Many phenomenological models have been constructed in the
meanwhile (for an overview see for instance
\cite{Blumenhagen:2005mu, Ott:2003yv, Gorlich:2004zs,
MarchesanoBuznego:2003hp, Lust:2004ks}), and in all these
constructions (9+1)-dimensional spacetime of type II string theory
is factorized into a direct product of a (3+1)-dimensional
external space $\mathbb{R}^{3,1}$ and a 6-dimensional internal
space $\mathcal{M}^6$,i.e.
\begin{equation}\label{eq:alter_spactime_ansatz}
\mathcal{X}=\mathbb{R}^{3,1}\times \mathcal{M}^6\ .
\end{equation}
$\mathbb{R}^{3,1}$ is the usual flat Minkowski space (also warped
compact versions have been discussed) and $\mathcal{M}^6$ is a
compact Calabi-Yau 3-fold or an orbifolded/ orientifolded $T^6$,
where the size of this internal manifold is small enough compared
to $\mathbb{R}^{3,1}$ in order not to conflict with any gravity
experiments.

Although the ansatz (\ref{eq:alter_spactime_ansatz}) seems
phenomenologically obvious, it is nevertheless highly
unsatisfying. If string theory is to be a fundamental theory that
is valid for all energy scales, it should even be able to provide
for an explanation why spacetime splits into such a product
(\ref{eq:alter_spactime_ansatz}). It is more natural to assume a
very isotropic and homogeneous spacetime in the very beginning of
the universe, where all spatial dimensions (leaving time aside)
are of the same type and size. That they should be of the same
type can be concretized in the demand that they should all be
either compact or non-compact. Effective compactification or
decompactification of certain dimensions then could be understood
dynamically within the theory. In this line of thought, a product
structure (\ref{eq:alter_spactime_ansatz}) at best will be valid
for late times\footnote{It is not known what `late times'
precisely means, as on the one hand the order of the string scale
is unknown and on the other, a completely satisfying dynamical
string (field) theory is missing.}. For this reason, it seems
questionable if a model with an a priori spacetime product
structure (\ref{eq:alter_spactime_ansatz}) does correctly describe
early cosmology (like inflation). If inflation indeed happens at a
scale $\sim 10^{10}$ GeV and the compactification scale is around
the GUT scale, the ansatz (\ref{eq:alter_spactime_ansatz}) would
already be sufficient to describe it.

Indeed, a picture with initially similar compact spatial
dimensions has been discussed already as early as 1988 by
Brandenberger and Vafa \cite{Brandenberger:1988aj}, where it was
assumed that space in the beginning can be described by a $T^9$.
It is argued there that a (3+1)-dimensional large spacetime torus
might be favored for late times by dimension counting arguments
and considerations about Kaluza-Klein and winding modes, but these
arguments were rather qualitative than quantitative.

We will follow a similar spirit in this paper for intersecting
branes: we will not a priori require a spacetime structure
(\ref{eq:alter_spactime_ansatz}), but start with a more general
spacetime
\begin{equation}\label{eq:neuer_spactime_ansatz}
\mathcal{X}=\text{time}\times \mathcal{M}^9\ ,
\end{equation}
where $\mathcal{M}^9$ is a 9-dimensional compact manifold. In order to be
able to use conformal field theory methods, we will describe the
simplest case, the 9-torus, $\mathcal{M}^9=T^9$. Concretely, we
will discuss the factorization
\begin{equation}\label{eq:neuer_spactime_ansatzT9}
\mathcal{M}^9=T^3\times T^3 \times T^3.
\end{equation}
The reasons for this choice are of different nature:
\begin{enumerate}
\item The ansatz (\ref{eq:neuer_spactime_ansatzT9})
seems appealing as it is the most symmetrical way for three large
dimensions to emerge, i.e. one $T^3$ might exactly give rise to
one large spatial dimension of the 4-dimensional external
spacetime ($9=3\times 3$).

\item The most successful models of recent years employ only
D6-branes for such constructions, as they typically intersect
along (3+1) common dimensions, see for instance
\cite{Sagnotti:1987tw, Blumenhagen:1999db, Blumenhagen:1999ev,
Blumenhagen:2000wh, Blumenhagen:2000ea, Angelantonj:2000hi,
Aldazabal:2000dg, Forste:2000hx, Forste:2001gb, Kokorelis:2002ns, Cvetic:2001nr, Kachru:1999vj,
Uranga:2003pz, Ott:2003yv,Honecker:2004kb,
Ott:2005sa,Gorlich:2004zs, Angelantonj:2002ct}. In all the
constructions, it is assumed that the D6-branes completely fill
out the external space $\mathbb{R}^{3,1}$, and wrap 3-cycles on
the internal space $\mathcal{M}^6$ that generically intersect in
one (or several) point(s). The topological intersection number on
the compact internal space in this case is interpreted as the
multiplicity of fermion representations, or in other words, the
number of generations. The string one-loop amplitude can be
calculated with CFT methods exactly if the $T^6$ torus is
factorized as $T^6=T^2 \times T^2 \times T^2$. In this case, the
3-cycle that the D-brane wraps is factorized into three 1-cyles
$\pi^{(1)}$, each wrapping one of the $T^2$, i.e.
$\pi^{(3)}=\pi^{(1)}\otimes\pi^{(1)}\otimes\pi^{(1)}$.

In the alternative ansatz (\ref{eq:neuer_spactime_ansatzT9}),
these computations can be generalized by simply requiring the
D6-branes to wrap 2-cycles on every $T^3$, i.e.
$\pi^{(6)}=\pi^{(2)}\otimes\pi^{(2)}\otimes\pi^{(2)}$. The
2-cycles $\pi^{(2)}_a$ and $\pi^{(2)}_b$, describing different
stacks of branes, will then generically intersect along a 1-cycle
on every $T^3$. This intersection 1-cycle $\pi^{(1)}_I$ is the
place within each $T^3$ where the massless bifundamental fermions
are located. The three intersection 1-cycles together (coming from
the different $T^3$) form a 3-cycle,
$\pi^{(3)}_I=\pi^{(1)}_I\otimes\pi^{(1)}_I\otimes\pi^{(1)}_I$, and
together with time this is the effective 4-dimensional `spacetime'
for the fermion under consideration. This understanding is valid
as long as the fermions are massless compared to the Planck mass.
The new picture of D6-branes on the $T^9$ is illustrated in direct
comparison to the old one on the $T^6$ in figure
\ref{Fig:all_tori_old_new}.
\begin{figure}[h]
\begin{center}
\begin{overpic}[scale=.82,unit=1mm]%
      {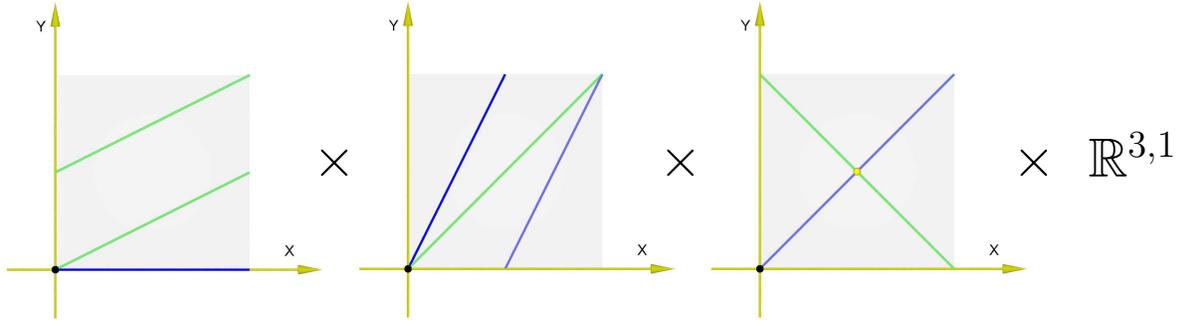}
  \put(42,19){\LARGE $\times$}
  \put(88,19){\LARGE $\times$}
  \put(135,19){\LARGE $\times\ \ \mathbb{R}^{3,1}$}
\end{overpic}
\subfigure{(a) The old picture on the $T^6=T^2\times T^2 \times T^2$}
\begin{overpic}[scale=.82,unit=1mm]%
      {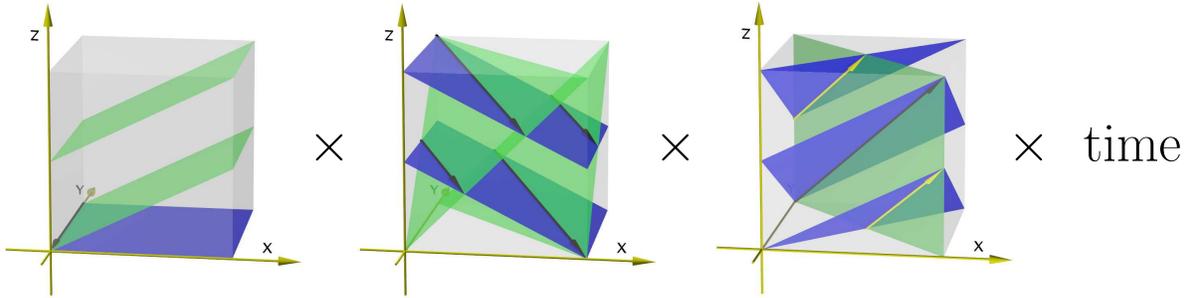}
  \put(42,19){\LARGE $\times$}
  \put(88,19){\LARGE $\times$}
  \put(135,19){\LARGE $\times\ \ $time}
\end{overpic}
\subfigure{(b) The new picture on the $T^9=T^3\times T^3 \times
T^3$} \caption{The two pictures of intersecting D6-branes. The
total intersection number is two in both examples, meaning for (a)
the number counting distinct intersection points, for (b) the
number counting the distinct intersection 1-cycles (black and
yellow arrows in the picture). The explicit wrapping numbers of
the two branes in the lower figure are stated in table
\ref{tab:wrap_picture}.} \label{Fig:all_tori_old_new}
\end{center}
\end{figure}

Of course, the real closed string background and
therefore gravity is still (9+1)-dimensional, but this is the case
as in any other brane construction in type IIA/B string theory.
Such a picture can only make sense in the present state of the
universe if six dimensions are small. It has been pointed out by
Arkani-Hamed, Dimopoulos and Dvali in 1998 that they might be even
as large as one millimeter, if only gravity is able to propagate
through them \cite{Arkani-Hamed:1998rs, Antoniadis:1998ig}.

The wrapping of the D-branes is topological. This means that for
every $T^3$, the size of the intersection 1-cycle $\pi^{(1)}_I$
is dependent on the sizes of the fundamental torus radii. If for
(\ref{eq:neuer_spactime_ansatzT9}) indeed two compactification
radii within every $T^3$ are getting small and one is getting
large, then the intersection 1-cycle $\pi^{(1)}_I$ is also getting
large provided that it wraps the large torus cycle. This
then implies that `spacetime' for the massless fermions
effectively is (3+1)-dimensional. This property is illustrated in
figure \ref{Fig:radius_growing}.
\begin{figure}[h]
\begin{center}
\begin{overpic}[scale=.7,unit=1mm]%
      {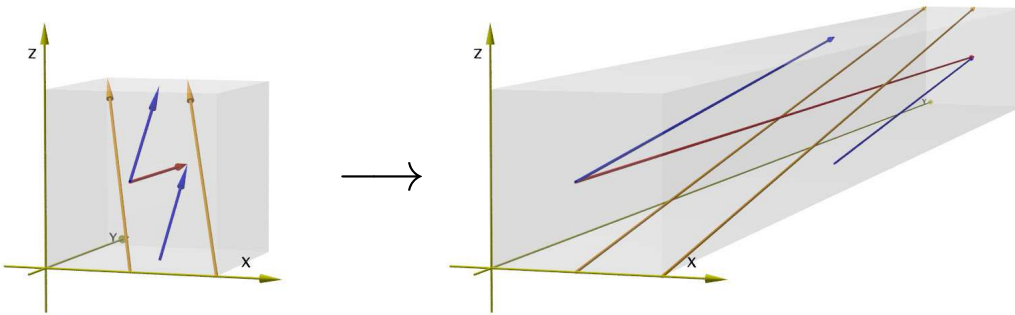}
  \put(48,20){\LARGE $\longrightarrow$}
\end{overpic}
\caption{One radius of the $T^3$ growing in time. The arrows
depict the intersections between different stacks of branes.
Global Poincare invariance on the left-hand-side is broken, but
gets restored effectively on the right-hand-side in the late time
picture.} \label{Fig:radius_growing}
\end{center}
\end{figure}
\item The torus is flat. This not just allows for trustable world
sheet string computations, but also agrees with the picture of
cosmology of the present universe which spatially seems to be
flat. This can be seen for instance in the measurements of the
total energy density by the so far highest precision experiment
WMAP \cite{Spergel:2003cb}, which has yielded the result
$\Omega_\text{tot}=1.02\pm 0.02$, where $\Omega_\text{tot}=1$
describes a flat universe within standard cosmology. Of course,
this immediately poses one question: we only know that the
universe is flat since the radiation of CMB, before this time the
picture is not so clear. Within standard cosmology the flat
universe is unstable, meaning that any initial deviation from
flatness is blown up extremely during the evolution of the
universe. Inflation was created particularly to solve this
finetuning problem by smoothing out any initial perturbation. On
the other hand, we know that standard cosmology cannot be valid
for very early cosmology anyway and that string theory might
replace it. This leaves the possibility that the universe could
have been flat even in the very early universe and that the
spatial curvature stays negligible during the evolution. This is
the working hypothesis of this paper and it is possible to check
the validity of this approach by calculating the backreaction of
the D-branes onto spacetime explicitly.

Even if such a picture finally cannot be maintained for the whole
evolution of the universe, it still can be valid for the evolution
after inflation and after supersymmetry and electroweak breaking.
This assumption is inherit to all non-supersymmetric intersecting
brane constructions even in the context of particle physics. We
know that both supersymmetry breaking and electroweak breaking
surely takes place at energies above 1 TeV. Therefore, below this
energy scale in the evolution of the universe, the particle
spectrum should not have changed anymore and the constructed
non-supersymmetric intersecting brane models should stay valid
after this time.
\end{enumerate}
Thus, we will discuss intersecting D6-branes that wrap general
2-cycles on every $T^3$ of a background
(\ref{eq:neuer_spactime_ansatzT9}). Intersecting D6-branes in type
IIA string theory can be obtained from several stacks of D9-branes
in type IIB with a constant but different F-flux. To see this, one
has to perform three T-dualities. If we apply one T-duality in
every $T^3$ and allow for a constant F-flux having no component in
the direction in which the T-duality is performed (this is merely
a technical requirement), then we indeed obtain D6-branes with
vanishing F-flux. Also from this point of view, the ansatz
(\ref{eq:neuer_spactime_ansatzT9}) might be favorable. Every stack
of D6-branes wraps a general 2-cycle on every $T^3$ being at a
general angle within the $T^3$. These angles are quantized in
terms of three coprime wrapping numbers $m'$, $n'$ and $p'$, as we
will see later, it is completely analogue to the case on the
$T^2$. In this approach it is possible to construct both
non-supersymmetric and supersymmetric models.

The string background $T^9=T^3\times T^3\times T^3$ is not
the only one that can be described exactly by CFT within this
framework. This is very pleasant as D-branes alone on a compact
space cannot fulfill R-R tadpole cancellation which is needed for
a string model free of gauge anomalies. The $T^9$ can be orientifolded
and/or orbifolded, again in close analogy to the picture on the
$T^6$. In this paper, we will discuss the simplest case of a
worldsheet $\Omega$-orientifold in type IIB. In the type IIA
picture, this translates to an $\Omega R$-orientifold that gives
rise to a geometric O6-plane. This O6-plane also wraps a geometric
2-cycle on every $T^3$ and $R$ can be understood as a geometrical
reflection along one of the three coordinates on every $T^3$. Such
a background conserves 16 supersymmetries.
\subsection{Some consequences of the new picture}
This picture of generally intersecting D6-branes has
some interesting but unusual features which have to be addressed:
\begin{enumerate}
\item It is clear that all the successful models which have been
constructed using the $\Omega R$-orientifold on the $T^6=T^2\times
T^2 \times T^2$ \cite{Blumenhagen:2000wh, Blumenhagen:2000ea} are
embedded in this larger class of models, especially the prominent
model with the standard model spectrum has to be mentioned
\cite{Ibanez:2001nd}. In the new picture, these models are
described by D6-branes that all have one common direction, such
that all intersection lines are parallel on each $T^3$.
Subsequently, this common direction has to be decompactified.
Parallel intersection 1-cycles conserve 1-dimensional Poincare
invariance, therefore (3+1)-dimensional Poincare invariance is
conserved. The internal space can be described by any
perpendicular plane to the intersection 1-cycles in each $T^3$.
Every D-brane wraps a 1-cycle in this plane and these 1-cycles
generically intersect along a point. One of the problems of the
model \cite{Ibanez:2001nd} has been the impossibility to cancel
the NS-NS tadpole. There always remains an instability in the
complex structure \cite{Blumenhagen:2001te, Blumenhagen:2001mb},
implying that the brane configurations completely break
supersymmetry if the tori are not degenerate. Furthermore, also
the dilaton tadpole remains.

\item But in a generic model on the $T^9$ consisting of more than
two stacks\footnote{This is surely the case for the standard model
which consists of three gauge factors.}, it might happen that the
D6-branes are aligned in such a way that the intersection 1-cycles
between different pairs of brane stacks are not all parallel to
each other on one or several $T^3$. In this case, global Poincare
invariance is broken in more than six spatial directions down to
(2+1)- or (1+1)-Poincare invariance, in the worst case in all of
them. Nevertheless, (3+1)-Poincare invariance for just one type of
particles (so for one pair of intersecting branes) by itself is
always conserved. The strength of global Poincare invariance
breaking surely is connected to the size proportions between the
three radii within one $T^3$. If one radius is very large in
comparison to the others, then different intersection 1-cycles
(along which distinct types of massless fermions are located)
parallelize effectively. Unbroken Poincare invariance corresponds
to completely parallel 1-cycles. Thus one has an effective
restoration of (3+1)-dimensional Poincare invariance. It is easy
to give a rough estimate that this restoration is very well off
with current experimental bounds. The so-called Hughes-Drever
tests\footnote{The deviations from the local Lorentz invariance
are inferred from possible anisotropies of the inertial mass $M_I$
in measurements of the quadrupole splitting time dependence of
nuclear Zeeman levels along the orbit of the earth, $\delta=|M_I
c^2/\sum_A E^A-1|$, where the sum goes over all forms of internal
energy of a chosen nucleus.} so far have given the best bounds on
the violation of Lorentz symmetry\cite{Chupp:1989ty}, being
$\delta< 3\cdot10^{-21}$. The diameter of the visible universe is
approximately 13.7 light years. The size of the highest possible
compactification scale for the small dimensions is at the
millimeter scale, implying that the ratio between the largest
radii and the smallest radii has to be at least $7\cdot10^{-30}$.
This effect of Lorentz invariance restoration is illustrated in
figure \ref{Fig:radius_growing}, too.

\item The same reasoning is true for supersymmetry, the D6-branes
can break the initial 16 closed string supersymmetries down to any
fraction or even completely. For the one-loop string amplitude, a
complete breaking results in non-vanishing NS-NS tadpoles,
generally in our case in one dilaton tadpole and nine radion
tadpoles. It has been demonstrated that generally, NS-NS tadpoles
do not indicate that the theory is inconsistent but rather that we
perturb around the wrong spacetime vacuum \cite{Fischler:1986ci,
Fischler:1986tb,Polchinski:1987tu, Angelantonj:1999qg,
Angelantonj:2002ct, Dudas:2004nd}. In order to correct this and
redefine a background with vanishing NS-NS tadpole, the
Fischler-Susskind mechanism (which has been invented for the
dilaton tadpole in the bosonic string) should be applied
\cite{Fischler:1986ci, Fischler:1986tb}, but unfortunately there
is no utilizable formulation for the NSR superstring so far. An
alternative approach is to include the tadpole in the effective
field theory equations of motion and then search for a new
classical background \cite{Dudas:2000ff,
Blumenhagen:2000dc,Dudas:2003wp}. This could allow one to really
see the temporal and spatial evolution of the different closed
string moduli, this would indeed be desirable for our model. But
one has to be aware of the fact that this procedure only cures the
problem of the tadpole in leading order (classically). Moreover,
if the redefined geometry is highly curved, then the non-linear
sigma model on the worldsheet cannot be solved exactly anymore.

The NS-NS tadpoles can be written as partial derivatives of a
scalar potential which can be derived directly from the
Born-Infeld action \cite{Fradkin:1985qd, Leigh:1989jq}, too. This
scalar potential has been interpreted in the past as a dynamical
potential for the scalar moduli on which it depends
\cite{Blumenhagen:2001te, Blumenhagen:2001mb, Burgess:2001vr,
Blumenhagen:2002ua}. This step is not trivial as the potential is
actually derived for a static brane configuration, in other words,
it is derived from a perturbative CFT computation which actually
requires a constant spacetime background, thus one has to be
careful with interpretation. Furthermore, the Born-Infeld
potential includes all orders in $\alpha'$, but not in the string
coupling, it arises already at the open string tree-level. In the
later discussion in chapter \ref{Sec:scalar_ptential_gen}, we will
derive also the string one-loop quantum corrections (including all
orders in $\alpha'$). One also has to carefully check if some
bifundamental scalars during the evolution become tachyonic which
is generally possible for non-supersymmetric models.


\item The most urgent question will be: can we still define our
string theoretical model in a consistent way if the intersecting
1-cycles of different (3+1)-dimensional massless fermions are not
always parallel on every $T^3$, i.e. we globally conserve only
(2+1)- or (1+1)-Poincare invariance\footnote{The case of
(0+1)-dimensional Poincare invariance breaking might be
problematic in this approach due to the use of light cone gauge.}.
At the level of the one-loop amplitude, this translates into the
question if the R-R tadpole still can be cancelled for a general
wrapping, as this would imply an anomaly free effective gauge
theory. The R-R tadpole condition will be derived in chapter
\ref{Sec:RRtp_genchapter} and from the result it is indeed
possible. This can be understood easily from the fact that the
R-R-tadpole can be expressed purely homologically. An explicit
example is discussed in detail in chapter \ref{Sec:Example}.

\item In the discussed limit that all intersection 1-cycles are
parallel on each $T^3$,  we restore the factorization of
10-dimensional spacetime (\ref{eq:alter_spactime_ansatz}) and
thereby global (3+1)-dimensional Poincare invariance. The internal
space can be split from the external space in a unique way for all
particles simultaneously. The external space then is given by the
longitudinal direction to the 1-cycles, the internal space by the
transverse (orthogonal) plane. But how shall we do this splitting
in the case that the intersection 1-cycles are not all parallel?
To understand this, we take another look at figure
\ref{Fig:radius_growing}. In the case on the left hand side, the
radii all are of a similar size, a unique external 6-dimensional
space cannot be defined. The picture therefore is intrinsically
10-dimensional and cannot be described by a factorization of
spacetime as (\ref{eq:alter_spactime_ansatz}). Whereas if one
radius is much bigger than the two other ones, like on the right
hand side of figure \ref{Fig:radius_growing}, then the different
transversal planes all nearly coincide with the closed string
plane orthogonal to the growing radius. In the case of a complete
decompactification along this growing direction, they all do
exactly coincide. This leads us to the apparent conjecture that
the closed string moduli actually determine the internal
6-dimensional space, and not the 1-dimensional intersections
between the branes themselves. If one direction within the $T^3$
is very much bigger than the two orthogonal ones, then these
orthogonal ones span the internal space. By the use of this plane,
we can immediately get the 4-dimensional massless fermion
spectrum.

\item One of the most direct effects of those solutions breaking
Poincare invariance further down from (3+1) dimensions, is that
Yukawa couplings obtain a small spatial dependence at the
tree-level. This can be understood by the following argument: the
size of the Yukawa couplings on the tree level is determined by
the area of the triangle between the intersection points of the
three relevant particles on the $T^2$, see \cite{Cremades:2003qj, Cvetic:2003ch}.
In the new picture on the $T^3$, this statement will still be
valid in the plane which defines the internal space (see last
point). If we move this plane along the large direction of the
$T^3$, then the chiral spectrum does not change (it is
topological), but the distance between the particles does change.
This implies that also the area of the triangle and therefore the
Yukawa couplings change. In the late time picture, this locally is
of course an effect which will be proportional to the ratio
between the biggest and the small radii (smaller than $10^{-30}$
at the present state), but for the early universe it might have
many new consequences which have to be considered in future work.
Depending on how large the compactification scale of the large
dimension compared to the visible universe actually is, there
might be even visible phenomenological consequences today on a
cosmological scale.

\item The first goal to see if the present construction can work
is to obtain directions in the scalar moduli potential, along
which certain radii grow very much compared to the other two ones
on a particular $T^3$. This problem is best formulated using shape
and volume moduli, as will be defined in section
\ref{Sec:kineticTerms}. One shape modulus (of the two) per $T^3$
should have a runaway potential, the other one should be
stabilized dynamically. The overall volume of the $T^9$ in the
late time picture should scale like the size of the present
universe (an acceleration with a very small effective cosmological
constant) and the relative volume ratios between the three
different $T^3$ should have been stabilized much earlier in order
to account for the isotropy of the present universe. For the very
early cosmology (like before or during inflation), the picture
could be very different, with many moduli not yet being
stabilized. In the end it will be a very quantitative question if
all of this is possible.
\end{enumerate}
There are many more features of this construction which have to be
thought through, we will address some of them in sections
\ref{Sec:Example} and \ref{Sec:Conclusions}.

\vspace{12pt} The organization of the paper is as follows. The
main sections \ref{Sec:fluxquant} to \ref{Sec:oneloop}
deal with the precise construction of the discussed orientifold
model, they are of rather technical nature and can be skipped at
first reading. In section \ref{Sec:fluxquant}, we will derive the
F-flux quantization of D9-branes in type IIB string theory in
detail for an abelian and non-abelian flux. The results will be
related to the T-dual picture of intersecting D6-branes in section
\ref{Sec:branes_at_angles}. Also the extension to tilted 3-tori
(which correspond to an additional discrete NS-NS 2-form flux)
will be discussed in section \ref{Sec:nsns2formflux}. The one loop
amplitude and the R-R tadpole condition will be derived in section
\ref{Sec:RRtp_genchapter}. Subsequently, we shortly derive the
supersymmetry conditions and discuss conditions for the absence of
tachyons in section \ref{Sec:SUSYcond}.

The scalar moduli potential will be discussed in section
\ref{Sec:scalar_ptential_gen}, including both the tree level
potential in section \ref{Sec:tree_level_pot} and the one-loop
potential in section \ref{Sec:oneloopPotential} together with a
discussion of the kinetic terms in Einstein frame in section
\ref{Sec:kineticTerms}.

Finally, one concrete toy model fulfilling the R-R tadpole
cancellation condition will be investigated in detail in section
\ref{Sec:Example}. The conclusions and prospects are stated in
chapter \ref{Sec:Conclusions}.
\newpage
\section{Generalized magnetic flux quantization on D9-branes}\label{Sec:fluxquant}
At first, we will consider the simple case of just one D9-brane on
a $T^9=T^3\times T^3 \times T^3$, carrying a $U(1)$ gauge field. Afterwards we generalize to the
non-abelian $U(N)$ case of $N$ D9-branes on top of each other.
\subsection{The Abelian case}\label{Sec:D9branes_abelian}
For simplicity, we define a right-handed coordinate system $X_i$,
$Y_i$ and $Z_i$ on every one of the $T_{i}^3$ ($i=1,2,3$).
A general constant magnetic $U(1)$ flux can then be written as
\begin{equation}\label{eq:Fab_form}
F_{ab}=\bigotimes_{i=1,2,3}\left(\begin{array}{ccc} 0 & -B_z^i &
B_y^i \\ B_z^i & 0 & -B_x^i \\-B_y^i & B_x^i &0
\end{array}\right)\ ,
\end{equation}
where $a,b\in\{1,\ldots,9\}$. Of course, by a orthogonal coordinate
transformation in a non-compact space we can transform this matrix
into one with only one non-vanishing component of $B$, i.e.
\begin{equation}
\bigotimes_{i=1,2,3}\left(\begin{array}{ccc} 0 & -B^i & 0 \\ B^i& 0 & 0 \\0&0 &0
\end{array}\right)\ . \nonumber
\end{equation}
This just means that a B-field always can be represented as a
vector in three dimensions and we can use a coordinate system where
one coordinate axis corresponds to the direction of the B-field.

However, notice that our present case is more difficult because
such a transformation has to be made simultaneously for all stacks
of branes. If for different stacks the direction of the B-vector
is not identical, we cannot transform the whole system of branes
into a coordinate system with only one component of $B$. Secondly,
our 3-dimensional space is compact and we cannot a priori ensure
that the necessary coordinate transformation respects the lattice
symmetry.

In order to be able to make a connection with the T-dual picture
containing only intersecting D6-branes and vanishing B-fluxes, we
will assume that there is one direction in every $T^3$ common to
all branes, along which the B-flux is vanishing. This then will be
the direction along which we perform T-duality. We make the
explicit choice
\begin{equation}\label{eq:Bzvanishing}
B_z^i\equiv 0\qquad\text{for } i=1,2,3.
\end{equation}
This is a necessary technical requirement in order to ensure that
starting from D9-branes with F-fluxes, in the D6-brane picture
after T-duality there is no F-flux remaining.

In other words, we insist on a unique direction
within every $T^3$ along which we can perform a T-duality and
always get D6-branes with vanishing F-flux from the original
D9-branes. If we do not make this assumption, then this might not
be possible, we clearly want to avoid such a complication at this
time. Choosing (\ref{eq:Bzvanishing}), it is always possible to
achieve our goal by performing the T-duality in the
$Z_i$-direction.

In the familiar case of F-fluxes on a $T^6$, the flux is quantized
and the constant magnetic field on the side of D9-branes can
directly be related to the angle on the side of D6-branes via the
prominent equation
\begin{equation}
2\pi \alpha'B=\tan \varphi\ . \nonumber
\end{equation}
One expects a similar result in the generalized case of equation
(\ref{eq:Fab_form}). This will be derived in this section. At
first we will derive the flux quantization and subsequently make
the connection with the branes at angles side. One has to find a
vector potential $A$ that by the equation
\begin{equation}
F_{ab}=\partial_a A_b-\partial_b A_a
\end{equation}
reproduces the field strength (\ref{eq:Fab_form}) with
(\ref{eq:Bzvanishing}). Such a vector potential for instance is
given by the simple choice
\begin{equation}
A_x^i=-B_y^i Z_i,\qquad A_y^i=B_x^i Z_i,\qquad A_z^i=0\ .
\end{equation}
The vector potential $A$, translated by one of the three
fundamental lengths on the 3-torus, has to be gauge equivalent to
the original vector potential (again for i=1,2,3), i.e.
\begin{align}
&A_x^i(X_i,Y_i,Z_i)\sim A_x^i(X_i+2\pi L_x^i,Y_i,Z_i)\sim A_x^i(X_i,Y_i+2\pi L_y^i,Z_i)\sim A_x^i(X_i,Y_i,Z_i+2\pi L_z^i),\nonumber\\
&A_y^i(X_i,Y_i,Z_i)\sim A_y^i(X_i+2\pi L_x^i,Y_i,Z_i)\sim A_y^i(X_i,Y_i+2\pi L_y^i,Z_i)\sim A_y^i(X_i,Y_i,Z_i+2\pi L_z^i),\nonumber\\
&A_z^i(X_i,Y_i,Z_i)\sim A_z^i(X_i+2\pi L_x^i,Y_i,Z_i)\sim
A_z^i(X_i,Y_i+2\pi L_y^i,Z_i)\sim A_z^i(X_i,Y_i,Z_i+2\pi
L_z^i).\nonumber\\
\end{align}
Therefore, a specific gauge transformation $\chi_i(X_i,Y_i,Z_i)$ has
to exist that acts as an equivalence relation. For our case this
requires
\begin{equation}
\partial_{X_i} \chi_i=-B_y^i 2\pi L_z^i,\qquad \partial_{Y_i} \chi_i=B_x^i 2\pi L_z^i, \qquad \partial_{Z_i} \chi_i=0.
\end{equation}
The gauge transformation then is given by
\begin{equation}\label{eq:chi_gtrafo}
\chi_i=-B_y^i 2\pi L_z^i X_i+B_x^i 2 \pi L_z^i Y_i + const.
\end{equation}
The gauge parameter $U_i(X_i,Y_i,Z_i) \equiv \exp(i \chi_i)$ has
to stay constant if one translates by any lattice vector, i.e.
\begin{equation}
U(X_i,Y_i,Z_i)=U(X_i+2\pi L_x^i,Y_i,Z_i)=U(X_i,Y_i+2\pi
L_y^i,Z_i)=U(X_i,Y_i,Z_i+2\pi L_z^i)\ .
\end{equation}
This induces
\begin{align}
&\chi_i(X_i+2\pi L_x^i,Y_i,Z_i)-\chi_i(X_i,Y_i,Z_i)=-2\pi m_i,\nonumber\\
&\chi_i(X_i,Y_i+2\pi L_y^i,Z_i)-\chi_i(X_i,Y_i,Z_i)=2\pi n_i,\\
&\chi_i(X_i,Y_i,Z_i+2\pi L_z^i)-\chi_i(X_i,Y_i,Z_i)=2\pi r_i,\nonumber
\end{align}
with $m_i, n_i, r_i \in \mathbb{Z}$. The fact that $\chi_i$ does not
depend on $Z_i$ forces that $r_i \equiv 0$. Writing down the first
two equations explicitly using (\ref{eq:chi_gtrafo}) gives us the
quantization condition for the magnetic field components
\begin{align}\label{eq:bflux3dim_quant}
&B_x^i=\frac{n_i}{2\pi L_y^i L_z^i}\ ,\\
&B_y^i=\frac{m_i}{2\pi L_x^i L_z^i}\ .\nonumber
\end{align}
The 3-dimensional $B$-vector on every 3-torus thus is given by
\begin{equation}
\vec{B}_i=\frac{1}{2\pi}\left(\begin{array}{c} n_i/(L_y^i L_z^i)\\m_i/(L_x^i L_z^i) \\ 0 \end{array}\right)
\qquad \text{with}\qquad n_i, m_i \in \mathbb{Z}.
\end{equation}
It is easy to see the quantization of the flux through the three
different planes spanned by two of the three fundamental $T^3$
coordinate axes using the definition
\begin{equation}
\Phi=\int_\mathcal{S}\vec{n}\cdot\vec{B}da\ .
\end{equation}
One simply obtains:
\begin{equation}\label{eq:flux_quant_abelian}
\Phi_{xy}^i=0, \qquad \Phi_{yz}^i=2\pi n_i, \qquad \Phi_{xz}^i=2\pi m_i \qquad \text{with}\qquad n_i, m_i \in \mathbb{Z}.
\end{equation}
\subsection{The U(N) case}\label{Sec:D9branes_nonabelian}
In this section, we will generalize the result
(\ref{eq:bflux3dim_quant}) to the non-abelian case of $N$
D9-branes in a slightly more mathematical language using fiber
bundles, an introduction to this approach can be found in
\cite{Hashimoto:1997gm, Taylor:1997dy}. In order to describe a
$U(N)$ bundle on a compact space (being here the direct product of
three $T^3$), one has to chose coordinate patches on the manifold.
It can be characterized by the so-called transition functions
$\Omega_j$, making the transition between different patches. In
the toroidal case, one transition function per compact direction
$j$ is sufficient. These functions in general can depend on all
the other compact dimensions (but not on the one for that it makes
the transition), i.e. on one $T^3$ we have to regard
\begin{equation}
\Omega_1(y,z),\ \Omega_2(x,z),\ \Omega_3(x,y).
\end{equation}
To make sure that these functions describe a well-defined bundle,
they have to fulfill cocycle conditions. These ensure that for
every possible path in the $T^3$ ending at the starting point,
being described by a chain of different transitions, the result is
the identity. For each $T^3$ one obtains three conditions (one
for every pair of dimensions)
\begin{align}\label{eq:Omegas_map}
\Omega_1(y, 2\pi L_z)\ \Omega_3(0,y)\ \Omega_1^{-1}(y,0)\ \Omega_3^{-1}(2\pi L_x,y)&=\openone \nonumber\\
\Omega_1(2\pi L_y,z)\ \Omega_2(0,z)\ \Omega_1^{-1}(0,z)\ \Omega_2^{-1}(2\pi L_x,z)&=\openone\\
\Omega_2(x,2\pi L_z)\ \Omega_3(x,0)\ \Omega_2^{-1}(x,0)\ \Omega_3^{-1}(x,2\pi L_y)&=\openone
\nonumber
\end{align}
On general tori, $U(N)$ bundles are completely classified by a first Chern number
per pair of dimensions, i.e. in our case
\begin{align}\label{eq:def_chern_number}
&C_1^{xy}\equiv\int  dx\, dy\ c_1^{xy}=\frac{1}{2\pi}\int dx\, dy\ \text{Tr} \mathcal{F}=r \in \mathbb{Z},\\
&C_1^{yz}=\frac{1}{2\pi}\int dy\, dz\ \text{Tr} \mathcal{F}=n \in \mathbb{Z},\qquad
C_1^{xz}=\frac{1}{2\pi}\int dx\, dz\ \text{Tr} \mathcal{F}=m \in \mathbb{Z}. \nonumber
\end{align}
In this equation, $\mathcal{F}$ stands for a general non-abelian
magnetic flux. This corresponds exactly to the flux quantization
condition (\ref{eq:flux_quant_abelian}) of the preceding section.
At this point we restrict again to $r=0$ in order to have
a good direction for performing T-duality afterwards. The gauge
group $U(N)$ can be decomposed into its abelian and non-abelian
components, $U(N)=(U(1)\times SU(N))/\mathbb{Z}_N$. Due to the
trace which is taken in equation (\ref{eq:def_chern_number}), only the abelian
part plays an important role for the flux quantization and we
obtain for the $U(1)$ part of $\mathcal{F}$
\begin{equation}\label{eq:abelian_part_flux}
\mathcal{F}^{U(1)}_{xy}=\mathbf{0}, \qquad \qquad\mathcal{F}^{U(1)}_{yz}=\frac{n}{2\pi N L_y L_z} \openone ,
\qquad \qquad \mathcal{F}^{U(1)}_{xz}=\frac{m}{2\pi N L_x L_z} \openone .
\end{equation}
In the following, we will see that we can indeed find concrete
transition functions $\Omega$ and a connection $A$ describing
such a flux for a general choice of $n$ and $m.$ One set of
transition functions fulfilling the cocycle conditions
(\ref{eq:def_chern_number}) is given by
\begin{equation}
\Omega_1=e^{2\pi i \frac{z}{2\pi L_z}\mathbf{T}_{xz}} \mathbf{V}^{m},
\qquad \qquad \Omega_2=e^{2\pi i \frac{z}{2\pi L_z}\mathbf{T}_{yz}} \mathbf{V}^{n},
\qquad \qquad\Omega_3=\openone ,
\end{equation}
where
\begin{align}
&\mathbf{T}_{yz}=\text{diag}\left(\frac{n-\tilde n}{N},\ldots, \frac{n-\tilde n}{N},\frac{n-\tilde n}{N}+1,\ldots,\frac{n-\tilde n}{N}+1\right),\nonumber\\
&\mathbf{T}_{xz}=\text{diag}\left(\frac{m-\tilde m}{N},\ldots, \frac{m-\tilde m}{N},\frac{m-\tilde m}{N}+1,\ldots,\frac{m-\tilde m}{N}+1\right),\\
&\mathbf{V}=\left(\begin{array}{ccccc} 0 & 1 & & & \\ & 0& 1 & &\\ & & 0& \ddots & \\ & & & \ddots &  1\\ 1 & & &&0
\end{array}\right),\qquad \tilde n\equiv n \left(\text{mod}\ N\right),\qquad \tilde m\equiv m\left(\text{mod}\ N\right). \nonumber
\end{align}
A connection on the bundle has to fulfill the boundary conditions
\begin{equation}\label{eq:bcons_bundle}
A'=\Omega \cdot A \cdot \Omega^{-1}-i\ d\Omega \cdot \Omega^{-1}.
\end{equation}
The prime denotes a transition along any of the three directions
$L_x$, $L_y$ or $L_z$. A constant curvature background that fulfills all the nine equations (\ref{eq:bcons_bundle}) is given by
\begin{align}
    &A_1^0=A_2^0=0,\\ &A_3^0=\mathcal{F}^{U(1)}_{yz}\ y+ \mathcal{F}^{U(1)}_{xz}\ x
    +\frac{4\pi}{L_z}\text{diag}\left(0,1/N,\ldots ,(N-1)/N\right)\nonumber
\end{align}
Therefore, the abelian part of the flux indeed is given by
equation (\ref{eq:abelian_part_flux}). The only important
difference as compared to the purely abelian case is an additional
factor $N$ for the gauge group $U(N)$ as compared to the numerator
of the two equations (\ref{eq:bflux3dim_quant}).

In the subsequent chapter we will see what the flux quantization
in the picture of $N$ D9-branes carrying a gauge group $U(N)$
implies for the T-dual picture of D6-branes. One technicality is
still needed to understand this connection.
 The boundary conditions for
open strings that end on the D9-brane carrying the magnetic
field can be written as
\begin{equation}\label{eq:gen_bcond_openstring}
\partial_\sigma X_a-2\pi \alpha' {F}^{U(1)}_{ab}\partial_\tau X^b=0, \qquad \sigma=0, \pi\ .
\end{equation}
We again simplify notation by defining
\begin{align}\label{eq:B_simp_notation}
&\mathcal{B}_x\equiv 2\pi \alpha' B_x=-2\pi \alpha'{F}^{U(1)}_{yz}=-\frac{\alpha' n}{N L_y L_z} ,\\
&\mathcal{B}_y\equiv 2\pi \alpha' B_y=2\pi \alpha'{F}^{U(1)}_{xz}=\frac{\alpha' m}{N L_x L_z}.\nonumber
\end{align}
For the field strength (\ref{eq:abelian_part_flux}), equation
(\ref{eq:gen_bcond_openstring}) explicitly can be written as
\begin{align}\label{eq:bc_xyz}
&\partial_\sigma X_i - \mathcal{B}_y^i \partial_\tau Z_i=0\ , \nonumber\\
&\partial_\sigma Y_i + \mathcal{B}_x^i\partial_\tau Z_i=0\ ,\\
&\partial_\sigma Z_i + \mathcal{B}_y^i\partial_\tau X_i-\mathcal{B}_x^i \partial_\tau Y_i=0\ ,\nonumber
\end{align}
for $i=1,2,3$. Passing to the worldsheet light cone derivatives defined by
\begin{align}
&\partial_+\equiv \frac{1}{2}\left(\partial_\tau+\partial_\sigma\right)\ ,\\
&\partial_-\equiv \frac{1}{2}\left(\partial_\tau-\partial_\sigma\right)\ ,\nonumber
\end{align}
one can rewrite the boundary conditions as follows
\begin{equation}\label{eq:lc3dim_Fab_matrix}
\partial_+ \left(\begin{array}{c} X_i\\ Y_i \\ Z_i \end{array}\right)
=\left(\begin{array}{ccc} \frac{1+{\mathcal{B}_x^i}^2-{\mathcal{B}_y^i}^2}{1+{\mathcal{B}_x^i}^2+{\mathcal{B}_y^i}^2}
& \frac{2{\mathcal{B}_x^i}{\mathcal{B}_y^i}}{1+{\mathcal{B}_x^i}^2+{\mathcal{B}_y^i}^2} & \frac{2{\mathcal{B}_y^i}}{1+{\mathcal{B}_x^i}^2+{\mathcal{B}_y^i}^2}\\
\frac{2{\mathcal{B}_x^i}{\mathcal{B}_y^i}}{1+{\mathcal{B}_x^i}^2+{\mathcal{B}_y^i}^2}& \frac{1-{\mathcal{B}_x^i}^2+{\mathcal{B}_y^i}^2}{1+{\mathcal{B}_x^i}^2+{\mathcal{B}_y^i}^2}
& \frac{-2{\mathcal{B}_x^i}}{1+{\mathcal{B}_x^i}^2+{\mathcal{B}_y^i}^2}\\
\frac{-2{\mathcal{B}_y^i}}{1+{\mathcal{B}_x^i}^2+{\mathcal{B}_y^i}^2}&\frac{2{\mathcal{B}_x^i}}{1+{\mathcal{B}_x^i}^2+{\mathcal{B}_y^i}^2}
&\frac{1-{\mathcal{B}_x^i}^2-{\mathcal{B}_y^i}^2}{1+{\mathcal{B}_x^i}^2+{\mathcal{B}_y^i}^2}
\end{array}\right)\partial_- \left(\begin{array}{c} X_i\\ Y_i \\ Z_i \end{array}\right),
\end{equation}
for $i=1,2,3.$ This matrix will be needed in the next section for
a comparison with the corresponding matrix in the T-dual picture
of branes at angles.

\section{Generalized intersecting D6-branes}\label{Sec:branes_at_angles}
In this chapter it is shown that a magnetic F-flux on D9-branes
corresponds to T-dual D6-branes at general angles (within every
$T^3$), if one performs a T-duality along one of the three axes of
every $T^3$. This is a direct generalization of the common
picture of intersecting D6-branes on a $T^6$.

A general rotation $\mathbf{O}$ in three dimensions can be
described by the three Euler angles, $\phi$, $\theta$ and $\psi$,
i.e. by the matrix
\begin{equation}
\mathbf{O}=\mathbf{B}\mathbf{C}\mathbf{D}
\end{equation}
with the three subsequently performed rotations along the three Euler angles
\begin{equation}
\mathbf{B}= \left( \begin {array}{ccc} \cos \psi &\sin \psi &0\\\noalign{\medskip}-\sin \psi &\cos
\psi &0\\\noalign{\medskip}0&0&1\end {array} \right),
\ \mathbf{C}= \left( \begin {array}{ccc} 1&0&0\\\noalign{\medskip}0&\cos
\theta &\sin \theta \\\noalign{\medskip}0&-\sin
 \theta &\cos \theta \end {array}
 \right),
\ \mathbf{D}= \left( \begin {array}{ccc} \cos \phi &\sin \phi
 &0\\\noalign{\medskip}-\sin \phi &\cos \phi &0\\\noalign{\medskip}0&0&1\end {array} \right).
\end{equation}
This leads to the general rotation
\begin{equation}\label{eq:gen_rotationmatrix}
\mathbf{O}= \left( \begin {array}{ccc} \cos \psi \cos \phi
 -\sin \psi \cos \theta \sin
 \phi &\cos \psi \sin \phi +\sin \psi \cos \theta \cos
 \phi &\sin \psi \sin \theta
 \\\noalign{\medskip}-\sin \psi \cos \phi -\cos \psi \cos \theta \sin
 \phi &-\sin \psi \sin \phi +\cos \psi \cos \theta \cos
 \phi &\cos \psi \sin \theta
 \\\noalign{\medskip}\sin \theta \sin \phi &-\sin \theta \cos \phi &
\cos \theta \end {array} \right)
\end{equation}
In the frame where a D6-brane (described by a 2-cycle on every
$T^3$) spans the two coordinate axes $X_i'$ and $Y_i'$, the
boundary conditions on one $T^3$ are simply given by two Neumann
(along $X_i'$ and $Y_i'$) and one Dirichlet (along $Z_i'$)
boundary conditions, i.e.
\begin{equation}\label{eq:ND_bc_ang1}
\partial_+ \left(\begin{array}{c} X_i'\\ Y_i' \\ Z_i' \end{array}\right)
=\left(\begin{array}{ccc} 1& 0& 0\\0&1&0\\0&0&-1\end{array}\right)
\partial_- \left(\begin{array}{c} X_i'\\ Y_i' \\ Z_i' \end{array}\right).
\end{equation}
If a rotation matrix $\mathbf{O}$ takes the coordinates $X_i$,
$Y_i$, $\tilde Z_i$ to the primed ones,
\begin{equation}
\left(\begin{array}{c} X_i'\\ Y_i' \\ Z_i' \end{array}\right)
=\mathbf{O} \left(\begin{array}{c} X_i\\ Y_i \\\tilde Z_i \end{array}\right)\ ,
\end{equation}
then (\ref{eq:ND_bc_ang1}) can be rewritten for the unprimed
coordinates and one obtains mixed Neumann/Dirichlet boundary conditions
\begin{equation}\label{eq:ND_bc_ang2}
\partial_+ \left(\begin{array}{c} X_i\\ Y_i \\ \tilde Z_i \end{array}\right)
=\mathbf{O}^{-1} \left(\begin{array}{ccc} 1& 0& 0\\0&1&0\\0&0&-1\end{array}\right) \mathbf{O}
\ \partial_- \left(\begin{array}{c} X_i\\ Y_i \\ \tilde Z_i \end{array}\right).
\end{equation}
In order to compare with the last section, one has to
include another matrix on the right hand side of this equation,
taking care of the T-duality transformation on the coordinate,
taking the coordinate $\tilde Z_i$ into $Z_i$, such that the
boundary conditions become
\begin{equation}\label{eq:ND_bc_ang3}
\partial_+ \left(\begin{array}{c} X_i\\ Y_i \\ Z_i \end{array}\right)
=\mathbf{O}^{-1} \left(\begin{array}{ccc} 1& 0& 0\\0&1&0\\0&0&-1\end{array}\right) \mathbf{O}
\left(\begin{array}{ccc} 1& 0& 0\\0&1&0\\0&0&-1\end{array}\right) \partial_- \left(\begin{array}{c} X_i\\ Y_i \\ Z_i \end{array}\right).
\end{equation}
Multiplying all the matrices, one obtains
\begin{align}\label{eq:ND_bc_ang4}
&\partial_+ \left(\begin{array}{c} X_i\\ Y_i \\ Z_i \end{array}\right)
=\mathbf{G}\ \partial_- \left(\begin{array}{c} X_i\\ Y_i \\ Z_i \end{array}\right)\ ,\\
 &\small\mathbf{G}\equiv\left( \begin {array}{ccc} 2\cos^2 \theta+2\cos^2\phi-2\cos^2\phi
 \cos^2\theta-1 &2 \sin\theta \sin\phi \cos \phi&2\sin \theta \sin \phi \cos \theta \\
\noalign{\medskip}2\sin \theta \sin \phi \cos \phi &2\, \cos^2 \phi
 \cos^2 \theta
-2 \cos^2\phi+1 &-2\cos \theta \sin \theta \cos \phi
\\\noalign{\medskip}-2\sin \theta \sin \phi
 \cos \theta  &2\cos \theta
\sin \theta \cos \phi &2 \cos^2 \theta-1\end {array} \right).\nonumber
\end{align}
This matrix does not depend anymore on the third angle $\psi$,
meaning that after T-duality one obtains a 2-dimensional compact
plane in every $T^3$, as expected. A rotation along $\psi$
corresponds to one within the plane. So, the choice of Euler
coordinates is indeed a very useful one. If we substitute the flux
quantization (\ref{eq:bflux3dim_quant}) together with
(\ref{eq:B_simp_notation}) into the matrix of equation
(\ref{eq:lc3dim_Fab_matrix}), one can compare the resulting matrix
element by element with $\mathbf{G}$. There is a one-to-one
correspondence between the two matrices if one assumes that
\begin{align}
&\phi_i=\arccos\left(\frac{n_i L_x^i}{\sqrt{{n_i}^2{L_x^i}^2+{m_i}^2
{L_y^i}^2}}\right)\\
&\theta_i=\arccos\left(\frac{L_x^i L_y^i L_z^i}{\sqrt{{L_x^i}^2
{L_y^i}^2 {L_z^i}^2+\frac{{n_i}^2}{N^2} {L_x^i}^2 {\alpha'}^2+\frac{{m_i}^2}{N^2}
{L_y^i}^2 {\alpha'}^2}}\right)\nonumber
\end{align}
Now we have to rewrite the two angles for the correct lattice length
in the D6-branes picture, $\tilde L_z^i=\alpha'/L_z^i$.
Furthermore, we will rewrite the general quantum numbers $n_i$ and
$m_i$ and also the number of D9-branes $N$ into a combination of
their coprime part $n_i'$, $m_i'$ and $p_i'$ and their greatest
common divisor, $q_i \equiv\gcd(n_i, m_i, N)$, such that
\begin{equation}\label{eq:umschreibung_nm1}
n_i=n_i' q_i\qquad\qquad m_i=m_i' q_i\qquad\qquad N= p_i' q_i\ .
\end{equation}
In our approach $N$ is positive for a D-brane, but it is also
possible to formally include the case $N<0$ and thus allow also
for anti-D-branes. This statement will be made more precise later.
Thereafter, we rewrite
$q_i$ for all $i$ in the following way
\begin{equation}
    q_i=\gcd\left(q_1,q_2,q_3\right) k_i=\tilde{N} k_i \qquad
\text{with}\qquad \tilde{N}\equiv \gcd\left(q_1,q_2,q_3\right).
\end{equation}
The $k_i$ and also $N$ are completely fixed by the third equation in
(\ref{eq:umschreibung_nm1}) as $N$ does not carry any torus index, i.e.
\begin{align}\label{eq:weitere_umdef}
&k_1=p_2' p_3',& &k_2=p_1' p_3',& &k_3=p_1' p_2',& &N=p_1' p_2' p_3' \tilde N,
\end{align}
if we assume that the $p_i'$ is a completely arbitrary integer,
$\tilde N$ is a positive integer. The two angles then take the new
form
\begin{align}\label{eq:anglesT9_endform}
&\phi_i=\arccos\left(\frac{n_i' L_x^i}{\sqrt{{n_i'}^2{L_x^i}^2+{m_i'}^2
{L_y^i}^2}}\right),\\
&\theta_i=\arccos\left(\frac{p_i'L_x^i L_y^i}{\sqrt{\left(p_i'\right)^2{L_x^i}^2
{L_y^i}^2+{n_i'}^2 {L_x^i}^2{(\tilde L_z^i)}^2+{m_i'}^2{L_y^i}^2 {(\tilde L_z^i)}^2}}\right).\nonumber
\end{align}
Interestingly, the positive integer $\tilde N$ drops out of the
two angles that now depend on nine integers $n_i'$, $m_i'$ and
$p_i'$ (for the three 3-tori $i$). They will have a very simple
geometrical interpretation as we will soon see.

First we are getting reminded of the usual case of D6-branes on
the $T^6$, see for instance \cite{Blumenhagen:2000wh}. The
analogue formula for a quantized angle there is simply given by
\begin{equation}\label{eq:std_angles}
    \tan \phi_i=\frac{m_i L_y^i}{n_i L_x^i}\nonumber
\end{equation}
and the two coprime integers $n_i$ and $m_i$ have the
interpretation of wrapping numbers on a 2-torus ($i=1,2,3$). A
reflection at the origin of only one $T$, i.e. the application of the map
\begin{equation}\label{eq:antibrane2d}
    n_i\rightarrow -n_i,\qquad\qquad m_i\rightarrow -m_i,
\qquad\qquad \text{for}\ i=1,2\ \text{or}\ 3,\nonumber
\end{equation}
has the interpretation of exchanging a brane with its
corresponding anti-brane. Geometrically, it can be understood as a
reversion of the orientation on one $T^2$ and by this also of the
total orientation on the $T^6$. From the sign of a certain brane
contribution in the R-R-tadpole equations, one can read off that
branes with a product $n_1 n_2 n_3<0$ are actually anti-D-branes,
whereas whose with $n_1 n_2 n_3\geq 0$ correspond to D-branes.
This can be seen by a comparison of this sign to that of the
orientifold brane (that has a negative R-charge) within the same
equation.

A similar interpretation holds in our present case: a compact
two-dimensional plane within a $T^3$ can be uniquely specified by three
integers that are mutually coprime. These integers have the
interpretation of a vector in the reciprocal lattice of the
$T^3$, this is demonstrated in appendix \ref{AppSec:topology}.
Thus we can describe any D6-brane on one $T^3_i$ by a vector
\begin{equation}
\vec n_i=m_i'\mathbf{e}_1^{(i)*}+n_i'\mathbf{e}_2^{(i)*} +p_i'\mathbf{e}_3^{(i)*} ,
\end{equation}
where the reciprocal lattice vectors $\mathbf{e}_j^{(i)*}$ are defined
by equation (\ref{eq:reciprocal_torus}) in appendix
\ref{AppSec:Basis3Tori}.
 Simply speaking, this vector is similar to a normal vector, with
the only difference that is is defined on the reciprocal lattice
and not on the lattice itself. The direction of this vector gives the
orientation, reversing the orientation on one $T^3$ by applying the map
\begin{equation}
n'_i\rightarrow -n'_i,\qquad\qquad m'_i\rightarrow
-m'_i,\qquad\qquad p'_i\rightarrow -p'_i, \qquad\qquad \text{for}\
i=1,2\ \text{or}\ 3,
\end{equation}
also here means that we make the transition from a brane to its
corresponding anti-brane. Here the tadpole equation
(\ref{eq:Rtadpole1}) which will be derived in section \ref{Sec:rtp_expl}
reveals, that wrapping numbers $p_1' p_2' p_3'\geq 0$ correspond
to D-branes and wrapping numbers $p_1' p_2' p_3'< 0$ to
anti-D-branes.

The additional positive integer $\tilde N$ on the other hand has
the interpretation of the stacksize of D6-branes (or
anti-D6-branes) in the type IIA picture (branes on top of each
other). This can be understood by the fact that the angles do not
depend on $\tilde N$ but it is still present as a factor in the
number of original D9-branes (the last equation in
(\ref{eq:weitere_umdef})).

The wrapping numbers of the D6-branes shown in the lower part of
figure \ref{Fig:all_tori_old_new} are stated as an instructive
example in table \ref{tab:wrap_picture}.
\begin{table}
\centering
\sloppy
\renewcommand{\arraystretch}{1.3}
\begin{tabular}{|l|ccc|ccc|ccc|}

  \hline
   & $m_1'$& $n_1'$ & $p_1'$ & $m_2'$& $n_2'$ & $p_2'$ & $m_3'$& $n_3'$ & $p_3'$\\
  \hline\hline
    brane $a$ (green)& -1 & 0 & -2 & -1 & -1 & 1 & 1 & 1 &0\\
\hline
     brane $b$ (blue)& 0 & 0 & -1 & -1 & 0 & -2 & -1 & 1 & 2\\
\hline
\end{tabular}
\caption{The wrapping numbers of the model shown in the lower part
of figure \ref{Fig:all_tori_old_new}.} \label{tab:wrap_picture}
\end{table}

\subsection{Introducing Tilted $T^3$}\label{Sec:nsns2formflux}
In the picture of D9-branes on the $T^6$, it is possible to switch
on an additional background NS-NS 2-form flux $\mathcal{B}$, such
that the total magnetic flux is given by $\mathcal{B}+2\pi
\alpha'\mathcal{F}$ \cite{Kakushadze:1998bw, Angelantonj:1999jh}.
By applying T-duality, the $\mathcal{B}$-flux translates into
tilted $T^2$ in the picture of intersecting D6-branes
\cite{Blumenhagen:2000ea}. The $\mathcal{B}$-flux for the
D9-branes is discrete, and on the side of branes at angles this is
manifest in the fact that the geometric part of the orientifold
projection $\Omega R$ allows only for certain discrete tilts of
the $T^2$. This is because the $R$ projection has to map any
lattice point of the torus onto another lattice point in order to
respect the torus symmetry. Therefore, this fixes the angle
between the two elementary radii in the type IIA picture.
\begin{figure}[t]
\begin{center}
\begin{overpic}[scale=.8,unit=1mm]%
      {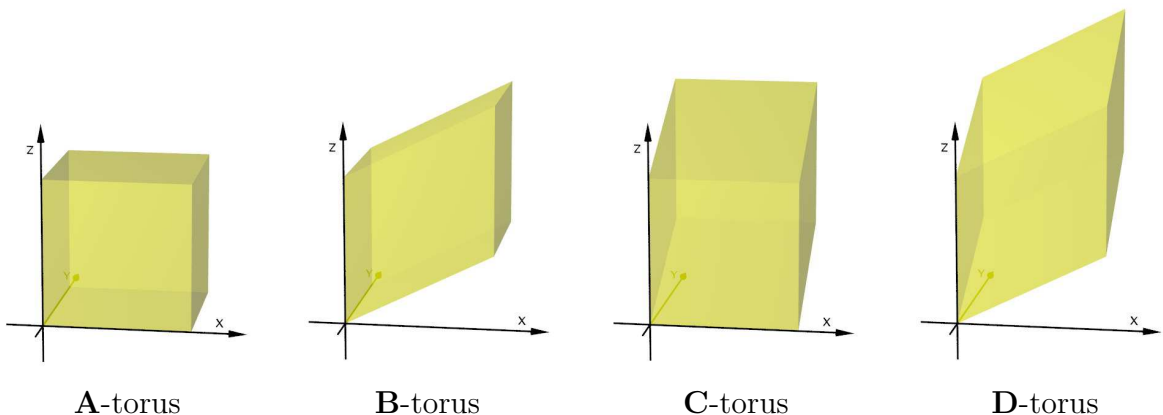}
  \put(13,0){$\bf A$-torus}
  \put(53,0){$\bf B$-torus}
  \put(94,0){$\bf C$-torus}
  \put(135,0){$\bf D$-torus}
\end{overpic}
\caption{The four different untilted and tilted $T^3$.} \label{Fig:different_tori}
\end{center}
\end{figure}
This fact also allows for a simple derivation of the quantization
of $\mathcal{B}$-flux for the $T^3$ which will be described in the
following. We assume that the tilted $T^3$ has a basis consisting
of the three fundamental lattice vectors $2\pi L_x \mathbf{e}_1$,
$2\pi L_y \mathbf{e}_2$ and $2\pi \tilde L_z \mathbf{e}_3$ with
\begin{equation}\label{eq:ansatzTiltedToriBasis}
\mathbf{e}_1=\left(\begin{array}{c} c_1\\ 0 \\ c_2\end{array}\right),\qquad
\mathbf{e}_2=\left(\begin{array}{c} 0\\ c_3 \\ c_4\end{array}\right),\qquad
\mathbf{e}_3=\left(\begin{array}{c} 0\\ 0 \\ c_5 \end{array}\right).
\end{equation}
The reciprocal basis, being defined by equation
(\ref{eq:reciprocal_torus}), then is given by
\begin{equation}
\mathbf{e}_1^*=\left(\begin{array}{c}1/c_1 \\ 0\\ 0 \end{array}\right),\
\mathbf{e}_2^*=\left(\begin{array}{c} 0\\ 1/c_3\\ 0 \end{array}\right),\
\mathbf{e}_3^*=\left(\begin{array}{c} - c_2/(c_1 c_5)\\-c_4/(c_3 c_5) \\ 1/c_5 \end{array}\right).
\end{equation}
The reflection $R$ reflects the $X_i^*$- and $Y_i^*$-coordinates
on every $T^3_i$. It maps a point of the reciprocal lattice onto
another one if and only if for any $m', n', p' \in \mathbb{Z}$
three arbitrary integers $u, v, w \in \mathbb{Z}$ can be found such
that the following equation holds
\begin{equation}
R\left(m'\mathbf{e}_1^*+n'\mathbf{e}_2^* +p'\mathbf{e}_3^*\right)=
u\mathbf{e}_1^*+v\mathbf{e}_2^* +w\mathbf{e}_3^* .
\end{equation}
This implies
\begin{equation}
u_i=-m_i'+2p_i'\frac{L_x c_2}{\tilde L_z c_5},
\qquad\qquad v_i=-n_i'+2p_i'\frac{L_y c_4}{\tilde L_z c_5}, \qquad\qquad w_i=p_i',
\end{equation}
and the right hand side of these equations will generally only be
integer if
\begin{equation}\nonumber
2\frac{L_x c_2}{\tilde L_z c_5}\in \mathbb{Z}, \qquad\text{and}\qquad
2\frac{L_y c_4}{\tilde L_z c_5}\in \mathbb{Z}.
\end{equation}
We obtain four dissimilar possibilities
\begin{equation}\nonumber
c_2=b_1\frac{\tilde L_z}{L_x}{c_5},\qquad\qquad c_4=b_2\frac{\tilde L_z}{L_y}{c_5},
\end{equation}
where $b_1, b_2\in \{0,1/2\}$. This result is again in direct
analogy to the case on the $T^6$ \cite{Blumenhagen:2000ea}.
Therefore we call the case $(b_1,b_2)=(1/2,0)$ $\bf B$-torus, the
case $(b_1,b_2)=(0,1/2)$ $\bf C$-torus and the case
$(b_1,b_2)=(1/2,1/2)$ $\bf D$-torus. The untilted torus formally
is also included by $(b_1,b_2)=(0,0)$ and will be called $\bf
A$-torus. All these possibilities are depicted in figure
\ref{Fig:different_tori}. After normalizing the lattice vectors
according to $(\mathbf{e}_i)^2=1$, we obtain the final lattice
vectors given by equations (\ref{eq:bcd_torus}) and
(\ref{eq:bcd_torus_rec}), as stated in appendix
\ref{AppSec:Basis3Tori}.  There will be $12=2\times 6$ distinct
choices for basis of the whole $T^9=T^3\times T^3\times T^3$.

For simplicity in computations, it is possible to reexpress all
formulae which can be derived for the tilted tori formally like
for the untilted torus and vice versa \cite{Lust:2003ky}. In order
to do so, one has to find new `effective' wrapping numbers $\tilde
m'$, $\tilde n'$ and $\tilde p'$ for the tilted torus such that $R$
acts on them like on the untilted torus, i.e. $R\ (\tilde
m_i',\tilde n_i',\tilde p_i')=(-\tilde m_i',-\tilde n_i',\tilde
p_i')$. It is easy to see that this is just the case for
\begin{equation}
\tilde m_i'=m_i'+b_1 p_i', \qquad\qquad \tilde n_i'=n_i'+b_2 p_i', \qquad \qquad \tilde p_i'=p_i'.
\end{equation}

To state it clearly, the orientifold projection fixes two of the
three possible angles between the elementary radii to discrete
values. This means that one angle (the one between the $x$- and
the $y$-axis) remains unfixed. This can be seen by repeating the
calculation above for two additional elements ($x$- and
$y$-components) within the basis vector $\mathbf{e}_3$ in the
ansatz \ref{eq:ansatzTiltedToriBasis}. These relative size of
these components is not getting fixed by the orientifold
projection.

\section{One-loop amplitude}\label{Sec:oneloop}
In this section, we will calculate the R-R and NS-NS tadpole
equations for the proposed model
\begin{equation}\label{eq:prop_modelIIA}
\frac{\text{type IIA on }T^9}{\Omega R}
\end{equation}
where the 9-torus is factorized as $T^9=T^3\times T^3 \times T^3$.
This is the picture of intersecting D6-branes with a vanishing
F-flux, as presented in chapter \ref{Sec:branes_at_angles}. We will
present the computations for all $T^3$ being untilted, but extend
the final R-R and NS-NS-tadpoles also to the case of tilted tori,
as described in section \ref{Sec:nsns2formflux}.

The action of the antiholomorphic involution $R$ will be given by
\begin{equation}\label{eq:orientifold_def}
R:\qquad \tilde Z_i \rightarrow -\tilde Z_i
\end{equation}
for $i=1,2,3$. This is in complete analogy to the usual
orientifolds on the $T^6$, where $R$ in most cases is taken to act
as a complex conjugation, also reflecting the three coordinates
along which the T-duality from the D9-branes beforehand has been
performed. For this simple choice (\ref{eq:orientifold_def}), the
orientifold plane is localized along the ($X_i$-$Y_i$)-plane of
every 3-torus.

The one-loop vacuum amplitude $\mathcal{Z}_\text{one-loop}$ is the
sum of the 4 contributions from the different $\chi=0$ worldsheets
\begin{equation}
\mathcal{Z}_\text{one-loop}=\mathcal{T}+\mathcal{K}+\mathcal{A}+\mathcal{M},
\end{equation}
where we will work with the Hamiltonian formalism for which every
worldsheet integral is written as a trace. The torus amplitude
$\mathcal{T}$ is modular invariant and therefore finite (or even
vanishing for a supersymmetric model). Hence it is irrelevant for
the calculation of tadpoles and will not be treated here, only the
three remaining worldsheets for which it cannot be guaranteed that
they do not contain divergencies spoiling the theory at the
quantum level \cite{Green:1984sg, Green:1985ed}. At first we will
calculate the R-R-tadpole equation in the tree channel, the NS-NS
tadpole is stated in appendix \ref{Sec:NSNStp_genchapter}.

\subsection{The R-R tadpole}\label{Sec:RRtp_genchapter}
\subsubsection{Klein bottle}
The R-R sector of the tree channel Klein bottle amplitude in the
loop channel corresponds to the (NS-NS,$-$)-sector,
\begin{equation}\label{eq:Kleinbottle_amplitude}
\mathcal{K}^\text{(NS-NS,$-$)}=\sqrt{2}c \int\limits_{0}^{\infty}\frac{dt}{t^{3/2}}
\text{Tr}_{\text{(NS-NS,$-$)}}\left(\frac{\Omega R}{2}\ \frac{1+(-1)^F}{2}e^{-2\pi t
\mathcal{H}_{\text{closed}}}\right),
\end{equation}
where $c={V_1}/{(8\pi^2 \alpha')^{1/2}}$. The trace still includes the
9-dimensional momentum integration over the compact space, whereas
the integration over time already has been carried out and $V_1$
is its regularized length. The closed string Hamiltonian for the
NS-NS loop sector is given by
\begin{multline}\label{eq:Hamilton_closed}
    {\mathcal{H}}^{\text{NS-NS}}_{\text{closed}} = ({p^\mu})^2
+\sum_{\mu}{ \left( \sum_{n=1}^\infty{\left
      ( \alpha_{-n}^\mu \alpha_{n}^\mu + \tilde{\alpha}_{-n}^\mu
      \tilde{\alpha}_{n}^\mu \right)} \right. } \\
    \left. + \sum_{r\in\mathbb{Z}+1/2,\ r>0}{\left( r \psi^\mu_{-r}
        \psi^\mu_{r}+r
  \tilde{\psi}^\mu_{-r} \tilde{\psi}^\mu_{r} \right)} \right)  +
  E_0^{\text{NS-NS}}+\mathcal{H}_{\text{lattice, cl.}}^{\mathcal{K}}\ .
\end{multline}
The zero point energy in the NS-NS sector is determined by the
number of complex fermions and bosons and simply given by
$E_0^{\text{NS-NS}}=E_0^{\text{NS}, L}+E_0^{\text{NS}, R}=2\cdot4
\left(-{1}/{12}-{1}/{24}\right)=-1$. The lattice contribution is
calculated in appendix \ref{AppSec:KB_lattice}, equation
(\ref{eq:closedStr_latOR}). The trace over the oscillator modes
just gives the standard NS-NS sector theta functions, altogether
we obtain
\begin{equation}\label{eq:Kleinbottle_amplitude_OR_loop}
\mathcal{K}^{\text{(NS-NS,$-$)}}=c \frac{\sqrt{2}}{4}
\int\limits_{0}^{\infty}\frac{dt}{t^{3/2}}\frac{-\tfkto{0}{1/2}^4}{\eta^{12}}
\prod_{i=1}^{3}\left[\sum_{s_1^i,s_2^i,r_3^i} e^{-\pi t
\left(\frac{\alpha'}{{L_x^{i}}^2}{s_1^i}^2+\frac{\alpha'}{{L_y^{i}}^2}{s_2^i}^2+\frac{({\tilde L_z^{i}})^2
}{\alpha'}{r_3^i}^2\right)}\right].
\end{equation}
The argument of the $\vartheta$ and $\eta$ functions is given by $q=\exp(-4\pi
t)$. After the transformation into the tree channel via
$t=1/(4l)$, the Klein bottle amplitude reads
\begin{equation}\label{eq:Kleinbottle_amplitude_OR_tree}
\widetilde{\mathcal{K}}^{\text{(R-R,$+$)}}=c\frac{16
\sqrt{2}}{{\alpha'}^{3/2}}
\int\limits_{0}^{\infty}{dl}\frac{\tfkto{1/2}{0}^4}{\eta^{12}}
\prod_{i=1}^{3}\left[\frac{L_x^i
L_y^i}{\tilde L_z^i}\left(\sum_{\tilde s^{i}_1,\tilde s^{i}_2,\tilde r^{i}_3}
e^{-4\pi
l\left(\frac{{L_x^i}^2}{\alpha'}({{\tilde s}^{i}_1})^2
+\frac{{L_y^i}^2}{\alpha'}({{\tilde s}^{i}_2})^2
+\frac{\alpha'}{({\tilde L_z^i})^2}({{\tilde r}^{i}_3})^2\right)}\right)\right],
\end{equation}
with an argument $q=\exp(-4\pi l)$ for the $\vartheta$ and $\eta$
functions. The R-R tadpole contribution from the Klein bottle,
being the zeroth order term in the $q$-expansion, eventually is
given by
\begin{equation}
c\frac{256
\sqrt{2}}{{\alpha'}^{3/2}}\prod_{i=1}^{3}\left(\frac{L_x^i
L_y^i}{\tilde L_z^i}\right).
\end{equation}
\subsubsection{Annulus}
The (R,+)-tree channel sector corresponds to the (NS,-)-sector in
the loop channel which we will now calculate. It contains four
different contributions for just one stack of branes
\begin{equation}\label{eq:cylinder_amp_OR_allcontrib}
    \mathcal{A}=\mathcal{A}_{jj}+\mathcal{A}_{j'j'}+\mathcal{A}_{jj'}+\mathcal{A}_{j'j} \ .
\end{equation}
The first two contributions correspond to strings going from one
stack of branes to itself, whereas the third and fourth
contributions represent strings that go from one brane to its
mirror image, which generically is located at a non-vanishing
angle. For this reason, the two types of contributions will be
treated separately, we will start with the amplitude
$\mathcal{A}_{ii}$. It is given by
\begin{equation}\label{eq:Annulus_aii_general}
    \mathcal{A}_{jj}^\text{(NS,$-$)}=c\int\limits_{0}^{\infty}\frac{dt}{t^{3/2}}
\text{Tr}_{\text{D6j-D6j}}^\text{(NS,$-$)}\left(\frac{1}{2}\frac{1+(-1)^F}{2}e^{-2\pi t
\mathcal{H}_{\text{open}}^{\mathcal{A}_{jj}}}\right)\ .
\end{equation}
The open string Hamiltonian in light cone gauge for the case of
coinciding branes is given by
\begin{multline}\label{eq:Hamilton_open}
    {\mathcal{H}}_{\text{open}}^{\mathcal{A}_{jj}}=\frac{({p^\mu})^2}{2} +\sum_{\mu}{ \left( \sum_{n=1}^\infty{\left
      ( \alpha_{-n}^\mu \alpha_{n}^\mu\right)}+ \sum_{r\in\mathbb{Z}+\nu,\ r>0}{\left( r \psi^\mu_{-r}
        \psi^\mu_{r}\right)}\right)}+ E_0+\mathcal{H}_{\text{lattice, op.}}^{\mathcal{A}_{jj}}\ ,
\end{multline}
The lattice contribution comprises some subtleties and is
calculated in detail in appendix \ref{AppSec:A_lattice}, equation
(\ref{eq:openStr_latAnn}). The trace in
(\ref{eq:Annulus_aii_general}) can be evaluated, leading to the
standard theta functions in this sector from the oscillator part
and a lattice sum from the Kaluza-Klein and winding contributions.
After the transformation to the tree channel (see
\ref{AppSec:A_lattice} for details), using $t=1/(2l)$, we obtain
the following amplitude for a stack of $\tilde N$ branes:
\begin{multline}\label{eq:Annulus_amplitudeii_OR_tree}
\widetilde{\mathcal{A}}_{jj}^\text{(R,$+$)}=\frac{\sqrt{2} c \tilde N^2}{64\alpha^{\prime 3/2}}
\int\limits_{0}^{\infty}{dl}\frac{-\tfkto{1/2}{0}^4}{\eta^{12}}\\
\prod_{i=1}^{3}\left[\frac{{n'_i}^2 {L_x^i}^2{(\tilde L_z^i)}^2+{m'_i}^2 {L_y^i}^2{(\tilde L_z^i)}^2
+{p'_i}^2{L_x^i}^2{L_y^i}^2}{L_x^i L_y^i \tilde L_z^i}\sum_{w_i, k_i, l_i}
e^{-2\pi l \mathcal{H}_{\text{lattice, cl.}}^{\mathcal{A}_{jj}}}\right].
\end{multline}
The closed string lattice Hamiltonian is given in equation
(\ref{eq:closedStr_latAnn}). The $\vartheta$ and $\eta$ functions
have an argument $q=\exp(-4\pi l)$. An expansion in $q$ leads to
the tadpole contribution
\begin{equation}\label{eq:tp_contrib_Aii}
\frac{\sqrt{2} c \tilde N^2}{4\alpha^{\prime 3/2}}\prod_{i=1}^{3}\left[\frac{{n'_i}^2 {L_x^i}^2{(\tilde L_z^i)}^2+{m'_i}^2 {L_y^i}^2{(\tilde L_z^i)}^2
+{p'_i}^2{L_x^i}^2{L_y^i}^2}{L_x^i L_y^i \tilde L_z^i}\right].
\end{equation}
Next, we will calculate the annulus tadpole contributions coming
from the sectors $\mathcal{A}_{jj'}$ and $\mathcal{A}_{j'j}$.
These two sectors correspond to strings in between branes at a
non-vanishing angle, necessitating a different treatment, also the
angles appear as a summand in the moding of the oscillator sums.
Nevertheless, one formally obtains exactly the same theta
functions for the oscillator part as in the standard case of
$T^6=T^2\times T^2 \times T^2$ if one simply understands the
intersection angles $\kappa_i$ as the angles between the normal
vectors of the involved branes on the specific $T^3$.

The starting equation in the loop channel then is the following
\begin{multline}\label{eq:Annulus_amplitude_OR_loopij}
\mathcal{A}_{ab}^\text{(NS,$-$)}=-\frac{c}{4} \tilde N_a \tilde N_b I_{ab}
\int\limits_{0}^{\infty}\frac{dt}{t^{3/2}}e^{-\frac{3}{2} \pi
i}\frac{\tfkto{0}{1/2}\tfkto{\kappa_1}{1/2}\tfkto{\kappa_2}{1/2}\tfkto{\kappa_3}{1/2}}
{\tfkto{\kappa_1-1/2}{1/2}\tfkto{\kappa_2-1/2}{1/2}\tfkto{\kappa_3-1/2}{1/2} \eta^{3}}\\
\cdot\prod_{i=1}^3 \sum_{\tilde w_i}e^{-2\pi t \mathcal{H}_{\text{lattice, op.}}^{\mathcal{A}_{ab}}}
\end{multline}
The oscillator part is derived for instance in \cite{Ott:2003yv},
but here we have an additional lattice contribution, originating
from the fact, that one of the spacetime dimensions is a
one-dimensional compact subset of every $T^3$. $\kappa_i$ is the
angle between the two involved branes on the $i$th $T^3$, the open
string lattice Hamiltonian $\mathcal{H}_{\text{lattice,
op.}}^{\mathcal{A}_{ab}}$ is derived in appendix
\ref{AppSec:A_lattice_nonvanangle}, equation
(\ref{eq:openStr_latAnnij}). $I_{ab}$ is a multiplying factor of
the amplitude which has been identified in the case of the
$T^6=T^2\times T^2\times T^2$ with the topological intersection
number of the corresponding one-cycles of every brane on every
$T^2$ . We will soon see that a similar meaning can be given to
this factor in the case of the $T^9$, there being the topological
intersection number of corresponding two-cycles. But the explicit
form will be more complicated. The transformation to the tree
channel leads to the following amplitude
\begin{multline}\label{eq:Annulus_amplitude_ij_OR_tree}
\widetilde{\mathcal{A}}_{ab}^\text{(R,$+$)}=-\frac{c \sqrt{2}}{8 \alpha^{'3/2}}\tilde N_a \tilde N_b I_{ab}
\int\limits_{0}^{\infty}{dl}\frac{\tfkto{1/2}{0}\tfkto{1/2}
{-\kappa_1}\tfkto{1/2}{-\kappa_2}\tfkto{1/2}{-\kappa_3}}{\tfkto{1/2}{1/2-\kappa_1}
\tfkto{1/2}{1/2-\kappa_2}\tfkto{1/2}{1/2-\kappa_3}\eta^{3}}\\
\cdot\prod_{i=1}^3 \left[\frac{1}{\Lambda_i}\sqrt{\left({p^{b\prime}_i}{n^{a\prime}_i}-{p^{a\prime}_i}{n^{b\prime}_i}\right)^2{L_x^i}^2
+\left({m^{b\prime}_i}{p^{a\prime}_i}-{m^{a\prime}_i}{p^{b\prime}_i}\right)^2{L_y^i}^2
+\left({m^{b\prime}_i}{n^{a\prime}_i}-{m^{a\prime}_i}{n^{b\prime}_i}\right)^2{(\tilde L_z^i)}^2}\right.\\
\left. \cdot\sum_{w_i}e^{-2\pi l \mathcal{H}_{\text{lattice, cl.}}^{\mathcal{A}_{ab}}}\right],
\end{multline}
where
\begin{equation}
\Lambda_i\equiv\gcd\left({m^{b\prime}_i}{p^{a\prime}_i}-{m^{a\prime}_i}{p^{b\prime}_i},\
{p^{b\prime}_i}{n^{a\prime}_i}-{p^{a\prime}_i}{n^{b\prime}_i},\
{m^{b\prime}_i}{n^{a\prime}_i}-{m^{a\prime}_i}{n^{b\prime}_i}\right).
\end{equation}
Again expanding in $q=\exp(-4\pi l)$ leads to a tadpole contribution
\begin{equation}
\frac{\sqrt{2} c \tilde N_a \tilde N_b I_{ab}}{4\alpha^{\prime
3/2}}\prod_{i=1}^{3}\frac{1}{\Lambda_i}\left[\frac{{n'_a}{n'_b} {L_x^i}^2{(\tilde
L_z^i)}^2+{m'_b}{m'_b} {L_y^i}^2{(\tilde L_z^i)}^2
+{p'_a}{p'_b}{L_x^i}^2{L_y^i}^2}{L_x^i L_y^i \tilde L_z^i}\right].
\end{equation}
This expression is exactly the same contribution for the limit of
coinciding branes $a=b$, equation (\ref{eq:tp_contrib_Aii}), if
one assumes that
\begin{equation}\label{eq:intersectionnumber}
    I_{ab}=\prod_{i=1}^3 \Lambda_i.
\end{equation}
It is indeed shown in appendix \ref{AppSec:top_intnr} that
$I_{ab}$ is the unoriented topological intersection number between
two D6-branes which correspond to 2-cycles on every $T^3$. The
fact that an unoriented intersection number (and not an oriented
one as in the case of the $T^2$) appears in the amplitude is due
to the fact that also the angle appearing in the theta
functions is unoriented (in the sense that it is the 'smaller' angle
between the two (oriented) normal vectors of the two involved branes,
therefore $\leq \pi$). Further implications of this fact will be
seen later.
\subsubsection{R-R tadpole cancellation}\label{Sec:rtp_expl}
We are now able to calculate the complete annulus tadpole [sum of
all contributions from (\ref{eq:cylinder_amp_OR_allcontrib})] for
one stack of branes, where we have to use that a $\Omega R$-mirror
brane is given by the map
\begin{equation}\label{eq:OR_mirrormap}
    \Omega R:\qquad\qquad {n^{\prime}_i}\rightarrow
    {-n^{\prime}_i}, \qquad\qquad {m^{\prime}_i}\rightarrow
    {-m^{\prime}_i}, \qquad\qquad {p^{\prime}_i}\rightarrow {p^{\prime}_i}.
\end{equation}
The R-R-tadpole contributions of the Klein bottle and the annulus
together are sufficient to write down the full cancellation
condition. Different radii factors have to cancel independently
and one finally obtains the following set of tadpole
cancellation conditions (after generalizing to $k$ stacks of
branes with a stacksize $\tilde N_l$)
\begin{align}
&\sum_{l=1}^{k}{\tilde N_l}\,{p^{(l)\prime}_1}{p^{(l)\prime}_2}{p^{(l)\prime}_3}=16,& \label{eq:Rtadpole1} \\
&\sum_{l=1}^{k}{\tilde N_l}\,{p^{(l)\prime}_I}\left(n^{(l)\prime}_J+b_2^J p^{(l)\prime}_J\right)\left(n^{(l)\prime}_K+b_2^K p^{(l)\prime}_K\right)=0& &\text{with\quad } I\neq J\neq K\neq I,\label{eq:Rtadpole2}\\
&\sum_{l=1}^{k}{\tilde N_l}\,{p^{(l)\prime}_I}\left(n^{(l)\prime}_J+b_2^J p^{(l)\prime}_J\right)\left(m^{(l)\prime}_K+b_1^K p^{(l)\prime}_K\right)=0& &\text{with\quad } I\neq J\neq K\neq I,\label{eq:Rtadpole3}\\
&\sum_{l=1}^{k}{\tilde N_l}\,{p^{(l)\prime}_I}\left(m^{(l)\prime}_J+b_1^J p^{(l)\prime}_J\right)\left(m^{(l)\prime}_K+b_1^K p^{(l)\prime}_K\right)=0& &\text{with\quad } I\neq J\neq K\neq I,\label{eq:Rtadpole4}
\end{align}
and $I,J,K\in \{1,2,3\}$. These equations have already been
generalized to the case of tilted $T^3$ by introducing $b_1^i$ and
$b_2^i$. On every $T^3_i$ they independently can take the values
$0$ or $1/2$.
In general, these are $1+3+6+3=13$ different
equations. Several immediate comments can be given. At first, the
statement that a negative product of
${p^{(l)\prime}_1}{p^{(l)\prime}_2}{p^{(l)\prime}_3}$ corresponds
to an anti-D-brane is confirmed by the first tadpole equation
(\ref{eq:Rtadpole1}), as the $R$-charge then switches sign as
compared to the O-plane that contributes the term 16 on the other
side of the equation.

The tadpole equations on the standard $T^6$ (that have been
derived in \cite{Blumenhagen:2000wh}) can be obtained by setting
${n^{(l)\prime}_J}=0$ (for all $l,J$) and then performing the
decompactification limit $L_y^i\rightarrow\infty$. Equivalently,
one can set ${m^{(l)\prime}_J}=0$ (for all $l,J$) and then let
$L_x^i\rightarrow\infty$. In both cases, exactly four equations
out of the set (\ref{eq:Rtadpole1})-(\ref{eq:Rtadpole4}) remain
and indeed are the same equations as in
\cite{Blumenhagen:2000wh}(after renaming the wrapping numbers to
the conventions used in that paper).

These equations should also be obtainable in a very different way,
namely from the topological equation
\begin{equation}\label{eq:RRtadpole_top}
\sum_{l=1}^k {\tilde N}_l\ \left(\pi^{(l)}+\pi^{\prime {(l)}}\right)-4\pi_{\rm O6}=0\ ,
\end{equation}
as it has been suggested in \cite{Blumenhagen:2002wn}, see also
\cite{Ibanez:2001nd, Blumenhagen:2001te}. A D6-brane which
factorizes in three 2-cycles on a $T^9=T^3\times T^3 \times T^3$
can be written as
\begin{equation}
\pi^{(l)}=\pi^{(l)}_1\otimes \pi^{(l)}_2\otimes \pi^{(l)}_3\ ,
\end{equation}
where
\begin{equation}
\pi^{(l)}_i=p_i^{\prime (l)} \left(\pi_{x_i}\otimes \pi_{y_i}\right)+m_i^{\prime
(l)} \left(\pi_{y_i}\otimes \pi_{z_i}\right)+n_i^{\prime (l)} \left(\pi_{z_i}\otimes \pi_{x_i}\right),
\end{equation}
and $i$ is the torus index (this case corresponds to an $\bf
AAA$-torus). The O6-plane then is located at
\begin{equation}
\pi_{\rm O6}=8 \left(\pi_{x_1}\otimes
\pi_{y_1}\right)\otimes\left(\pi_{x_2}\otimes
\pi_{y_2}\right)\otimes\left(\pi_{x_3}\otimes \pi_{y_3}\right)
\end{equation}
and the $\Omega R$-mirror brane can be obtained by an application of
the map (\ref{eq:OR_mirrormap}). One immediately obtains the set
(\ref{eq:Rtadpole1})-(\ref{eq:Rtadpole4}) for $b_1=b_2=0$.
\subsection{Supersymmetry conditions, absence of tachyons}\label{Sec:SUSYcond}
The O6-brane breaks half of the supersymmetry of type IIA string
theory, so in the closed string sector, there are 16 unbroken
supercharges left, corresponding to $\mathcal{N}=1$ supersymmetry
in 10 dimensions. In the open string sector, the considered
D6-branes in general can break this supersymmetry further down,
depending on where they are located. For intersecting D-branes on
the $T^6=T^2\times T^2\times T^2$, several calculations have been
performed in order to classify supersymmetry (see for instance
\cite{Berkooz:1996aa, SheikhJabbari:1997cv, Ohta:1997fr}), the general result
was that an angle criterion of the type
\begin{equation}
\phi_1\pm \phi_2\pm \phi_3=0\nonumber
\end{equation}
guarantees for some unbroken supersymmetry. $\phi_i$ is the
oriented angle between two stacks of D6-branes on the $i$th
2-torus, where one of the stacks preserves half of the type IIA
closed string supersymmetry. In an orientifold background, the
angle criterion still holds if we understand the angles as those
in between the O6-plane and a certain stack of D-branes on every
2-torus. This condition then can be interpreted as a calibration
condition, requiring that both the special Lagrangian cycles (that
the D-branes wrap) and the orientifold brane are calibrated with
respect to $\Re(\Omega_3)$, where $\Omega_3$ is a holomorphic
$3$-form \cite{Blumenhagen:2002wn}.

Following the worldsheet approach of \cite{SheikhJabbari:1997cv},
it is possible to characterize spacetime supersymmetry in an
elegant way by just looking at the one-loop annulus amplitudes
between different branes. Some supersymmetry is conserved when the
NS and corresponding R amplitudes of that particular sector
exactly cancel against each other, i.e.
\begin{equation}
\widetilde{\mathcal{A}}_{ab}^\text{(R,$+$)}-\left(\widetilde{\mathcal{A}}_{ab}^\text{(NS,$+$)}
+\widetilde{\mathcal{A}}_{ab}^\text{(NS,$-$)}\right)=0.
\end{equation}
The amplitude $\widetilde{\mathcal{A}}_{ab}^\text{(R,$+$)}$ is
explicitly given in equation
(\ref{eq:Annulus_amplitude_ij_OR_tree}), the corresponding
NS-amplitude only differs by the theta equations in the numerator
of the first fraction (for a detailed discussion see \cite{Ott:2003yv}).
Therefore, one immediately obtains the following supersymmetry
condition
\begin{multline}
\tfkto{1/2}{0}\tfkto{1/2}{-\kappa_1}\tfkto{1/2}{-\kappa_2}\tfkto{1/2}{-\kappa_3}
-\tfkto{0}{0}\tfkto{0}{-\kappa_1}\tfkto{0}{-\kappa_2}\tfkto{0}{-\kappa_3}\\
+\tfkto{0}{1/2}\tfkto{0}{1/2-\kappa_1}\tfkto{0}{1/2-\kappa_2}\tfkto{0}{1/2-\kappa_3}=0,
\end{multline}
where $\kappa_i$ is the smaller angle ($<\pi$) between the normal
vectors between one stack of branes $a$ and another one $b$ on the
$i$th 3-torus, as explained earlier. Expanding the theta functions
in $q$ leads to the leading contribution equation
\begin{equation}
\cos^2(\kappa_1)+\cos^2(\kappa_2)+\cos^2(\kappa_3)-2
\cos(\kappa_1)\cos(\kappa_2)\cos(\kappa_3)=1.
\end{equation}
The possible solutions to this equation are given by
\begin{equation}\label{eq:sol_susy_angle}
\kappa_1\pm \kappa_2 \pm \kappa_3=0
\end{equation}
This angle criterion at first sight coincides with the angle
criterion on the $T^6$. Again the case of an orientifold
background can formally be included in this analysis by taking the
angles $\kappa$ to be those between the orientifold plane normal
vector and the D-brane\footnote{although there is no annulus
amplitude of this kind, as the orientifold brane itself only
contributes to the Klein-bottle and M\" obius amplitudes. On the
other hand our formal treatment is equivalent to calculating the
annulus amplitudes between a brane and its $\Omega R$ mirror
image, $\tilde{\mathcal{A}}_{jj'}$.}.

But there is one very important subtlety as compared to the case
on the $T^6$: the angles are lying in a three-dimensional space.
The overall annulus amplitude for several stacks of branes will
only vanish (and therefore the whole system will be
supersymmetric), if all possible annulus amplitude sectors between
different branes and between different branes and the O6-plane
vanish separately. In the case of the $T^6$ it is guaranteed that
overall SUSY is preserved if every one of the branes is aligned
supersymmetrically to the O6-plane. If for two different D6-branes
the angle criterion is fulfilled separately (with the angle taken
in between the D-brane and the O-plane), i.e.
\begin{equation}
\phi_1^a+\phi_2^a+\phi_3^a=0,\qquad \phi_1^b+\phi_2^b+\phi_3^b=0,\nonumber
\end{equation}
then these two D-branes are also aligned supersymmetrically
against each other,
\begin{equation}\nonumber
(\phi_1^a-\phi_1^b)+(\phi_2^a-\phi_1^b)+(\phi_3^a-\phi_1^b)=0,
\end{equation}
simply because the rotations on every 2-torus are commutative and
are performed in the same 2-plane (being the $T^2$ itself).
This is not the case for the $T^9=T^3\times T^3\times T^3$,
equation (\ref{eq:sol_susy_angle}), as these angles are taken in
between 3-dimensional vectors and therefore do not necessarily lie
within the same 2-plane. Consequently, we really have to require
(\ref{eq:sol_susy_angle}) for all different possible sectors
between all stacks of branes and in addition between all these
stacks and the O-plane in order to conserve some supersymmetry. It
is possible to count the conserved supersymmetries by looking at
the potential bifundamental scalars\footnote{if they are really
there depends on the topological intersection number $I_{ab}$ between the
branes under consideration.} in the NS-sector in between two stacks
of branes \cite{Aldazabal:2000dg, Ibanez:2001nd}. Their masses are
controlled by the angles between the branes, in our case those
that appear in equation (\ref{eq:sol_susy_angle}). Their masses
are given by
\begin{align}\label{eq:tachyons}
&\alpha'{M_1}^2={1 \over 2}\left(-\kappa_1+\kappa_2+\kappa_3\right),&
&\alpha'{M_2}^2=\frac{1}{2}\left(\kappa_1-\kappa_2+\kappa_3\right),&\\
&\alpha'{M_3}^2=\frac{1}{2}\left(\kappa_1+\kappa_2-\kappa_3\right),&
&\alpha'{M_4}^2=1-\frac{1}{2}\left(\kappa_1+\kappa_2+\kappa_3\right).&\nonumber
\end{align}
For a supersymmetric configuration fulfilling the angle criterion
(\ref{eq:sol_susy_angle}), some of these four equations will be
zero, the remaining ones positive and this lets us directly
determine the amount of unbroken supersymmetry. For a
non-supersymmetric configuration, it might happen that some of
these masses become negative and thus tachyonic. This signals an
instability and according to \cite{Douglas:1999vm} on a Calabi-Yau
can be interpreted in a way that the joining process of the branes
is energetically favored. There are attempts to understand these
tachyons as Higgs field in the effective theory (see for instance
\cite{Blumenhagen:2001te, Blumenhagen:2002ua} and references
within), but the validity of this approach seems rather
questionable as it is an off-mass-shell process. The point of view
taken in this article is to avoid tachyons if possible. It
carefully has to be checked for any explicit model, if this is
indeed possible\footnote{Note that the angles vary with a change of the
torus radii.}. We will come back to this question in the explicit
example in section \ref{Sec:Example}.

\section{The scalar potential}\label{Sec:scalar_ptential_gen}
The (perturbative) scalar potential for the moduli is contained in
the perturbative vacuum partition function. The open string
partition function contains all diagrams of a certain world sheet
topology, weighted with a factor of the open string coupling
constant, $\exp(-\chi \tilde\phi^{(10)}_0)$. In this factor,
$\chi$ is the Euler number, $\chi=2-2g-b-c$ and
$\tilde\phi^{(10)}_0$ is the expectation value of the
10-dimensional dilaton within type IIA string theory\footnote{A
similar result holds in type IIB.}. The lowest contributing open
string diagrams are the disc and the projective plane, which both
have an Euler number one and will be called open string tree level
diagrams. The open string next to leading order diagrams are the
annulus (cylinder) and M\"obius strip diagrams which have an Euler
number zero, they will be called open string one loop diagrams.
We can schematically write down the open string vacuum partition function as
\begin{equation}\label{eq:complete_partfct_sf}
\mathcal{Z}=e^{-\tilde\phi^{(10)}_0}\left(\text{disc + projective plane}\right)+e^{0}\left(\mathcal{A}
+\mathcal{M}\right)+\text{higher loop contributions}.
\end{equation}
The scalar potential for the moduli then can also be written as
\begin{equation}
V_\text{scalar}=V_\text{tree}+V_\text{1-loop}+\text{higher loop contributions},
\end{equation}
and every scalar potential contribution arises from the
corresponding partition function contribution. The tree-level
potential can be derived from the Born Infeld potential in type
IIB (and a similar reasoning is true for the orientifold plane).
Alternatively, the disc and projective plane diagrams can be
normalized from the one-loop partition function. The one-loop
potential has to be calculated within the worldsheet approach.

Both potentials give contributions that are independent of
$\alpha'$, but also $\alpha'$ corrections. The latter ones are
specific to string theory and might play an important role for the
understanding of the discussed model as we will see.

There are two further subtleties that one has to take into account
in order to speak about the physical meaning of the scalar
potential within our universe: the potential has to be transformed
into the Einstein frame. After this procedure, one can simplify
the physical picture in two more ways. Instead of the radii, one
can equally well switch to shape and volume moduli (similar to the
imaginary part of the complex and Kaehler structures in the
$T^2$). Furthermore, for convenience, one can make a field
redefinition into new scalar variables with canonically normalized
kinetic terms as in common field theory.

All these procedures require a very careful treatment of the
prefactors of the different potential contributions, including
both powers of $\alpha'$ and the dilaton expectation value, in
order to judge their relative importance. Therefore, we will
discuss the relevant kinetic terms in the following subsection.

One technical problem arises if $\alpha'$ corrections indeed play
an important role. In this case it is hard to get an understanding
of the spacetime picture, because also higher derivative kinetic
terms most likely have to be taken into account in the Einstein
frame in order to integrate the equations of motion, which is
generally not well understood up till now. We will come back to
this point later.
\subsection{The kinetic terms in effective field theory}\label{Sec:kineticTerms}
We are interested in the kinetic terms of the metric components
and the dilaton. All these closed string scalar moduli are
contained in the NS-sector of type IIA string theory, for which
the effective action can be written as\cite{Polchinski:1998rr}
\begin{equation}\label{eq:supergravityNS}
    \emph{S}_{\mathrm{NS}}=\frac{1}{2{\kappa_{10}}^2}\int d^{10}x \sqrt{-G} e^{-2\tilde\phi_{10}}\left[{\mathcal{R}_{10}}
    +4\, \partial_\mu \phi_{10}\, \partial^\mu \phi_{10}+\mathcal{O}(\alpha')\right]\ ,
\end{equation}
where we have assumed that $H_3\equiv 0$. Writing the torus metric
in the diagonal form\footnote{This is the case for the case where all
$b_1^i$ and $b_2^i$ are identically zero.} and assuming only a time dependence,
\begin{equation}\label{eq:eff_metric}
G_{\mu \nu}=\text{diag}\left(-g_{00}(t),{L_x^1(t)}^2,{L_y^1(t)}^2,{\tilde L_z^1(t)}^2,{L_x^2(t)}^2,
{L_y^2(t)}^2,{\tilde L_z^2(t)}^2,{L_x^3(t)}^2,{L_y^3(t)}^2,{\tilde L_z^3(t)}^2\right),
\end{equation}
one can write the action as an one-dimensional one. The
10-dimensional dilaton has to be dimensionally reduced to the one
dimensional one by using
\begin{equation}\label{eq:dilaton_dimred}
e^{-\tilde\phi^{(10)}}=\frac{e^{-\tilde\phi^{(1)}}\alpha'^{9/4}}{\prod_{i=1}^3 \sqrt{{L_x^i}
{L_y^i}\tilde L_z^i}}.
\end{equation}
Subsequently, we can go into the 1-dimensional Einstein frame. In
order to do so, we first define the dilaton $\phi^{(1)}_\text{E}$ with a
vanishing expectation value,
\begin{equation}\label{eq:phi1Einstein}
\phi^{(1)}_\text{E}(t)=\tilde\phi^{(1)}(t)-\tilde\phi^{(1)}_0,
\end{equation}
and then rescale the metric by
\begin{equation}\label{eq:EinstMetric}
G_{\mu\nu}^\text{E}=e^{-\frac{4}{D-2}\phi^{(D)}_\text{E}(t)}G_{\mu\nu}=e^{4\phi^{(1)}_\text{E}(t)}G_{\mu\nu}.
\end{equation}
This leads to the effective action [after assuming $g_{00}(t)\equiv 1$ in (\ref{eq:eff_metric})] for the kinetic terms
\begin{multline}\label{eq:kinTerms}
\emph{S}_{\mathrm{NS}}^{\text{E}}=
\frac{16\pi^2{\alpha'}^{1/2}}{e^{2\tilde\phi^{(1)}_0}}\int dt \Bigg[\sum_{i=1}^9
\left(-\frac{1}{2}\left(\frac{\dot{L}^i_\text{E}}{L^i_\text{E}}\right)^2
+9\frac{\dot{L}^i_\text{E}}{L^i_\text{E}}\dot\phi^{(1)}_\text{E}\right)
-\frac{1}{2}\sum_{i=1}^9\sum_{j>i}
\left(\frac{\dot{L}^i_\text{E}\dot{L}^j_\text{E}}{L^i_\text{E}L^j_\text{E}}\right)
\Bigg.\\ \Bigg.-73\left(\dot\phi^{(1)}_\text{E}\right)^2
+\frac{1}{4}e^{-4\phi^{(1)}_\text{E}}\mathcal{O}(\alpha')
\Bigg],
\end{multline}
where the dots denote the derivative with respect to $t$. The term
$9\ddot\phi^{(1)}_\text{E}$ does not play any role for the
equations of motion (as it is a surface term) and has been
omitted. For convenience of writing, the different torus radii in
the metric have been renamed from 1 to 9. We will later factor out
the same prefactors for the action of the tree-level and one-loop
potential.

Later we will see that it is very useful to reformulate both
potentials in terms of volume and (dimensionless) shape moduli of
the $T^3$ in order to get a better physical understanding. Such a
procedure is analogous to the change of variables from the radii
to complex structure and K\"ahler moduli in case of the $T^2$,
their imaginary parts corresponding to shape and volume
moduli. We thus define the following moduli for each $T^3_i$:
\begin{equation}\label{eq:shape_volume_moduli}
\qquad q_{xz}^i=\frac{L_{x,\text{E}}^i}{\tilde{L}_{z,\text{E}}^i}\, ,
\qquad q_{yz}^i=\frac{L_{y,\text{E}}^i}{\tilde{L}_{z,\text{E}}^i}\, ,
\qquad V_\text{tot}= V^1_\text{E} V^2_\text{E} V^3_\text{E}\, ,
\qquad V_{13}=\frac{V^1_\text{E}}{V^3_\text{E}}\, ,
\qquad V_{23}=\frac{V^2_\text{E}}{V^3_\text{E}}\, ,
\end{equation}
where
$V^i_\text{E}={L_{x,\text{E}}^i}{L_{y,\text{E}}^i}{\tilde{L}_{z,\text{E}}^i}$
is the volume of the $i$th 3-torus, $q_{xz}$ is the shape
parameter in the ($x$,$z$)- and $q_{yz}$ in the ($y$,$z$)-plane,
this is for the specific $T^3$ with orthogonal axes.

Instead of the original 9 radii moduli, we now have 6 shape moduli
$q_{xz}^i$ and $q_{yz}^i$ (which describe the ratios between two
tori axes on each $T^3_i$), the overall $T^9$ volume
$V_\text{tot}$ and the two ratios between the volume of two of the
$T^3_i$, $V_{13}$ and $V_{23}$. This new choice will prove to be
very useful for a better physical understanding, as will be seen
in the subsequent sections. The kinetic terms (\ref{eq:kinTerms})
take a simpler form in terms of these new moduli:
\begin{multline}\label{eq:kinTerms_shapevolume}
\emph{S}_{\mathrm{NS}}^{\text{E}}=
\frac{16\pi^2{\alpha'}^{1/2}}{e^{2\tilde\phi^{(1)}_0}}\int dt \Bigg(\sum_{i=1}^9
\left[
-\frac{1}{6}\left(\frac{\dot q_{xz}^i}{q_{xz}^i}\right)^2
-\frac{1}{6}\left(\frac{\dot q_{yz}^i}{q_{yz}^i}\right)^2
+\frac{1}{6}\frac{\dot q_{xz}^i \dot q_{yz}^i}{q_{xz}^i q_{yz}^i}\right]
-\frac{5}{18}\left(\frac{\dot V_\text{tot}}{V_\text{tot}}\right)^2\Bigg.\\
 \Bigg.
-\frac{1}{18}\left(\frac{\dot V_{13}}{V_{13}}\right)^2
-\frac{1}{18}\left(\frac{\dot V_{23}}{V_{23}}\right)^2
+\frac{1}{18}\left(\frac{\dot V_{13}\dot V_{23}}{V_{13}V_{23}}\right)
-73\left(\dot\phi^{(1)}_\text{E}\right)^2+9\frac{\dot{V}_\text{tot}}{V_\text{tot}}\dot\phi^{(1)}_\text{E}
+\frac{1}{4}e^{-4\phi^{(1)}_\text{E}}\mathcal{O}(\alpha')
\Bigg).
\end{multline}
The contributions of the shape moduli can be separated into terms
for the three different $T^3_i$, no mixing occurs between
different tori. Additionally, there are simple kinetic terms for
the overall volume and the volume ratios. The dilaton mixes only
with the overall volume, and the two volume ratios mix with one
another. In order to get canonically normalized kinetic terms, one
can make field redefinitions,
\begin{align}\label{eq:can_var_einst}
&\tilde V_\text{tot}=\sqrt{\frac{5}{9}}\ln V_\text{tot} \, ,&
&\tilde V_{13}={\frac{1}{3}}\ln V_{13} \, ,&
&\tilde V_{23}={\frac{1}{3}}\ln V_{23} \, ,&\\
&\tilde q_{xz}^i=\sqrt{\frac{1}{3}}\ln q_{xz}^i\, ,&
&\tilde q_{yz}^i=\sqrt{\frac{1}{3}}\ln q_{yz}^i\, ,&
&\phi^{(1)}_\text{c}=\sqrt{146}\phi^{(1)}_\text{E}\, ,&\nonumber
\end{align}
such that the standard kinetic terms get the canonical form, but
the mixing terms of course still occur and not all of them have canonical
prefactors.
\subsection{The tree-level potential}\label{Sec:tree_level_pot}
The Born-Infeld action of type IIB string theory for $N_l$
coinciding D9-brane(s) with a constant $U(1)$ or $U(N_l)$ flux, as
discussed in sections \ref{Sec:D9branes_abelian} and
\ref{Sec:D9branes_nonabelian}, respectively, in general is given
by
\begin{equation}\label{eq:Born_Infeld}
{\cal S}_{\rm DBI} =
\text{T}_9 \int_{D9_l}{d^{10}x\ e^{-\phi^{(10)}} \sqrt{ {\rm det} \left ( G_{\mu\nu} +2\pi \alpha'
F_{\mu\nu}^{(l)} \right) } }\ ,
\end{equation}
where $\text{T}_p$ stands for the Dp-brane tension
\begin{equation}\label{eq:tension}
    \text{T}_p = \frac{\sqrt{\pi}}{\kappa_{(10)}}(4\pi^2\alpha')^\frac{3-p}{2}
\end{equation}
and $\kappa_{(10)}^2=1/2(2\pi)^7\alpha'^4$. Therefore, the tension
has a dependence $\text{T}_9\sim \alpha'^{-5}$ and the action is
dimensionless.

By inserting the metric $G_{\mu
\nu}=\delta_{\mu\nu}$ and the F-flux (\ref{eq:B_simp_notation})
for all three $T^3$ into this equation, we obtain the square root
of the determinant
\begin{equation}\label{eq:det_Born_Infeld}
\sqrt{{\rm det} \left ( G_{\mu\nu} +2\pi \alpha' F_{\mu\nu}^{(l)}
\right) } =\prod_{i=1}^{3}\sqrt{1+\frac{{\alpha'}^2
n_i^2}{N_l^2{L_y^i}^2{L_z^i}^2}+\frac{{\alpha'}^2
m_i^2}{N_l^2{L_x^i}^2{L_z^i}^2}}\ .
\end{equation}
We can now integrate over the 9-dimensional compact space, where
we assume that the 10-dimensional dilaton is spatially constant,
as is the F-flux. This implies also that the compactification
radii $L_i$ are assumed to be spatially constant (they do not
depend on each other).

One has to take into account that
$N_l$ D9-branes wrap the 9-dimensional torus together $N_l$ times,
and that this in the picture of the T-dual D6-branes can be
written as $N_l=p_1' p_2' p_3' \tilde N_l$, according to the last
equation in (\ref{eq:weitere_umdef}). We also rewrite the other
quantum numbers using (\ref{eq:umschreibung_nm1}). Furthermore, we
have to rewrite the three radii into the T-dual type IIA ones, $\tilde
L_z^i=\alpha'/L_z^i$, and rewrite the dilaton in the T-dual picture,
\begin{equation}\label{eq:dilaton_tdual}
    e^{-\phi^{(10)}}=\frac{\tilde L_z^1 \tilde L_z^2 \tilde L_z^3}{{\alpha'}^{3/2}}e^{-\tilde\phi^{(10)}},
\end{equation}
and then rewrite the tension using
\begin{equation}\label{eq:tension_tdual}
    T_9=\frac{T_6}{(2\pi \sqrt{\alpha'})^3}.
\end{equation}
One finally obtains
\begin{equation}\label{eq:Born_Infeld_tdual10dimdilaton}
{\cal S}_{\rm DBI} =T_6(2\pi)^6\tilde N_l\int dt\
e^{-\tilde\phi^{(10)}}\prod_{i=1}^{3}\sqrt{{n'_i}^2
{L_x^i}^2{(\tilde L_z^i)}^2+{m'_i}^2 {L_y^i}^2{(\tilde L_z^i)}^2
+{p'_i}^2{L_x^i}^2{L_y^i}^2},
\end{equation}
As all the spatial dimensions are compact and integrated out, one
can equally well dimensionally reduce the dilaton to only the
dimension of time using equation (\ref{eq:dilaton_dimred}) and
then write (\ref{eq:Born_Infeld_tdual10dimdilaton}) as
\begin{equation}\label{eq:Born_Infeld_tdual}
{\cal S}_{\rm DBI} =T_6(2\pi)^6\alpha'^{9/4}\tilde N_l\int dt\
e^{-\tilde\phi^{(1)}}\prod_{i=1}^{3}\sqrt{{n'_i}^2
\frac{{L_x^i}{\tilde L_z^i}}{L_y^i}+{m'_i}^2 \frac{{L_y^i}{\tilde L_z^i}}{L_x^i}
+{p'_i}^2\frac{{L_x^i}{L_y^i}}{\tilde L_z^i}}.
\end{equation}
This expression can be understood as the tree level scalar
potential for one stack of $\tilde N_l$ D6-branes
\cite{Aldazabal:2000dg,Blumenhagen:2001te,Blumenhagen:2001mb,Blumenhagen:2002ua},
i.e. ${\cal S}_{\rm DBI}=\int dt\
V_\text{tree}[\tilde\phi^{(1)}(t),{L_x^i}(t),{L_y^i}(t),\tilde
{L}_z^i(t)]$.\footnote{If one does not want to understand the
potential as a time dependent one, one can just integrate time
formally out and thus write an additional factor$(2\pi V_1)$ in
front, where $V_1$ is the regularized time.} Together with the
$\Omega R$ mirror branes and the orientifold plane, one obtains
the following total scalar potential in the string frame
\begin{multline}\label{eq:totalNSpotD6} {\cal S}_{\rm DBI}=\int dt\,2\, T_6(2\pi)^6\alpha'^{9/4}
e^{-\tilde\phi^{(1)}}\\\cdot\left[\sum_{l=1}^k \tilde N_l
\prod_{i=1}^{3}\sqrt{\left(n^{(l)\prime}_i+b_2^i
p^{(l)\prime}_i\right)^2 \frac{{L_x^i}{\tilde
L_z^i}}{L_y^i}+\left(m^{(l)\prime}_i+b_1^i p^{(l)\prime}_i\right)^2
\frac{{L_y^i}{\tilde L_z^i}}{L_x^i} +{p^{(l)
\prime}_i}^2\frac{{L_x^i}{L_y^i}}{\tilde L_z^i}}\right.\\
\left.-16\prod_{i=1}^{3}\sqrt{\frac{{L_x^i}{L_y^i}}{\tilde L_z^i}}\right].
\end{multline}
We have extended this equation to the case of tilted $T^3$ by including
$b_1^i$ and $b_2^i$, $b_1^i,b_2^i\in \{0,1/2\}$, corresponding to an additional discrete
B-flux in the Born-Infeld potential (\ref{eq:Born_Infeld}) as
discussed in section \ref{Sec:nsns2formflux}.
In the limit ${n^{(l)\prime}_i}=0$ (for all $l,i$)
and subsequently, $L_y^i\rightarrow\infty$, this potential does
exactly agree with the one derived in
\cite{Blumenhagen:2001te,Blumenhagen:2001mb} for the
$T^6=T^2\times T^2\times T^2$. One only has to rewrite the
1-dimensional dilaton in terms of the 4-dimensional one (taking
into account that in this limit the three $L_y^i$ are the three
additional spatial dimensions of 4-dimensional spacetime). This
is unproblematic as these dimensions have been
integrated out in both cases ($T^3$ and $T^2$).

Going to the 1-dimensional Einstein frame by using equations
(\ref{eq:phi1Einstein}) and (\ref{eq:EinstMetric}), the tree level
potential finally takes the form
\begin{equation}\label{eq:totalNSpotD6Einstein}
{\cal S}_{\rm DBI}^\text{E}=
\frac{16\pi^2{\alpha'}^{1/2}}{e^{2\tilde\phi^{(1)}_0}}\int dt\ V_\text{tree}^\text{E},
\end{equation}
where
\begin{multline}
V_\text{tree}^\text{E}=\frac{e^{\tilde\phi^{(1)}_0}}{8\alpha'^{7/4}\pi^2}\ e^{-4\phi^{(1)}_\text{E}}\\
\cdot \left[\sum_{l=1}^k \tilde N_l
\prod_{i=1}^{3}\sqrt{\left(n^{(l)\prime}_i+b_2^i
p^{(l)\prime}_i\right)^2 \frac{{L_{x,\text{E}}^i}{\tilde
L_{z,\text{E}}^i}}{L_{y,\text{E}}^i}+\left(m^{(l)\prime}_i+b_1^i p^{(l)\prime}_i\right)^2
\frac{{L_{y,\text{E}}^i}{\tilde L_{z,\text{E}}^i}}{L_{x,\text{E}}^i} +{p^{(l)
\prime}_i}^2\frac{{L_{x,\text{E}}^i}{L_{y,\text{E}}^i}}{\tilde L_{z,\text{E}}^i}}\right.\\
\left.-16\prod_{i=1}^{3}\sqrt{\frac{{L_{x,\text{E}}^i}{L_{y,\text{E}}^i}}{\tilde L_{z,\text{E}}^i}}\right].
\end{multline}
It is most instructive to rewrite the tree-level potential in
terms of the shape and volume moduli, defined by (\ref{eq:shape_volume_moduli}). The
result for all twists
$b_1^i=b_2^i=0$ is given by
\begin{multline}\label{eq:treepot_kanvar_einst}
V_\text{tree}^\text{E}=\frac{e^{\tilde\phi^{(1)}_0}}{8\alpha'^{7/4}\pi^2}\ e^{-4\phi^{(1)}_\text{E}}
V_\text{tot}^{{1}/{6}}\\
\cdot \left[\sum_{l=1}^k \tilde N_l
\prod_{i=1}^{3}\left(q_{xz}^i q_{yz}^i\right)^{1/3} \sqrt{\frac{{n^{(l)\prime}_i}^2}{{q_{yz}^i}^2}
+\frac{{m^{(l)\prime}_i}^2}{{q_{xz}^i}^2} +{p^{(l)
\prime}_i}^2}
-16\prod_{i=1}^{3}\left(q_{xz}^i q_{yz}^i\right)^{1/3}\right].
\end{multline}

\noindent Some important remarks regarding this potential have to be given:
\begin{enumerate}
\item The tree level potential depends only on the total $T^9$
volume $V_\text{tot}$ in a way $\sim {V_\text{tot}}^{1/6}$, but
it does not depend on the volume ratios between the different $T^3_i$. In terms
of the canonical variables (\ref{eq:can_var_einst}), this implies
a dependence $\sim \exp[{1}/{\sqrt{60}}\,\tilde V_\text{tot}]$.
In terms of the volumes of the different $T^3_i$, they all trivially have
a similar potential $\sim {V^i_\text{E}}^{1/6}$. This behavior is
independent of the wrapping numbers and thus general. Disregarding
the kinetic terms, the overall volume (and at the same time the
volume of every $T^3_i$) cannot be stabilized, it tends to zero.

\item The dependence of the potential on the time-dependent
one-dimensional dilaton is $\sim\exp[-4\phi^{(1)}_\text{E}]$,
this is again a general statement. The dilaton shows a runaway
behavior on the tree-level.

\item This is different for the shape moduli ${q_{xz}^i}$ and
${q_{yz}^i}$. Depending on the wrapping numbers of the different
stacks of branes, the behavior of every stack contribution can be
either $\sim (q^i)^{1/3}$ or $\sim (q^i)^{-2/3}$. For the
canonical variables, this translates into a behavior $\sim
\exp[\frac{1}{\sqrt{3}}\tilde q^i]$ or $\sim
\exp[-\frac{2}{\sqrt{3}}\tilde q^i]$. In the interplay between
different stack contributions and the orientifold plane, it is
even possible to stabilize some of the $q^i$ on the tree-level.
This can be seen for instance in the explicit example of section
\ref{Sec:Example}.
\end{enumerate}
\subsection{The one-loop potential}\label{Sec:oneloopPotential}
As explained in the introduction of this chapter, the one-loop
potential contribution arises from the annulus and M\"obius partition functions
of the worldsheet calculation. If there is a NS-NS tadpole, these
amplitudes then formally are diverging due to the lowest (zeroth)
order contribution in the modular function expansion parameter
$q=\exp(-4\pi l)$. This divergence is simply the NS-NS tadpole
itself. In \cite{Blumenhagen:2002ua} the suggestion has been made
to regularize every diagram simply by subtracting the NS-NS tadpole,
indeed this seems to make sense as the higher orders in $q$ do not
lead to any additional divergence. Therefore, this somewhat crude
procedure will be used here, too.

For a given number $k$ of stacks
of D6-branes, the complete annulus amplitude is given by
\begin{align}\label{eq:comp_ann_amp}
\tilde{\mathcal{A}}_\text{tot}=&\sum_{l=1}^{k}\left(\tilde{\mathcal{A}}_{ll}+\tilde{\mathcal{A}}_{l'l'}
+\tilde{\mathcal{A}}_{ll'}+\tilde{\mathcal{A}}_{l'l}\right)\\
+&\sum_{l<j}\left(\tilde{\mathcal{A}}_{lj}+\tilde{\mathcal{A}}_{jl}+\tilde{\mathcal{A}}_{l'j'}+\tilde{\mathcal{A}}_{j'l'}
+\tilde{\mathcal{A}}_{lj'}+\tilde{\mathcal{A}}_{jl'}+\tilde{\mathcal{A}}_{l'j}+\tilde{\mathcal{A}}_{j'l}\right)
\ . \nonumber\end{align} and every amplitude can be directly
obtained from the two general formulas in the R-sector
(\ref{eq:Annulus_amplitude_ij_OR_tree}) and
\ref{eq:Annulus_amplitudeii_OR_tree}). The additional NS-sector
amplitudes only differ in the usual way in the arguments of the
$\vartheta$-functions. The complete M\"obius amplitude consists of
the contributions
\begin{align}\label{eq:comp_moeb_amp}
\tilde{\mathcal{M}}_\text{tot}=&\sum_{l=1}^{k}\left(\tilde{\mathcal{M}}_{l}+\tilde{\mathcal{M}}_{l'}\right)\ ,
\end{align}
and every contribution can be obtained already from the
Klein-Bottle and annulus amplitudes, this is for instance
explained in \cite{Ott:2003yv}. The regularization can be done
independently for every term. In order to compare with the result
at the tree-level, one again has to transform into the
1-dimensional Einstein frame and write the potential as a
time-dependent one. This will be done exemplary in the remainder
of this section for an amplitude of the type
$\tilde{\mathcal{A}}_{ab}$. From Jacobi's abstruse identity,
\begin{equation}\label{eq:abstruse_identity}
\tfktvier{0}{0}{q}-\tfktvier{0}{1/2}{q}-\tfktvier{1/2}{0}{q}=0\ ,
\end{equation}
it follows that the amplitude vanishes when brane $a$ and brane
$b$ coincide, i.e. for all three angles
$\kappa_i\equiv 0$. This is surely the case for
$\tilde{\mathcal{A}}_{ll}$ and $\tilde{\mathcal{A}}_{l'l'}$ in the
total annulus amplitude (\ref{eq:comp_ann_amp}), but can be the
case for even more contributions. All non-vanishing contributions
therefore have at least one non-vanishing angle $\kappa_i$ on any
$T^3_i$. Exemplary, we will discuss the specific case where
$\kappa_1\neq 0$ and $\kappa_2\neq 0$, but $\kappa_3=0$ in the
remainder of this section.
In Einstein frame, the potential takes the following form:
\begin{equation}\label{eq:AregVpot}
\widetilde{\mathcal{A}}_{ab}^\text{reg,E}=\frac{16\pi^2{\alpha'}^{1/2}}{e^{2\tilde\phi^{(1)}_0}}\int dt\ V_\text{1-loop}^{ab, \text{E}},
\end{equation}
where
\begin{multline}\label{eq:Annulus_amplitude_ij_OR_treeEinstein}
V_\text{1-loop}^{ab, \text{E}}=\ e^{2\tilde\phi^{(1)}_0-6\phi^{(1)}_\text{E}}
\frac{\tilde N_a \tilde N_b }{512\pi^3 \alpha^{'5/2}}
\cdot\frac{{n'_3}^2 {(L_{x,\text{E}}^3)}^2{(\tilde L_{z,\text{E}}^3)}^2+{m'_3}^2 {(L_{y,\text{E}}^3)}^2{(\tilde L_{z,\text{E}}^3)}^2
+{p'_3}^2{(L_{x,\text{E}}^3)}^2{(L_{y,\text{E}}^3)}^2}{L_{x,\text{E}}^3 L_{y,\text{E}}^3 \tilde L_{z,\text{E}}^3}\\
\cdot\prod_{i=1}^2 \left[\sqrt{\left({p^{b\prime}_i}{n^{a\prime}_i}-{p^{a\prime}_i}{n^{b\prime}_i}\right)^2{(L_{x,\text{E}}^i)}^2
+\left({m^{b\prime}_i}{p^{a\prime}_i}-{m^{a\prime}_i}{p^{b\prime}_i}\right)^2{(L_{y,\text{E}}^i)}^2
+\left({m^{b\prime}_i}{n^{a\prime}_i}-{m^{a\prime}_i}{n^{b\prime}_i}\right)^2{(\tilde L_{z,\text{E}}^i)}^2}\right]\\
\shoveleft \ \ \cdot\int\limits_{0}^{\infty}{dl}\left[\left(\frac{-\tfktsq{\frac{1}{2}}{0}\tfkto{\frac{1}{2}}
{-\kappa_1}\tfkto{\frac{1}{2}}{-\kappa_2}}{\tfkto{\frac{1}{2}}{\frac{1}{2}-\kappa_1}
\tfkto{\frac{1}{2}}{\frac{1}{2}-\kappa_2}\eta^{6}}\right.\right.\\ \left. \left.
+\frac{\tfktsq{0}{0}\tfkto{0}{-\kappa_1}\tfkto{0}{-\kappa_2}
-\tfktsq{0}{\frac{1}{2}}\tfkto{0}{\frac{1}{2}-\kappa_1}\tfkto{0}{\frac{1}{2}-\kappa_2}}{\tfkto{\frac{1}{2}}{\frac{1}{2}-\kappa_1}
\tfkto{\frac{1}{2}}{\frac{1}{2}-\kappa_2}\eta^{6}}\right)\right.\\ \left.
\cdot\left(\sum_{w_1}e^{-2\pi l \mathcal{H}_{\text{lat, cl., $T^3_1$}}^{\mathcal{A}_{ab}, \text{E}}}\right)
\left(\sum_{w_2}e^{-2\pi l \mathcal{H}_{\text{lat, cl., $T^3_2$}}^{\mathcal{A}_{ab}, \text{E}}}\right)
\left(\sum_{w_3, k_3, l_3} e^{-2\pi l \mathcal{H}_{\text{lat, cl., $T^3_3$}}^{\mathcal{A}_{jj}, \text{E}}}\right)-1\right]\ .
\end{multline}
The Einstein frame lattice Hamiltonians on the first two $T^3$ are of the form
\begin{multline}\label{eq:closedStr_latAnnijEinstein}
    \mathcal{H}_{\text{lat, cl., $T^3_i$}}^{\mathcal{A}_{ab}, \text{E}}=
    \frac{e^{-4\phi^{(1)}_\text{E}}}{2\alpha'\Lambda_i^2}\left[
\left({p^{b\prime}_i}{n^{a\prime}_i}-{p^{a\prime}_i}{n^{b\prime}_i}\right)^2{(L_{x,\text{E}}^i)}^2
+\left({m^{b\prime}_i}{p^{a\prime}_i}-{m^{a\prime}_i}{p^{b\prime}_i}\right)^2{(L_{y,\text{E}}^i)}^2\right.\\
\left.+\left({m^{b\prime}_i}{n^{a\prime}_i}-{m^{a\prime}_i}{n^{b\prime}_i}\right)^2{(\tilde L_{z,\text{E}}^i)}^2\right]{w_i}^2\qquad \text{for}\ i=1,2\, ,
\end{multline}
and the one on the third torus is of the form
\begin{multline}\label{eq:closedStr_latEinstein}
    \mathcal{H}_{\text{lat, cl., $T^3_3$}}^{\mathcal{A}_{jj}, \text{E}}=\frac{e^{4\phi^{(1)}_\text{E}}\alpha'}{2}
    \left[\frac{{m'_3}^2}{{(L_{x,\text{E}}^3)}^2}+\frac{{n'_3}^2}{{(L_{y,\text{E}}^3)}^2}+\frac{{p'_3}^2}{{(\tilde L_{z,\text{E}}^3)}^2}\right]{w_3}^2\\
    +\frac{e^{-4\phi^{(1)}_\text{E}}}{2\alpha'}\Bigg[\left(\frac{{n'_3}^2}{{d_3}^2}{(L_{x,\text{E}}^3)}^2+\frac{{m'_3}^2}{{d_3}^2}{(L_{y,\text{E}}^3)}^2\right){k_3}^2
 +\left(\frac{n'_3 p'_3 x_3}{d_3}{(L_{x,\text{E}}^3)}^2-\frac{m'_3 p'_3 y_3}{d_3}{(L_{y,\text{E}}^3)}^2\right)2 k_3 l_3\Bigg.\\
\Bigg.+\left({p'_3}^2 {x_3}^2{(L_{x,\text{E}}^3)}^2+{p'_3}^2 {y_3}^2{(L_{y,\text{E}}^3)}^2+{d_3}^2{(\tilde L_{z,\text{E}}^3)}^2\right){l_3}^2\Bigg].
\end{multline}
The symbols $\Lambda_i$, $d_i$, $x_i$ and $y_i$ are some
number-theoretical constants (depending on the actual wrapping
numbers) and are defined in appendices \ref{AppSec:A_lattice} and
\ref{AppSec:A_lattice_nonvanangle}. Beware that the symbol $t$ in
equation (\ref{eq:AregVpot}) denotes time (and not the loop
channel modular parameter) but $l$ denotes the tree-channel
modular parameter. Interestingly, the actual angles $\kappa_i$
(and therefore the $\vartheta$-functions) do not depend on the
Einstein-rescaling of the metric as they are dimensionless. The
subtracted $1$ in (\ref{eq:Annulus_amplitude_ij_OR_treeEinstein})
is due to the regularization scheme. A general analysis of the
one-loop potential contribution
(\ref{eq:Annulus_amplitude_ij_OR_treeEinstein}) is much more
difficult. One has to perform a modular integration over $l$ and
within the integral there is a product of modular $\vartheta$ and
$\eta$-functions times the product of five infinite sums. The
fraction containing the $\vartheta$- and $\eta$-functions can be
$q$-expanded as stated in appendix \ref{AppSec:mod_fct}, equation
(\ref{eq:thetafctentw2winkel}), and different orders can be
integrated over $l$ separately (after the multiplication with the
Kaluza-Klein and winding sums). Due to the complicated structure
of the Hamiltonians, the Kaluza-Klein and winding summation later
in the example will performed only for a finite number of
summation terms. For each term, the integration over $l$ then is
performed analytically.

It is instructive to reformulate both the 1-loop potential
(\ref{eq:Annulus_amplitude_ij_OR_treeEinstein}) and the two
Hamiltonians (\ref{eq:closedStr_latAnnijEinstein}) and
(\ref{eq:closedStr_latEinstein}) in terms of the shape, overall
volume and volume ratio moduli, as defined in equation
(\ref{eq:shape_volume_moduli}). The following general observations can be
made:
\begin{enumerate}
\item The angles $\kappa_i$ do neither depend on the dilaton, the
overall volume or the volume ratios, only on the shape moduli.
Together with the terms in front of the modular integral, this
leads to a dilaton dependence $\sim\exp[-6\phi^{(1)}_\text{E}]$.
The remaining modular integral can be performed for every term of
the infinite sum separately. The structure of every single term
before integration is of the form
\begin{equation}
\sim \exp[-2\pi l (h_1 w_1^2+h_2 w_2^2+h_3 w_3^2+h_4 k_3^2 +h_5 l_3^2] \nonumber
\end{equation}
Every $h_i$ has a dependence on the dilaton of either $\sim
\exp[-4\phi^{(1)}_\text{E}]$ or $\sim \exp[4\phi^{(1)}_\text{E}]$.
After the integration has been performed (over all but the
constant divergent term which has been subtracted), the leading
order terms for large $\phi^{(1)}_\text{E}$ have a behavior $\sim
\exp[4\phi^{(1)}_\text{E}]$. Together with the terms in front,
this implies that the overall behavior of the one loop potential
(from the annulus amplitude) is at best
$\sim\exp[-2\phi^{(1)}_\text{E}]$, so the run-away behavior of the
dilaton cannot be cured at the one-loop level.

\item A similar reasoning can be performed for the overall volume
modulus $V_\text{tot}$. The dependence of all terms in front of
the modular integral together is $\sim V_\text{tot}^{1/3}$. Every
$h_i$ in the Hamiltonian of the infinite sums has a dependence of
either $\sim V_\text{tot}^{2/9}$ or $\sim V_\text{tot}^{-2/9}$.
After the integration, the leading order behavior (for large
$V_\text{tot}$) is $\sim V_\text{tot}^{2/9}$ for these terms and
the overall behavior of the potential is $\sim
V_\text{tot}^{5/9}$. Thus the overall volume tends to shrink to
zero also at the one-loop level (ignoring kinetic terms). One
seems to need non-perturbative effects of fluxes in order to
change this behavior.

\item The behavior of the other moduli is very much dependent on
the actual wrapping numbers of a model and is much more
promising as will be seen for the explicit example in the
following section.
\end{enumerate}

\section{An explicit example}\label{Sec:Example}
The most important consistency condition for any explicit string
theoretical model containing intersecting branes is the R-R
tadpole condition. It has been derived explicitly in terms of
wrapping numbers in section \ref{Sec:RRtp_genchapter} and is
stated in equations (\ref{eq:Rtadpole1})-(\ref{eq:Rtadpole4}). The
crucial question now is if there are models containing more than
two stacks of branes, where the intersection between different
stacks are not all parallel on a specific $T^3$. As such models
globally break Poincare invariance down to (2+1) or (1+1)
dimensions, it would most likely contain new features as compared
to the old picture on the $T^6$ (where (3+1)-dimensional Poincare
invariance is conserved for all times by definition).

It is much more involved to give a systematic analysis on all R-R
charge cancelling models in the new framework, simply because
there are nine instead of six wrapping numbers for every
stack\footnote{If we only take into account all integer wrapping
numbers from -3 up till 3, this in total implies $7^{l\cdot 9}$
different possibilities, compared to $7^{l\cdot 6}$ on the $T^6$,
where $l$ denotes the number of stacks.}. Especially for
extensions of the well-known standard model of
\cite{Ibanez:2001nd} this is problematic as it employs at least
$l=4$ stacks. The purpose of this section is not to give one
specific completely satisfying model already, but simply to get an
understanding of the general picture, some more detailed explicit
constructions will follow in the future\cite{Ott2:2005}. The model
which is presented here is a toy model.

A computer program was set up in order to search for such a
3-stack toy model which fulfills the R-R tadpole cancellation
condition on the most simple $\bf AAA$-torus. One completely
arbitrary model that has been found is stated explicitly in table
\ref{tab:explicit_3stackexample}. It fulfills the R-R tadpole
condition and the intersections between the different stacks are
not all parallel on the first and third $T^3$, but they are on the
second torus. This implies that total Poincare invariance is
broken down to (1+1)-dimensions.
\begin{table}
\centering
\sloppy
\renewcommand{\arraystretch}{1.3}
\begin{tabular}{|l|ccc|ccc|ccc|}
  \hline
   stacksize & $m_1'$& $n_1'$ & $p_1'$ & $m_2'$& $n_2'$ & $p_2'$ & $m_3'$& $n_3'$ & $p_3'$\\
  \hline\hline
    $N_a$=4 & 0 &  0 &  1 &  0 &  1 &  1 &  1 &  0 &  4\\
\hline
    $N_b$=2 & -1 & -1 & -1 & 0 &  1 &  0 &  1 &  -1 & 0\\
\hline
    $N_c$=1 & 3 &  0 &  2 &  0 &  1 &  0 &  -1 & -1 & 0\\
\hline
\end{tabular}
\caption{The wrapping numbers of the three stack model on the $\bf AAA$-torus discussed in
the text.} \label{tab:explicit_3stackexample}
\end{table}
Supersymmetry is not conserved, and the next step is to discuss
the complete scalar potential in Einstein frame, starting with the
tree-level potential.

The overall volume dependence of the tree-level potential is
generally $\sim {V_\text{tot}}^{1/6}$, and the dilaton dependence
is $\sim \exp[-4\phi^{(1)}_\text{E}]$. There is no dependence on
the volume ratio moduli $V_{13}$ and $V_{23}$ between the
different $T^3_i$, meaning that (if kinetic terms play no crucial
role), all three $T^3_i$ change volume uniformly.

\begin{figure}[t!]
\includegraphics[scale=0.8]{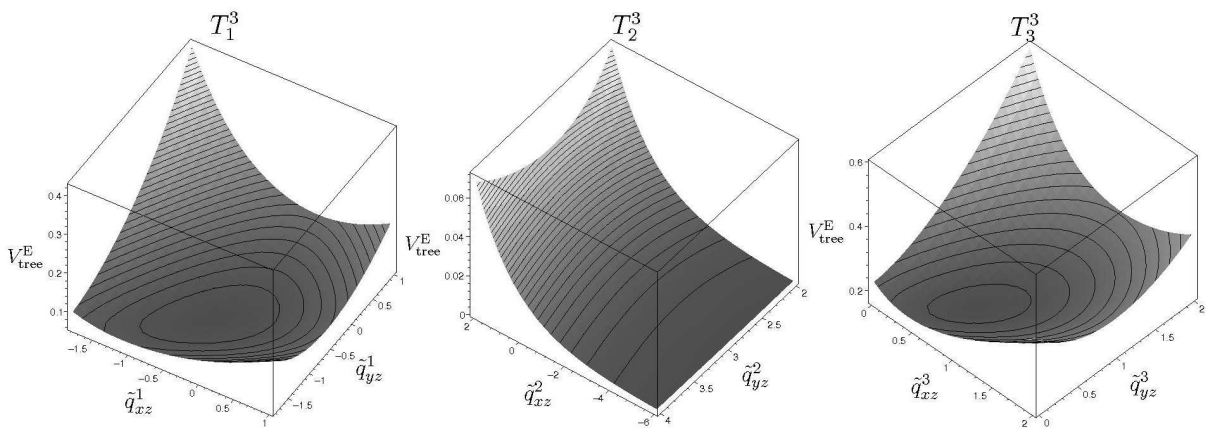}
\caption{Complete tree-level scalar potential dependence on
canonically normalized shape moduli for the explicit example. The
plots are shown on each $T^3_i$ separately, with the other shape
moduli fixed.} \label{Fig:BIpot_example}
\end{figure}

In order to be able to plot the potential for the remaining
moduli, we therefore fix both dilaton and overall volume terms
identically to one. Furthermore, we will fix both the string
coupling $e^{\tilde\phi^{(1)}_0}$ and $\alpha'$ identically to
one, the latter meaning that we measure all dimensionful scales in
terms of the string length. The total tree-level potential then
(containing the different terms from all stacks and the
orientifold plane) is shown in figure \ref{Fig:BIpot_example} for
the canonical variables (\ref{eq:can_var_einst}). For any of the
three plots (corresponding to one of the different $T^3_i$), the
shape moduli of the other two $T^3_i$ have been fixed to one. This
ignores the interaction between the different $T^3_i$, but seems
to be a good approximation locally close to the string length.
This is underlined by the fact that there are no kinetic mixing
terms between the shape moduli of the different $T^3_i$, see equation
(\ref{eq:kinTerms_shapevolume}).

It can be seen on the plots that a different kind of behavior is
possible on the three $T^3_i$, depending completely on the
particular chosen wrapping numbers on that torus. On the first
$T^3_1$ and the third $T^3_3$, the potential has indeed a minimum
in which both shape moduli $q_{xz}^i$ and $q_{yz}^i$ (for $i$=1,3)
can be stabilized. These local minima can be verified by partially
differentiating the potential (\ref{eq:treepot_kanvar_einst}).
They are lying at the numerical values $(q_{xz}^1\approx
0.520,q_{yz}^1\approx 0.249)$ and $(q_{xz}^3\approx
3.539,q_{yz}^3\approx 2.767)$ (note that in the figure the
canonically normalized variables have been plotted). The situation
is different on the second $T^3_2$. There is no local minimum for
both shape moduli $q_{xz}^2$ and $q_{yz}^2$ at the same time.
(which also can be checked numerically by differentiation). In
particular, the modulus $q_{xz}^2$ has no local minimum, but the
modulus $q_{yz}^2$ has one if we hold $q_{xz}^2$ constant. The
reason for this behavior is easy to see looking at the wrapping
numbers: $m_2'=0$ for all stacks of branes. No brane wraps around
this cycle in the reciprocal lattice. This implies that the
intersections between all branes on the second $T^3_2$ stretch
parallel in the $L_x^2$-direction, so this is the common
space-time direction on that torus. According to the second plot
in \ref{Fig:BIpot_example}, the radius $L_x^2$ shrinks compared to
the other two. This is a very interesting and general result, as
it means that by wrapping all cycles on a certain $T^3$, we at
least stabilize the ratios between the three radii at the
tree-level, such that no radius shrinks to zero compared to the
other two ones, this happens in the common picture that the
$T^3$ can be factorized into $T^3=T^2\times T^1$. This is of
course not yet enough, as we would actually like one radius to
grow and the other two to be stabilized compared to each other at
a constant scale. In other words, we would like to have one $q$ to
be growing appropriately and the other one to be stabilized. This
might still happen at the one-loop level, so we now take a look at
the one-loop potential.
\begin{figure}[t!]
\includegraphics[scale=0.8]{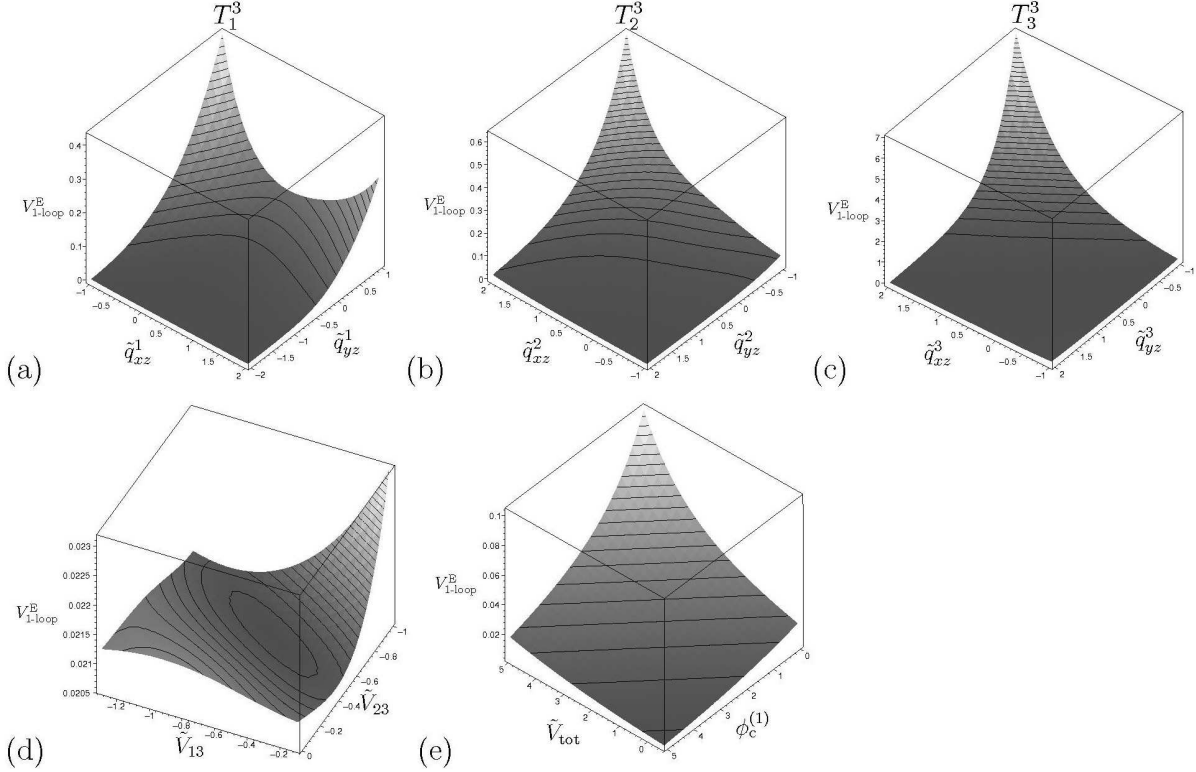}
\caption{One-loop scalar potential dependence on canonically
normalized moduli for the explicit example from annulus amplitude
$\mathcal{A}_{bc}$. The moduli not occurring in a certain plot are
fixed.} \label{Fig:1Loop_pot_example}
\end{figure}

The potential from the annulus amplitude
$\tilde{\mathcal{A}}_{bc}$ is calculated exemplarily, it is
exactly of the form as treated in chapter
\ref{Sec:oneloopPotential}. But at the same time, we have to be
aware that it only gives a part of the whole picture and that the
sum of all amplitudes might still change it even qualitatively
(especially as we disregard the first stack of branes).
Nevertheless, we will surely gain some understanding. The one loop
potential is calculated numerically, after an analytical
integration of the modular integral over $l$. The constant part in
the expansion of the $\vartheta$- and $\eta$- functions is folded
with the five series, for those only the terms from -1 up to 1 are
being taken into account. The NS-NS tadpole which is the only
divergent term in the integral is getting subtracted for
regularization. The resulting potential is plotted in figure
\ref{Fig:1Loop_pot_example} for several pairs of the scalar moduli
varying, with the others fixed. All the fixed moduli in every plot
are taken equal to one (besides the dilaton which is taken to be
$\exp[\phi^{(1)}_\text{E}]=1$) and also $\exp[\tilde\phi^{(1)}_0]$
and $\alpha'$ are set to one. Some comments immediately can be
given:
\begin{enumerate}
\item Plot (e) in figure \ref{Fig:1Loop_pot_example} shows the
potential dependence of the one-dimensional dilaton and the
overall volume which has been anticipated generally already in the
end of section \ref{Sec:oneloopPotential}. The dilaton still has a
runaway potential which is at least decaying as
$\sim\exp[-2\phi^{(1)}_\text{E}]$, the overall volume dependence
is decaying at least as $\sim V_\text{tot}^{5/9}$ (the plot shows
the canonically normalized variables).

\item Plot (d) shows the two volume ratio moduli $\tilde V_{13}$
and $\tilde V_{23}$ (which measure the relative ratio of each two
of the $T^3$ volumes). These two moduli have a flat potential on
the tree-level and it seems that at the one-loop level they can be
stabilized even at a local minimum. This is a very nice feature,
because it means that all three tori grow (or decay) in size
simultaneously, none of them changes size with a different scaling
from the other two. This seems to be a requirement for the
observed isotropy of the visual universe.

\item Plot (a)-(c) show the two shape moduli of every $T^3_i$ in
relation to each other. The results are quite different from those
at the tree level. On the first torus, figure (a), one modulus is
being stabilized and the other one runs towards zero size. At the
tree level, both generically are being stabilized (it was guessed
that this is the general behavior if all cycles on the $T^3_i$ are
being wrapped by some brane). So at the one-loop level, for one
modulus this behavior is changed. An even more interesting
behavior occurs on the second and third torus, figures (b) and
(c), one modulus is being stabilized at zero size, the other one
in all likelihood shows a runaway behavior. This is the behavior
we actually wanted from the start, as it means that one radius
grows in size compared to the other two and these are stabilized
compared to each other. It is of course far from being obvious
that this solves our problem, as the relative importance of
tree-level and one-loop potential depends on many different things,
mainly on both the vacuum expectation value of the dilaton and the
time-dependent Einstein frame dilaton. Nevertheless, it is very
encouraging to see that the scaling behavior of the potential
dependence on the time-dependent dilaton can be different for
tree-level and one-loop potential. A direct comparison between the
general dilaton dependence on the tree-level and the numerically
calculated dependence on the one-loop level (for the given
example) shows that for a larger overall volume, the one-loop
contributions can grow in importance compared to the tree-level
contributions. It is of course an intricate task to show that the
string perturbation series then still can converge. It can be
guessed that a working mechanism of this kind will require some
fine-tuning.
\end{enumerate}


%
\newpage
\section{Conclusions and prospects}\label{Sec:Conclusions}
In this paper, the general formalism to describe D6-branes on a
orientifolded background $T^3\times T^3\times T^3$ has been
developed, where the D-branes are allowed to wrap general 2-cycles
on every $T^3$. The one-loop partition function has been derived
and the R-R tadpole equations have been calculated. As expected,
the result for the R-R tadpole is homological and depends only on
the stacksizes and the nine topological wrapping numbers of every
stack of branes. This determines the chiral massless fermion
spectrum.

The complete scalar moduli potential on the tree-level has been
derived. The general results are that the 1-dimensional Einstein
frame dilaton has a runaway potential
$\sim\exp[-4\phi^{(1)}_\text{E}]$ and that the overall $T^9$
volume dependence is $\sim {V_\text{tot}}^{1/6}$. The two ratios
$V_{13}$ and $V_{23}$ between the relative volumes of the 3-tori
do not receive a potential contribution at this level. The picture
for the six shape moduli $q_{xz}^i$ and $q_{yz}^i$ (i=1,2,3)
depends on the actual wrapping numbers and stacksizes of all
D-branes. It seems that they can be stabilized if some D-brane
wraps around the corresponding elementary cycle. If no D-brane
wraps around certain cycles (as is the case for the old known
(3+1)-dimensional Poincare invariant solutions in this new
framework), the corresponding shape moduli shrink to zero size.

Also the one-loop potential from the annulus amplitudes is derived
in a general form, where the modular integration still has to be
performed. This makes it technically rather difficult to obtain
general (not model dependent) results for a complicated brane
configuration. The general results are that the 1-dimensional
dilaton in Einstein frame still has a runaway potential and that
the overall volume still shrinks to zero. On the other hand,
$V_{13}$ and $V_{23}$ are typically stabilized at the one-loop.
This is of course a very important feature, as it assures
anisotropy of the three large spatial directions of the universe
at late times (provided that the rest of the proposed mechanism
works). Furthermore, the shape moduli can actually run off to
infinity, run to zero or even stay stabilized (as on the tree
level), this is completely dependent on the actual wrapping
numbers.

So far only annulus amplitudes were considered, but it is well
possible that the M\"obius amplitude (involving the orientifold
planes) can give opposite sign contributions to the potential and
this could alter the general picture. One can also think about
other possibilities to stabilize the dilaton and to alter the
overall volume dependence, the two being the two most important
problems of the present construction. They could be solved for
example by taking into account non-perturbative effects or
by adding fluxes. It could also be that kinetic terms play an
important role in the evolution, but this most probably requires
the potential for the overall volume to be very flat.

Then it will be necessary to think more quantitatively about the
shape moduli. How large do the wrapping numbers have to be in order
to describe such an enormous size difference between the different
radii as by a factor $10^{30}$? It is interesting to note that in
the explicitly discussed model, the relative minima where the shape
moduli are stabilized at the tree-level are directly related to
the wrapping numbers. The canonically normalized shape moduli are
on a logarithmic scale, this suggests that a difference of 60
$e$-foldings (as described by inflation) could only need wrapping
numbers of a size like 60. Then the next question would be, if it
is possible to get phenomenologically interesting spectra (like
that of the standard model) for the same wrapping numbers in the
late time picture? From experience this seems very well possible,
see for instance \cite{Blumenhagen:2001te}. The global evolution
of the universe would be intimately related to the massless
fermion spectrum in a single model, this surely would make such a
model much more falsifiable (or verifiable) and bring string
theory closer to its original goal of describing very different
energy scales at the same time. The technical tools are there, but
the enormous number of potential models makes it rather difficult to
handle.

A general $T^3$ has six moduli, as shown in appendix
\ref{AppSec:modulisp_T3}, of which in the present ansatz only five
are present, the three radii and two angles which are fixed by the
orientifold projection to discrete values. It will be interesting
to extend this formalism in the future to include also the
additional angle modulus. Another future step could be to see if
orbifolding the $T^3$ could cure some of the present problems as
on the $T^2$, see for instance \cite{Blumenhagen:2001te,
Blumenhagen:2002gw, Honecker:2004kb}. Naively, one just has to
make the transition from the five 2-dimensional Bravais lattices
to the fourteen 3-dimensional ones. This could also bring $N=1$
supersymmetry back into the game (supersymmetric configurations
could be limiting cases that the system dynamically evolves to).

To go even further, other more general 3-manifolds could be
considered as a possible background, but this surely would require
extraordinary technical expenses.

There is even the possibility that the general philosophy proposed
in this paper even allows for a quantization of the moduli space
itself in a simple setting (like the $T^3\times T^3\times T^3$),
instead of just constructing string theory on this classical
background. But this surely is an even farther reaching goal.

\vspace{1cm}

\noindent {\large \bf Acknowledgments}
\vspace{0.2cm}

\noindent I would like to thank S.~Bruers, M.~Billo, A.~Keurentjes
and L.~Martucci for helpful and lengthy discussions. Then the very
helpful advices and also the interest from the mathematician Joost
van Hamel and furthermore Wouter Castryck and Hendrik Hubrechts
have to be mentioned with best thanks. Further nice discussions
include R.~ Blumenhagen, A.~Krause, A.~Miemiec, I.~Kirsch,
I.~Runkel, A.~Celi, F.~Passerini, J.~Adam, G.~Smet, P.~Smyth,
J.~Van den Bergh, D.~Van den Bleeken, W.~Troost, L.~G\"orlich,
L.~Huiszoon and A.~Sen. For proofreading of the article I would
like to thank J.~Rosseel and A.~van~ Proeyen and at the same time
B.~K\"ors for helpful comments on an earlier stage of the
manuscript.

This work is supported in part by the European Community's Human
Potential Programme under contract MRTN-CT-2004-005104
`Constituents, fundamental forces and symmetries of the universe'.
The work of T.O. is supported in part by the FWO - Vlaanderen,
project G.0235.05 and by the Federal Office for Scientific,
Technical and Cultural Affairs through the "Interuniversity
Attraction Poles Programme -- Belgian Science Policy" P5/27.

\begin{appendix}
\section{Topology of 2-cycles on the $T^3$}\label{AppSec:topology}
At the beginning of this section, we will show in a very simple
way that a 2-cycle on a $T^3$, being characterized by the two
angles $\phi$ and $\theta$ from equation
(\ref{eq:anglesT9_endform}) can be uniquely described by a
primitive vector in the reciprocal lattice of the 3-torus, eq.
(\ref{eq:reciprocal_torus}). The two angles include three quantum
numbers $n'$, $m'$ and $p'$ that are coprime. A 2-dimensional
cycle that spans the $(x,y)$-plane of the $T^3$ trivially can be
described by a perpendicular vector along the z direction. Two
orientations are possible. We assign to the plane facing 'upwards'
the vector $(0,0,1/L_z)$. If we now apply the rotation matrix
$\mathbf{O}$ from (\ref{eq:gen_rotationmatrix}) together with
(\ref{eq:anglesT9_endform}) on this vector, the resulting vector
is proportional to
\begin{equation}\label{eq:gen_D2brane}
\vec{n}\equiv\left(\frac{m'}{2\pi L_x},\frac {n'}{2\pi L_y}, \frac{p'}{2\pi \tilde L_z}\right).
\end{equation}
This is a general primitive vector in the reciprocal lattice,
because $n'$, $m'$ and $p'$ are arbitrary coprime integers.
Consequently, there is a one-to-one correspondence between
primitive vectors in the reciprocal lattice of $T^3$ and possible
2-cycles in the lattice of $T^3$.
\subsection{Lattice definitions}\label{AppSec:Basis3Tori}
Every $T^3$ has the three fundamental cycles $2\pi L_x
\mathbf{e}_1$, $2\pi L_y \mathbf{e}_2$ and $2\pi \tilde L_z
\mathbf{e}_3$, where the lattice vectors $\mathbf{e}_i$ are
normalized to $(\mathbf{e}_i)^2=1$. The reciprocal lattice is
defined by
\begin{equation}\label{eq:reciprocal_torus}
\mathbf{e}_i\ \mathbf{e}^\ast_j=\delta_{ij},
\end{equation}
where the basis takes the form $1/(2\pi L_x) \mathbf{e}_1^*$,
$1/(2\pi L_y) \mathbf{e}_2^*$ and $1/(2\pi \tilde L_z) \mathbf{e}_3^*$.

\subsubsection{The $\bf A$-torus}
The $\bf A$-torus can be defined by the following lattice vectors
\begin{equation}
\mathbf{e}_1=\mathbf{e}_1^*=\left(\begin{array}{c} 1\\ 0 \\ 0 \end{array}\right),\qquad
\mathbf{e}_2=\mathbf{e}_2^*=\left(\begin{array}{c} 0\\ 1 \\ 0 \end{array}\right),\qquad
\mathbf{e}_3=\mathbf{e}_3^*=\left(\begin{array}{c} 0\\ 0 \\ 1 \end{array}\right).
\end{equation}
The lattice and reciprocal lattice vectors can be chosen to take
the same form in this type of torus, as the axes are all
perpendicular. The reciprocal lattice vectors $\mathbf{e}_i^*$
therefore are also normalized to $(\mathbf{e}_i^*)^2=1$.
\subsubsection{The $\bf B$, $\bf C$ and $\bf D$-tori}
The other three discussed tori can be defined by the following
lattice vectors
\begin{equation}\label{eq:bcd_torus}
\mathbf{e}_1=\left(\begin{array}{c} \sqrt{1-\left(b_1\frac{\tilde L_z}{L_x}\right)^2}\\ 0 \\ b_1 \frac{\tilde L_z}{L_x} \end{array}\right),\qquad
\mathbf{e}_2=\left(\begin{array}{c} 0\\ \sqrt{1-\left(b_2\frac{\tilde L_z}{L_y}\right)^2} \\ b_2 \frac{\tilde L_z}{L_y}\end{array}\right),\qquad
\mathbf{e}_3=\left(\begin{array}{c} 0\\ 0 \\ 1 \end{array}\right).
\end{equation}
They lead to the following reciprocal lattice vectors
\begin{equation}\label{eq:bcd_torus_rec}
\mathbf{e}_1^*=\left(\begin{array}{c} 1/{\sqrt{{1-\left(b_1\frac{\tilde L_z}{L_x}\right)^2}}}\\ 0\\ 0 \end{array}\right),\
\mathbf{e}_2^*=\left(\begin{array}{c} 0\\1/{\sqrt{{1-\left(b_2\frac{\tilde L_z}{L_y}\right)^2}}}\\ 0 \end{array}\right),\
\mathbf{e}_3^*=\left(\begin{array}{c} \frac{-b_1{\tilde L_z}}{\sqrt{{L_x}^2-{b_1}^2{\tilde L_z}^2}}\\
 \frac{-b_2{\tilde L_z}}{\sqrt{{L_y}^2-{b_2}^2{\tilde L_z}^2}}\\ 1 \end{array}\right).
\end{equation}
In these definitions, $b_1$ and $b_2$ independently can take one of the
two values $0$ or $1/2$.
\subsection{Intersection numbers}\label{AppSec:top_intnr}
In this section, a mathematical derivation of the conjectured
topological intersection number of two 2-dimensional branes, equation (\ref{eq:intersectionnumber}), on a
$T^3$ will be given. The two 2-dimensional cycles (denoted by $a$
and $b$) can be understood as 2-dimensional subtori of the original $T^3$, i.e.
\begin{equation}
T^2_a,\ T^2_b \subset T^3.
\end{equation}
Then the intersection of the two is a 1-torus $T^1_I$ times the collection of torsion points $V_I$ of $T^3$,
\begin{equation}
T^2_a\cap T^2_b = T^1_I \cdot V_I.
\end{equation}
The number of these torsion points $\#V_I$ corresponds to the
topological intersection number (without orientation). A 2-torus
most easily can be represented by a primitive vector in the
reciprocal lattice of the original $T^3$, equation
(\ref{eq:gen_D2brane}). This means that the intersecting one-torus
$T^1_I$ corresponds to the simultaneous solution of the two linear
diophantine equations
\begin{align}
&m_a'x+n_a'y+p_a'z=0, \\
&m_b'x+n_b'y+p_b'z=0.  \nonumber
\end{align}
A solution is given by
\begin{equation}\label{eq:vector_int}
x=p_b'n_a'-p_a'n_b', \qquad y=m_b'p_a'-m_a'p_b', \qquad z=n_b'm_a'-m_b'n_a' .
\end{equation}
The number of torsion points therefore is given by
\begin{equation}\label{eq:number_torsionpoints}
\#V_I=\gcd\left(p_b'n_a'-p_a'n_b',\ m_b'p_a'-m_a'p_b',\
n_b'm_a'-m_b'n_a'\right)
\end{equation}
We define the (unoriented) topological intersection number between
two 2-cycles on the $T^3$ by
\begin{equation}
I_{ab}^\text{u}\equiv \#V_I.
\end{equation}
There is no canonical mathematical way to attach a sign to this
intersection number (although it is of course possible to do so by
hand). This is easily understandable: an intersection is actually
described by a vector (\ref{eq:vector_int}) that has a
well-defined direction in the 3-dimensional space (but no sign)
and it can be reduced by the greatest common divisor of its
components (\ref{eq:number_torsionpoints}).

The standard unoriented intersection number of two 1-cycles on a
2-torus in our notation corresponds to the case when the two
branes both wrap either the cycles $m_a'=m_b'=0$ or
$n_a'=n_b'=0$, meaning that both vectors orthogonal to the
branes lie either in the $(y,z)$- or $(x,z)$-plane of the
reciprocal lattice. In this case, the sign of the oriented
intersection number is simply established by choosing an oriented
'reference plane' and then by checking if the vector of the
intersection is 'ingoing' or 'outgoing'. This reference plane is just
given by the $(x,z)$- or $(y,z)$-plane itself, respectively.

Therefore, generally in the 3-dimensional case, we can only assign
a sign to the intersection number (it will determine for
instance the chirality of bifundamental representations) by
explicitly choosing a 2-dimensional reference plane. In other
words we have to specify what our internal space (2-dimensional
per 3-torus) as opposed to the 4-dimensional spacetime exactly is.
\subsection{Angle between branes}
An oriented 2-cycle is characterized by a normal vector $\vec{n}$ in the
3-dimensional reciprocal lattice. The angle between two such 2-cycles therefore can
simply be obtained by the standard linear algebra formula
\begin{equation}
\alpha=\arccos\left(\frac{\vec{n}_a\cdot\vec{n}_b}{\|\vec{n}_a\| \|\vec{n}_b\|}\right).
\end{equation}
This angle is the smaller angle $(\leq \pi)$ between the two
normal vector, therefore it is unoriented in the sense that
exchanging brane $a$ with brane $b$ does not change the angle in
between them. It takes the explicit form
\begin{equation}
\alpha_i=\arccos\left(\frac{n^{a\prime}_i
n^{b\prime}_i{L_x^i}^2{(\tilde L_z^i)}^2+m^{a\prime}_i
m^{b\prime}_i{L_y^i}^2{(\tilde L_z^i)}^2+p^{a\prime}_i
p^{b\prime}_i{L_x^i}^2{L_y^i}^2}{v_i^a v_i^b}\right),
\end{equation}
where
\begin{equation}
v_i^l\equiv \sqrt{{n^{l\prime}_i}^2{L_x^i}^2{(\tilde
L_z^i)}^2+{m^{l\prime}_i}^2{L_y^i}^2{(\tilde
L_z^i)}^2+{p^{l\prime}_i}^2{L_x^i}^2{L_y^i}^2}
\end{equation}
is the 2-dimensional volume of the brane $l$ on one 3-torus $T^3_i$.
\section{Moduli space of a general $T^3$}\label{AppSec:modulisp_T3}
A general $T^k$ can be written as $\mathbb{R}^k/\mathbb{Z}^k$.
This means that the dimension of the coset
\begin{equation}\label{eq:coset}
GL(k,\mathbb{R})/SO(k,\mathbb{R})
\end{equation}
gives the dimension of the general $T^k$ moduli space. The
$GL(k,\mathbb{R})$ can be written as $\mathbb{R}\times
SL(k,\mathbb{R})$, where $\mathbb{R}$ is simply the determinant.
According to the Iwasawa
decomposition\footnote{I would like to thank A.~Keurentjes for
pointing me to this decomposition.}, every element $g\in
SL(k,\mathbb{R})$ can be written as
\begin{equation}
g=k\, a\, n\, ,
\end{equation}
where $k$ is an element of the maximal subgroup, which is given by
$SO(k)$, $a$ is an element of the maximal abelian subgroup, and
$n$ is an element of the nilpotent group. The maximal subgroup
cancels against the $SO(k,\mathbb{R})$ of the coset
(\ref{eq:coset}), and remaining are only the determinant and the
abelian subgroup, the elements of those together concretely can be
understood as the elementary radii, and the nilpotent group, which
can be understood as the angles between the elementary radii. For
example, a $T^2$ has two elementary radii and one angle, a $T^3$
has three elementary radii and three angles, so altogether six
moduli.

For the case of an orientifolded  $T^2$, the third modulus
(explicitly the angle) is fixed by the orientifold plane to only
discrete values, so only the two radii remain. A similar thing
happens in the case of the $T^3$: the orientifold projection fixes
two of the three elementary torus angles to a discrete value. This
is explained in more detail in section \ref{Sec:nsns2formflux}.
This of course means that one angle remains unfixed. In the main
text this angle is set by hand to 90 degrees, but the whole ansatz
might be extended by also taking into account this additional
angle in the future.

\section{Klein bottle lattice contributions}\label{AppSec:KB_lattice}
The Kaluza-Klein momenta and winding modes generally take the following form
\begin{align}\label{eq:def_P}
&\mathbf{P}=s_1/L_x \mathbf{e}_1^*+s_2/L_y \mathbf{e}_2^*+s_3/\tilde L_z \mathbf{e}_3^*,\\
&\mathbf{L}=\frac{1}{\alpha'}\big(r_1 L_x \mathbf{e}_1+r_2 L_y \mathbf{e}_2+r_3 \tilde L_z \mathbf{e}_3\big).\label{eq:def_L}
\end{align}
The lattice contribution to the closed sting Hamiltonian is generally given by
\begin{equation}\label{eq:def_clstrHam}
\mathcal{H}_{\text{lattice, cl.}}=\frac{\alpha'}{2}\left(\mathbf{P}^2+\mathbf{L}^2\right).
\end{equation}
In the discussed toroidal orientifold, equation (\ref{eq:prop_modelIIA})
together with (\ref{eq:orientifold_def}), the worldsheet parity
transformation $\Omega$ acts as
\begin{equation}
\Omega : \qquad \mathbf{P} \xrightarrow{\Omega } \mathbf{P}, \qquad\qquad
\mathbf{L} \xrightarrow{\Omega} -\mathbf{L} \ ,
\end{equation}
whereas the reflection $R$ acts as:
\begin{equation}
R : \qquad P_x \xrightarrow{R } P_x,\ \ P_y \xrightarrow{R } P_y,\ \ P_z \xrightarrow{R } -P_z,\ \
L_x \xrightarrow{R} L_x,\ \ L_y \xrightarrow{R } L_y,\ \ L_z \xrightarrow{R } -L_z.
\end{equation}
Therefore, the combined action is given by:
\begin{equation}\label{eq:OmegaR_action}
\Omega R : \qquad P_x \xrightarrow{R } P_x,\ \ P_y \xrightarrow{R } P_y,\ \ P_z \xrightarrow{R } -P_z,\ \
L_x \xrightarrow{R} -L_x,\ \ L_y \xrightarrow{R } -L_y,\ \ L_z \xrightarrow{R } L_z.
\end{equation}
Keeping just the invariant terms under (\ref{eq:OmegaR_action}), leads to the lattice contribution
\begin{equation}\label{eq:closedStr_latOR}
    \mathcal{H}_{\text{lattice,
cl.}}^{\mathcal{K}}=\frac{\alpha'}{2}\left(\frac{s_1^2}{L_x^2}+\frac{s_2^2}{L_y^2}\right)+\frac{1}{2\alpha'}\left(r_3^2
\tilde L_z^2\right).
\end{equation}
\section{Annulus lattice contributions}\label{AppSec:KB_lattice}
\subsection{Case of vanishing angles between branes}\label{AppSec:A_lattice}
After mapping the boundary conditions (\ref{eq:bc_xyz}) for the
open string to the corresponding closed string boundary conditions
and additionally performing T-duality, one obtains
\begin{align}\label{eq:bc_xyz_closed}
&\partial_\tau X_i + \mathcal{B}_y^i \partial_\tau \tilde Z_i=0\ , \nonumber\\
&\partial_\tau Y_i - \mathcal{B}_x^i\partial_\tau \tilde Z_i=0\ ,\\
&\partial_\sigma \tilde Z_i - \mathcal{B}_y^i\partial_\sigma X_i+\mathcal{B}_x^i \partial_\sigma Y_i=0\ ,\nonumber
\end{align}
Using the fact that $\partial_\tau X_i=\alpha' {\mathbf{P}_x^{i}}$
and $\partial_\sigma X_i=\alpha' {\mathbf{L}_x^{i}}$(similarly for
$Y_i$ and $\tilde Z_i$) and the definitions (\ref{eq:def_P}) and (\ref{eq:def_L}),
we obtain the following set of equations in the D6-branes picture,
\begin{align}
&s_1^i+\frac{m_i}{N}r_3^i=0\ ,\label{eq:a_bc_1}\\
&s_2^i+\frac{n_i}{N}r_3^i=0\ ,\label{eq:a_bc_2}\\
&s_3^i-\frac{m_i}{N}r_1^i-\frac{n_i}{N}r_2^i=0\ .\label{eq:a_bc_3}
\end{align}
These equations are linear diophantine equations and one has to be
careful in order to find the most general integer solution and not
just a subset, we will do this in some detail. The first two
equations (\ref{eq:a_bc_1}) and (\ref{eq:a_bc_2}) have to be
solved simultaneously, whereas (\ref{eq:a_bc_3}) is independent
from them and can be solved apart.
\subsubsection{Solving (\ref{eq:a_bc_1}) and (\ref{eq:a_bc_2})}
Applying our redefinitions (\ref{eq:umschreibung_nm1}) of the main text, the first set
takes the form
\begin{align}
&p_i's_1^i+{m_i'}r_3^i=0,\label{eq:a_bc_umg1}\\
&p_i's_2^i+{n_i'}r_3^i=0,\label{eq:a_bc_umg2}
\end{align}
where $n_i'$, $m_i'$, $p_i'$ are coprime. This fact implies that after defining
\begin{equation}
d_1^i=\gcd(p_i',m_i'),\qquad\qquad d_2^i=\gcd(p_i',n_i'),
\end{equation}
that $\gcd(d_1^i,d_2^i)=1$. Furthermore, we define
\begin{equation}
\hat p_i=d_1^i p_i',\qquad\qquad \hat m_i=d_1^i m_i',
\end{equation}
and solve first for (\ref{eq:a_bc_umg1}) in terms of $\hat
p_i$ and $\hat m_i$. The result is given by
\begin{equation}\label{eq:resglg1}
s_1^i=\hat m_i k_i, \qquad\qquad r_3^i=-\hat p_i k_i,\qquad \text{where}\qquad k_i\in \mathbb{Z}.
\end{equation}
Similarly, after defining
\begin{equation}
\Hat{\Hat{p}}_i=d_2^i p_i',\qquad\qquad \Hat{\Hat{n}}_i=d_2^i n_i',
\end{equation}
we find for (\ref{eq:a_bc_umg1}) the solution
\begin{equation}\label{eq:resglg2}
s_2^i=\Hat{\Hat{n}}_i l_i, \qquad\qquad r_3^i=-\Hat{\Hat{p}}_i l_i,\qquad \text{where}\qquad l_i\in \mathbb{Z}.
\end{equation}
These two results (\ref{eq:resglg1}) and (\ref{eq:resglg2}) are
surely not the simultaneous solution of (\ref{eq:a_bc_umg1}) and (\ref{eq:a_bc_umg2}), but it should be contained as a
subset. The most general result for $r_3^i$ is given by the set
\begin{equation}
\Big\{ -\frac{p_i}{d_1^i} k_i \Big\} \cap \Big\{ -\frac{p_i}{d_2^i} l_i \Big\}\ .
\end{equation}
Therefore $d_1^i l_i=d_2^i k_i$ and thus
\begin{equation}
k_i=d_1^i w_i, \qquad\qquad l_i=d_2^i w_i,\qquad \text{where}\qquad w_i\in \mathbb{Z}.
\end{equation}
Therefore, the simultaneous solution for (\ref{eq:a_bc_umg1}) and
(\ref{eq:a_bc_umg2}) is given by
\begin{equation}
s_1^i=m_i' w_i, \qquad s_2^i=n_i' w_i, \qquad r_3^i=-p_i'w_i,\qquad \text{where}\qquad w_i\in \mathbb{Z}.
\end{equation}
\subsubsection{Solving (\ref{eq:a_bc_3})}
The third equation of the boundary conditions takes the following form,
\begin{equation}
p_i's_3^i-{m_i'}r_1^i-{n_i'}r_2^i=0,\label{eq:a_bc_umg3}
\end{equation}
where again $n_i'$, $m_i'$, $p_i'$ are coprime. This is a linear
diophantine equation in three unknowns, the general solution is
known to be\footnote{We thank the three mathematicians Joost van
Hamel, Wouter Castryck and Hendrik Hubrechts to provide this
result.}
\begin{align}
&r_1^i=k_i\frac{n_i'}{d_i}+l_i p_i x_i,\\
&r_2^i=-k_i\frac{m_i'}{d_i}+l_i p_i y_i,\nonumber\\
&s_3^i=l_i d_i,\nonumber
\end{align}
where $k_i,l_i \in \mathbb{Z}$ are arbitrary integers,
$d_i=\gcd(n_i',m_i')$ and $x_i,y_i\in \mathbb{Z}$ have to fulfill
the equation
\begin{equation}\label{eq:mx_ny_d}
m_i'x_i+n_i'y_i=d_i.
\end{equation}
$x_i$ and $y_i$ are not unique, but a different choice for them
does not change the lattice as it just shifts the first lattice
vector by a finite integer $r$, such that $k_i'=r+k_i$. On the other
hand, no explicit solution for $x_i$ and $y_i$ can be given, it
has to be determined case by case using for instance the Euclidean
Divison algorithm. For the tadpole
calculation this actually is no problem at all, because all terms
including $x_i$, $y_i$ and $d_i$ chancel against each other, as we will see later.

The closed string lattice Hamiltonian, being defined by equation
(\ref{eq:def_clstrHam}), takes the final form for the annulus
amplitude
\begin{multline}\label{eq:closedStr_latAnn}
    \mathcal{H}_{\text{lattice, cl.}}^{\mathcal{A}}=\frac{\alpha'}{2}
    \left(\frac{{m'_i}^2}{{L_x^i}^2}+\frac{{n'_i}^2}{{L_y^i}^2}+\frac{{p'_i}^2}{{(\tilde L_z^i)}^2}\right){w_i}^2
    +\frac{1}{2\alpha'}\left[\left(\frac{{n'_i}^2}{{d_i}^2}{L_x^i}^2+\frac{{m'_i}^2}{{d_i}^2}{L_y^i}^2\right){k_i}^2\right. \\
\Bigg. +\left(\frac{n'_i p'_i x_i}{d_i}{L_x^i}^2-\frac{m'_i p'_i y_i}{d_i}{L_y^i}^2\right)2 k_i l_i
+\left({p'_i}^2 {x_i}^2{L_x^i}^2+{p'_i}^2 {y_i}^2{L_y^i}^2+{d_i}^2{(\tilde L_z^i)}^2\right){l_i}^2\Bigg],
\end{multline}
where the sum in the trace of the annulus amplitude runs over the
three independent integers $w_i, k_i$ and $l_i$ (for every
$T^3_i$). The corresponding open string Hamiltonian can be
generally obtained via a modular transformation, using the Poisson
resummation formula
\begin{equation}\label{eq:poisson_res_normal}
\sum\limits_{n\in\mathbb{Z}}e^{-\frac{\pi(n-c)^2}{t}}=\sqrt{t}\sum\limits_{m\in\mathbb{Z}}e^{2\pi i c m}\ e^{-\pi m^2 t}\ .
\end{equation}
The second part of the Hamiltonian (\ref{eq:closedStr_latAnn})
comprises a technical complication because it contains a term that is
bilinear in $k_i$ and $l_i$. The solution is to write the second
part of the Hamiltonian as a matrix of the form
\begin{equation}
\frac{1}{2\alpha'}\left[\big(k_i, l_i\big)\left(\begin{array}{cc}c_1&c_2
\\c_2&c_3\end{array}\right)\left(\begin{array}{c}k_i\\l_i\end{array}\right)\right],\nonumber
\end{equation}
and then use the matrix generalization of the Poisson resummation,
\begin{equation}\label{eq:poisson_res_Matrix}
\sum\limits_{\vec{v}\in {\mathbb{Z}}^d}e^{-\frac{\pi}{t}\vec{v}^{\text{T}}\mathbf{S}\vec{v}}
={t}^{d/2}\sqrt{\det{\mathbf{S}^{-1}}}\sum\limits_{\vec{w}\in {\mathbb{Z}}^d}e^{-\pi t \vec{w}^{\text{T}}\mathbf{S}^{-1}\vec{w}}.
\end{equation}
In this way we obtain the open string lattice Hamiltonian for the annulus
amplitude
\begin{multline}\label{eq:openStr_latAnn}
    \mathcal{H}_{\text{lattice, op.}}^{\mathcal{A}}=\frac{1}{\alpha'}
    \left({\frac{{m'_i}^2}{{L_x^i}^2}+\frac{{n'_i}^2}{{L_y^i}^2}+\frac{{p'_i}^2}{{(\tilde L_z^i)}^2}}\right)^{-1}{\tilde w_i}^2
    +\frac{1}{4\alpha'\Delta}\left[\left({p'_i}^2 {x_i}^2 {L_x^i}^2+{p'_i}^2 {y_i}^2 {L_y^i}^2
    \right. \right.\\
\left.\left. +{d_i}^2 {(\tilde L_z^i)}^2\right){d_i}^2{\tilde k_i}^2
+\left(m'_i p'_i y_i {L_y^i}^2-n'_i p'_i x_i {L_x^i}^2\right)2{\tilde k_i}{\tilde l_i}{d_i}
+\left({n'_i}^2{L_x^i}^2+{m'_i}^2{L_y^i}^2\right){\tilde l_i}^2\right],
\end{multline}
with
\begin{equation}
\Delta\equiv{d_i}^2\left({n'_i}^2 {L_x^i}^2{(\tilde L_z^i)}^2+{m'_i}^2 {L_y^i}^2{(\tilde L_z^i)}^2+{p'_i}^2{L_x^i}^2{L_y^i}^2\right).
\end{equation}
The sum in the amplitude runs over the independent integers
$\tilde w_i,\tilde k_i,\tilde l_i\in\mathbb{Z}$. Most remarkable
is the fact that indeed all terms containing $x_i$, $y_i$ and
$d_i$ chancel against each other in $\sqrt{\det\mathbf{S}}$ that
enters the tree channel tadpole cancellation condition. This
happens after we have resubstituted equation (\ref{eq:mx_ny_d}).
\subsection{Case of non-vanishing angles between branes}\label{AppSec:A_lattice_nonvanangle}
In contrast to the $T^6$ case, we still obtain a contribution from
the lattice, simply because the intersection is one-dimensional
and compact in the case of the $T^9$. Compared to the case of a
vanishing angle between the two branes, instead of
(\ref{eq:a_bc_1})-(\ref{eq:a_bc_3}) we have to solve the following
set,
\begin{align}
&p^{a\prime}_i s_1^i+{m^{a\prime}_i}r_3^i=0,& &p^{b\prime}_i s_1^i+{m^{b\prime}_i}r_3^i=0, \nonumber\\
&p^{a\prime}_i s_2^i+{n^{a\prime}_i}r_3^i=0,& &p^{b\prime}_i s_2^i+{n^{b\prime}_i}r_3^i=0, \label{eq:aijbc}\\
&p^{a\prime}_i s_3^i-{m^{a\prime}_i}r_1^i-{n^{a\prime}_i}r_2^i=0,& &p^{b\prime}_i s_3^i-{m^{b\prime}_i}r_1^i-{n^{b\prime}_i}r_2^i=0. \nonumber
\end{align}
For now assume that the two primitive vectors defining
the two branes $a$ and $b$ are not coinciding.
Then, the most general solution to the set (\ref{eq:aijbc}) is given by
\begin{align}
&s_1^i=s_2^i=r_3^i=0,\nonumber\\
&r_1^i=\Lambda_i^{-1}\left({m^{b\prime}_i}{p^{a\prime}_i}-{m^{a\prime}_i}{p^{b\prime}_i}\right) w_i,\\
&r_2^i=\Lambda_i^{-1}\left({p^{b\prime}_i}{n^{a\prime}_i}-{p^{a\prime}_i}{n^{b\prime}_i}\right) w_i,\nonumber\\
&s_3^i=\Lambda_i^{-1}\left({m^{b\prime}_i}{n^{a\prime}_i}-{m^{a\prime}_i}{n^{b\prime}_i}\right) w_i,\nonumber
\end{align}
where
\begin{equation}
\Lambda_i\equiv\gcd\left({m^{b\prime}_i}{p^{a\prime}_i}-{m^{a\prime}_i}{p^{b\prime}_i},\
{p^{b\prime}_i}{n^{a\prime}_i}-{p^{a\prime}_i}{n^{b\prime}_i},\
{m^{b\prime}_i}{n^{a\prime}_i}-{m^{a\prime}_i}{n^{b\prime}_i}\right).
\end{equation}
Therefore, we obtain the closed string lattice Hamiltonian
\begin{multline}\label{eq:closedStr_latAnnij}
    \mathcal{H}_{\text{lattice, cl.}}^{\mathcal{A}_{ab}}=\frac{1}{2\alpha'\Lambda_i^2}\left[
\left({p^{b\prime}_i}{n^{a\prime}_i}-{p^{a\prime}_i}{n^{b\prime}_i}\right)^2{L_x^i}^2\right.\\
\left.+\left({m^{b\prime}_i}{p^{a\prime}_i}-{m^{a\prime}_i}{p^{b\prime}_i}\right)^2{L_y^i}^2
+\left({m^{b\prime}_i}{n^{a\prime}_i}-{m^{a\prime}_i}{n^{b\prime}_i}\right)^2{(\tilde L_z^i)}^2\right]{w_i}^2.
\end{multline}
The sum in the amplitude runs over all integers
$w_i\in\mathbb{Z}$. The open string Hamiltonian can be obtained in
a similar way like in the last section by a modular transformation
using equation(\ref{eq:poisson_res_normal}), the result is given by
\begin{multline}\label{eq:openStr_latAnnij}
    \mathcal{H}_{\text{lattice, op.}}^{\mathcal{A}_{ab}}=\alpha'\Lambda_i^{2}\left[
\left({p^{b\prime}_i}{n^{a\prime}_i}-{p^{a\prime}_i}{n^{b\prime}_i}\right)^2{L_x^i}^2\right.\\
\left.+\left({m^{b\prime}_i}{p^{a\prime}_i}-{m^{a\prime}_i}{p^{b\prime}_i}\right)^2{L_y^i}^2
+\left({m^{b\prime}_i}{n^{a\prime}_i}-{m^{a\prime}_i}{n^{b\prime}_i}\right)^2{(\tilde L_z^i)}^2\right]^{-1}{\tilde w_i}^2.
\end{multline}

\section{The NS-NS tadpole}\label{Sec:NSNStp_genchapter}
For completeness, we will state the NS-NS tadpoles in this
section. This can be done in two different ways: either directly
from the one loop amplitude, which is exactly analogue to the
treatment of the R-R-tadpole in section \ref{Sec:RRtp_genchapter},
or from the tree level scalar potential arising the Born-Infeld
action. This derivation is more straightforward and equivalent to
the other one, so it will be used here. The potential is derived
in section \ref{Sec:tree_level_pot}, equation
(\ref{eq:totalNSpotD6}). All the NS-NS tadpoles can be obtained by
simply differentiating the potential with respect to all the
scalar fields that it depends on. This procedure leads to ten
different tadpoles, the dilaton tadpole $\langle \phi \rangle_D
\sim {\partial V /
\partial \phi}$ and nine radion tadpoles, for instance $\langle L^i_x \rangle_D\sim {\partial V /\partial L^i_x}$.

\noindent Explicitly, they are given by
\begin{equation}\label{eq:NSNStadpole1}
\langle \phi \rangle_D=\sum_{l=1}^k \tilde N_l v_1^l v_2^l v_3^l
-16\sqrt{\frac{{L_x^1}{L_y^1}{L_x^2}{L_y^2}{L_x^3}{L_y^3}}{\tilde
L_z^1 \tilde L_z^2 \tilde L_z^3}},
\end{equation}
\begin{multline}
\langle L^I_x \rangle_D=\sum_{l=1}^k \tilde N_l\frac{v_J^l
v_K^l}{v_I^l}\left[\left(n^{(l)\prime}_I+b_2^I
p^{(l)\prime}_I\right)^2 \frac{\tilde L_z^I}{L_y^I}
-\left(m^{(l)\prime}_I+b_1^I p^{(l)\prime}_I\right)^2
\frac{{\tilde L_z^I} L_y^I}{{L_x^I}^2}
+{p^{(l)\prime}_I}^2 \frac{L_y^I}{L_z^I}\right]\label{eq:NSNStadpole2}\\
-16\sqrt{\frac{{L_y^I}{L_x^J}{L_y^J}{L_x^K}{L_y^K}}{{L_x^I\tilde
L_z^I \tilde L_z^J \tilde L_z^K}}},
\end{multline}
\begin{multline}
\langle L^I_y \rangle_D=\sum_{l=1}^k \tilde N_l\frac{v_J^l
v_K^l}{v_I^l}\left[\left(m^{(l)\prime}_I+b_1^I
p^{(l)\prime}_I\right)^2 \frac{\tilde L_z^I}{L_x^I}
-\left(n^{(l)\prime}_I+b_2^I p^{(l)\prime}_I\right)^2
\frac{{\tilde L_z^I} L_x^I}{{L_y^I}^2}
+{p^{(l)\prime}_I}^2 \frac{L_x^I}{\tilde L_z^I}\right]\label{eq:NSNStadpole3}\\
-16\sqrt{\frac{{L_x^I}{L_x^J}{L_y^J}{L_x^K}{L_y^K}}{{L_y^I\tilde
L_z^I \tilde L_z^J \tilde L_z^K}}},
\end{multline}
\begin{multline}
\langle \tilde L^I_z \rangle_D=\sum_{l=1}^k \tilde N_l\frac{v_J^l
v_K^l}{v_I^l}\left[\left(n^{(l)\prime}_I+b_2^I
p^{(l)\prime}_I\right)^2
\frac{L_x^I}{L_y^I}+\left(m^{(l)\prime}_I+b_1^I
p^{(l)\prime}_I\right)^2
\frac{L_y^I}{L_x^I}-{p^{(l)\prime}_I}^2 \frac{L_x^I L_y^I}{(\tilde L_z^I)^2}\right]\\
+16\frac{1}{\tilde
L_z^I}\sqrt{\frac{{L_x^I}{L_y^I}{L_x^J}{L_y^J}{L_x^K}{L_y^K}}{\tilde
L_z^I \tilde L_z^J \tilde L_z^K}},\label{eq:NSNStadpole4}
\end{multline}
with $I,J,K\in \{1,2,3\}$ and $I\neq J\neq K\neq I$ and where
$v_i^l$ is defined by
\begin{equation}
v_i^l\equiv \sqrt{\left(n^{(l)\prime}_i+b_2^i
p^{(l)\prime}_i\right)^2 \frac{{L_x^i}{\tilde
L_z^i}}{L_y^i}+\left(m^{(l)\prime}_i+b_1^i
p^{(l)\prime}_i\right)^2 \frac{{L_y^i}{\tilde L_z^i}}{L_x^i}
+{p^{(l) \prime}_i}^2\frac{{L_x^i}{L_y^i}}{\tilde L_z^i}}\ .
\end{equation}

\section{Modular function expansions}\label{AppSec:mod_fct}
In the main text, the expansion of the following combination of
modular functions in $q$ is needed.
\begin{align}\label{eq:thetafctentw2winkel}
&\frac{-\tfktsqb{\frac{1}{2}}{0}\tfktob{\frac{1}{2}}
{-\kappa_1}\tfktob{\frac{1}{2}}{-\kappa_2}
+\tfktsqb{0}{0}\tfktob{0}{-\kappa_1}\tfktob{0}{-\kappa_2}
-\tfktsqb{0}{\frac{1}{2}}\tfktob{0}{\frac{1}{2}-\kappa_1}
\tfktob{0}{\frac{1}{2}-\kappa_2}}{\tfktob{\frac{1}{2}}{\frac{1}{2}-\kappa_1}
\tfktob{\frac{1}{2}}{\frac{1}{2}-\kappa_2}\eta^{6}}\\ \nonumber
&=2\,{\frac{\cos^2\left(\pi{\kappa_1} \right)+\cos^2 \left( \pi{\kappa_2} \right)-2
\cos \left( \pi {\kappa_1} \right) \cos \left( \pi{
\kappa_2} \right) }{\sin \left( \pi{\kappa_1} \right) \sin
 \left( \pi{\kappa_2} \right) }}\\ \nonumber
 &\ \ \ +4{\frac {\cos^2\left( \pi { \kappa_1}
 \right)+\cos^2\left( \pi { \kappa_2}
 \right)+8 \cos^2 \left( \pi { \kappa_1}
 \right)\cos^2\left( \pi { \kappa_2}
 \right)-6\cos \left( \pi { \kappa_1} \right)
\cos \left( \pi { \kappa_2} \right)}{\sin \left( \pi
{ \kappa_1} \right) \sin \left( \pi { \kappa_2} \right) }}q+\mathcal{O}(q^2)
\end{align}
\end{appendix}


\begin{thebibliography}{10}

\bibitem{Berkooz:1996aa}
M.~Berkooz, M.~R. Douglas, and R.~G. Leigh.
\newblock \textsl{Branes Intersecting at Angles}.
\newblock Nucl. Phys. B 480, 265 (1996), \texttt{hep-th/9606139}.

\bibitem{Blumenhagen:2005mu}
Ralph Blumenhagen, Mirjam Cvetic, Paul Langacker, and Gary Shiu.
\newblock \textsl{Toward realistic intersecting D-brane models}.
\newblock (2005), \texttt{hep-th/0502005}.

\bibitem{Ott:2003yv}
Tassilo Ott.
\newblock \textsl{Aspects of stability and phenomenology in type IIA
  orientifolds with intersecting D6-branes}.
\newblock (2003), \texttt{hep-th/0309107}.

\bibitem{Gorlich:2004zs}
Lars Gorlich.
\newblock \textsl{N = 1 and non-supersymmetric open string theories in six and
  four space-time dimensions}.
\newblock (2004), \texttt{hep-th/0401040}.

\bibitem{MarchesanoBuznego:2003hp}
Fernando~G. Marchesano~Buznego.
\newblock \textsl{Intersecting D-brane models}.
\newblock (2003), \texttt{hep-th/0307252}.

\bibitem{Lust:2004ks}
Dieter Lust.
\newblock \textsl{Intersecting brane worlds: A path to the standard model?}
\newblock (2004), \texttt{hep-th/0401156}.

\bibitem{Brandenberger:1988aj}
Robert~H. Brandenberger and C.~Vafa.
\newblock \textsl{Superstrings In The Early Universe}.
\newblock Nucl. Phys. B316, 391 (1989).

\bibitem{Sagnotti:1987tw}
Augusto Sagnotti.
\newblock \textsl{Open strings and their symmetry groups}.
\newblock (1987), \texttt{hep-th/0208020}.

\bibitem{Blumenhagen:1999db}
Ralph Blumenhagen, Lars G{\"o}rlich, and Boris K{\"o}rs.
\newblock \textsl{A new class of supersymmetric orientifolds with D-branes at
  angles}.
\newblock (1999), \texttt{hep-th/0002146}.

\bibitem{Blumenhagen:1999ev}
Ralph Blumenhagen, Lars G{\"o}rlich, and Boris K{\"o}rs.
\newblock \textsl{Supersymmetric 4D orientifolds of type IIA with D6-branes at
  angles}.
\newblock JHEP 01, 040 (2000), \texttt{hep-th/9912204}.

\bibitem{Blumenhagen:2000wh}
Ralph Blumenhagen, Lars G{\"o}rlich, Boris K{\"o}rs, and Dieter L{\"u}st.
\newblock \textsl{Noncommutative compactifications of type I strings on tori
  with magnetic background flux}.
\newblock JHEP 10, 006 (2000), \texttt{hep-th/0007024}.

\bibitem{Blumenhagen:2000ea}
Ralph Blumenhagen, Boris K{\"o}rs, and Dieter L{\"u}st.
\newblock \textsl{Type I strings with F- and B-flux}.
\newblock JHEP 02, 030 (2001), \texttt{hep-th/0012156}.

\bibitem{Angelantonj:2000hi}
C.~Angelantonj, Ignatios Antoniadis, E.~Dudas, and A.~Sagnotti.
\newblock \textsl{Type-I strings on magnetised orbifolds and brane
  transmutation}.
\newblock Phys. Lett. B489, 223--232 (2000), \texttt{hep-th/0007090}.

\bibitem{Aldazabal:2000dg}
G.~Aldazabal, S.~Franco, Luis~E. Ibanez, R.~Rabadan, and A.~M. Uranga.
\newblock \textsl{D = 4 chiral string compactifications from intersecting
  branes}.
\newblock J. Math. Phys. 42, 3103--3126 (2001), \texttt{hep-th/0011073}.

\bibitem{Forste:2000hx}
Stefan Forste, Gabriele Honecker, and Ralph Schreyer.
\newblock \textsl{Supersymmetric Z(N) x Z(M) orientifolds in 4D with D-branes
  at angles}.
\newblock Nucl. Phys. B593, 127--154 (2001), \texttt{hep-th/0008250}.

\bibitem{Forste:2001gb}
Stefan Forste, Gabriele Honecker, and Ralph Schreyer.
\newblock \textsl{Orientifolds with branes at angles}.
\newblock JHEP 06, 004 (2001), \texttt{hep-th/0105208}.

\bibitem{Kokorelis:2002ns}
  C.~Kokorelis,
  JHEP {\bf 0211} (2002) 027
  [arXiv:hep-th/0209202].

\bibitem{Cvetic:2001nr}
  M.~Cvetic, G.~Shiu and A.~M.~Uranga,
  Nucl.\ Phys.\ B {\bf 615}, 3 (2001)
  [arXiv:hep-th/0107166].

\bibitem{Kachru:1999vj}
Shamit Kachru and John McGreevy.
\newblock \textsl{Supersymmetric three-cycles and (super)symmetry breaking}.
\newblock Phys. Rev. D61, 026001 (2000), \texttt{hep-th/9908135}.

\bibitem{Uranga:2003pz}
Angel~M. Uranga.
\newblock \textsl{Chiral four-dimensional string compactifications with
  intersecting D-branes}.
\newblock Class. Quant. Grav. 20, S373--S394 (2003), \texttt{hep-th/0301032}.

\bibitem{Honecker:2004kb}
Gabriele Honecker and Tassilo Ott.
\newblock \textsl{Getting just the supersymmetric standard model at
  intersecting branes on the Z(6)-orientifold}.
\newblock Phys. Rev. D70, 126010 (2004), \texttt{hep-th/0404055}.

\bibitem{Ott:2005sa}
Tassilo Ott.
\newblock \textsl{Catching the phantom: The MSSM on the Z6-orientifold}.
\newblock (2005), \texttt{hep-th/0505274}.

\bibitem{Angelantonj:2002ct}
Carlo Angelantonj and Augusto Sagnotti.
\newblock \textsl{Open strings}.
\newblock Phys. Rept. 371, 1--150 (2002), \texttt{hep-th/0204089}.

\bibitem{Arkani-Hamed:1998rs}
Nima Arkani-Hamed, Savas Dimopoulos, and G.~R. Dvali.
\newblock \textsl{The hierarchy problem and new dimensions at a millimeter}.
\newblock Phys. Lett. B429, 263--272 (1998), \texttt{hep-ph/9803315}.

\bibitem{Antoniadis:1998ig}
  I.~Antoniadis, N.~Arkani-Hamed, S.~Dimopoulos and G.~R.~Dvali,
  Phys.\ Lett.\ B {\bf 436}, 257 (1998)
  [arXiv:hep-ph/9804398].

\bibitem{Spergel:2003cb}
D.~N. Spergel et~al.
\newblock \textsl{First Year Wilkinson Microwave Anisotropy Probe (WMAP)
  Observations: Determination of Cosmological Parameters}.
\newblock Astrophys. J. Suppl. 148, 175 (2003), \texttt{astro-ph/0302209}.

\bibitem{Ibanez:2001nd}
Luis~E. Ibanez, F.~Marchesano, and R.~Rabadan.
\newblock \textsl{Getting just the standard model at intersecting branes}.
\newblock JHEP 11, 002 (2001), \texttt{hep-th/0105155}.

\bibitem{Blumenhagen:2001te}
Ralph Blumenhagen, Boris K{\"o}rs, Dieter L{\"u}st, and Tassilo Ott.
\newblock \textsl{The standard model from stable intersecting brane world
  orbifolds}.
\newblock Nucl. Phys. B616, 3--33 (2001), \texttt{hep-th/0107138}.

\bibitem{Blumenhagen:2001mb}
Ralph Blumenhagen, Boris K{\"o}rs, Dieter L{\"u}st, and Tassilo Ott.
\newblock \textsl{Intersecting brane worlds on tori and orbifolds}.
\newblock Fortsch. Phys. 50, 843--850 (2002), \texttt{hep-th/0112015}.

\bibitem{Chupp:1989ty}
T.~E. Chupp et~al.
\newblock \textsl{Results of a new test of local Lorentz invariance: A Search
  for mass anissotropy in Ne-21}.
\newblock Phys. Rev. Lett. 63, 1541--1545 (1989).

\bibitem{Fischler:1986ci}
W.~Fischler and Leonard Susskind.
\newblock \textsl{Dilaton Tadpoles, String Condensates and Scale Invariance}.
\newblock Phys. Lett. B171, 383 (1986).

\bibitem{Fischler:1986tb}
Willy Fischler and Leonard Susskind.
\newblock \textsl{Dilaton Tadpoles, String Condensates and Scale Invariance.
  2}.
\newblock Phys. Lett. B173, 262 (1986).

\bibitem{Polchinski:1987tu}
Joseph Polchinski and Yunhai Cai.
\newblock \textsl{Consistency of Open Superstring Theories}.
\newblock Nucl. Phys. B296, 91 (1988).

\bibitem{Angelantonj:1999qg}
Carlo Angelantonj and Adi Armoni.
\newblock \textsl{Non-tachyonic type 0B orientifolds, non-supersymmetric gauge
  theories and cosmological RG flow}.
\newblock Nucl. Phys. B578, 239--258 (2000), \texttt{hep-th/9912257}.

\bibitem{Dudas:2004nd}
E.~Dudas, G.~Pradisi, M.~Nicolosi, and A.~Sagnotti.
\newblock \textsl{On tadpoles and vacuum redefinitions in string theory}.
\newblock Nucl. Phys. B708, 3--44 (2005), \texttt{hep-th/0410101}.

\bibitem{Dudas:2000ff}
E.~Dudas and J.~Mourad.
\newblock \textsl{Brane solutions in strings with broken supersymmetry and
  dilaton tadpoles}.
\newblock Phys. Lett. B486, 172--178 (2000), \texttt{hep-th/0004165}.

\bibitem{Blumenhagen:2000dc}
Ralph Blumenhagen and Anamaria Font.
\newblock \textsl{Dilaton tadpoles, warped geometries and large extra
  dimensions for non-supersymmetric strings}.
\newblock Nucl. Phys. B599, 241--254 (2001), \texttt{hep-th/0011269}.

\bibitem{Dudas:2003wp}
E.~Dudas, J.~Mourad, and C.~Timirgaziu.
\newblock \textsl{On cosmologically induced hierarchies in string theory}.
\newblock JCAP 0403, 005 (2004), \texttt{hep-th/0309057}.

\bibitem{Fradkin:1985qd}
E.~S. Fradkin and A.~A. Tseytlin.
\newblock \textsl{Nonlinear Electrodynamics from Quantized Strings}.
\newblock Phys. Lett. B163, 123 (1985).

\bibitem{Leigh:1989jq}
R.~G. Leigh.
\newblock \textsl{Dirac-Born-Infeld Action from Dirichlet Sigma Model}.
\newblock Mod. Phys. Lett. A4, 2767 (1989).

\bibitem{Burgess:2001vr}
C.~P. Burgess, P.~Martineau, F.~Quevedo, G.~Rajesh, and R.~J. Zhang.
\newblock \textsl{Brane antibrane inflation in orbifold and orientifold
  models}.
\newblock JHEP 03, 052 (2002), \texttt{hep-th/0111025}.

\bibitem{Blumenhagen:2002ua}
Ralph Blumenhagen, Boris K{\"o}rs, Dieter L{\"u}st, and Tassilo Ott.
\newblock \textsl{Hybrid inflation in intersecting brane worlds}.
\newblock Nucl. Phys. B641, 235--255 (2002), \texttt{hep-th/0202124}.

\bibitem{Cremades:2003qj}
D.~Cremades, L.~E. Ibanez, and F.~Marchesano.
\newblock \textsl{Yukawa couplings in intersecting D-brane models}.
\newblock JHEP 07, 038 (2003), \texttt{hep-th/0302105}.

\bibitem{Cvetic:2003ch}
  M.~Cvetic and I.~Papadimitriou,
  Phys.\ Rev.\ D {\bf 68}, 046001 (2003)
  [Erratum-ibid.\ D {\bf 70}, 029903 (2004)]
  [arXiv:hep-th/0303083].

\bibitem{Hashimoto:1997gm}
Akikazu Hashimoto and IV~Taylor, Washington.
\newblock \textsl{Fluctuation spectra of tilted and intersecting D-branes from
  the Born-Infeld action}.
\newblock Nucl. Phys. B503, 193--219 (1997), \texttt{hep-th/9703217}.

\bibitem{Taylor:1997dy}
IV~Taylor, Washington.
\newblock \textsl{Lectures on D-branes, gauge theory and M(atrices)}.
\newblock (1997), \texttt{hep-th/9801182}.

\bibitem{Kakushadze:1998bw}
Zurab Kakushadze, Gary Shiu, and S.~H.~Henry Tye.
\newblock \textsl{Type IIB orientifolds with NS-NS antisymmetric tensor
  backgrounds}.
\newblock Phys. Rev. D58, 086001 (1998), \texttt{hep-th/9803141}.

\bibitem{Angelantonj:1999jh}
Carlo Angelantonj.
\newblock \textsl{Comments on open-string orbifolds with a non-vanishing
  B(ab)}.
\newblock Nucl. Phys. B566, 126--150 (2000), \texttt{hep-th/9908064}.

\bibitem{Lust:2003ky}
D.~L{\"u}st and S.~Stieberger.
\newblock \textsl{Gauge threshold corrections in intersecting brane world
  models}.
\newblock (2003), \texttt{hep-th/0302221}.

\bibitem{Green:1984sg}
Michael~B. Green and John~H. Schwarz.
\newblock \textsl{Anomaly Cancellation in supersymmetric d=10 Gauge Theory and
  Superstring Theory}.
\newblock Phys. Lett. B149, 117--122 (1984).

\bibitem{Green:1985ed}
Michael~B. Green and John~H. Schwarz.
\newblock \textsl{Infinity Cancellations in SO(32) Superstring Theory}.
\newblock Phys. Lett. B151, 21--25 (1985).

\bibitem{Blumenhagen:2002wn}
Ralph Blumenhagen, Volker Braun, Boris K{\"o}rs, and Dieter L{\"u}st.
\newblock \textsl{Orientifolds of K3 and Calabi-Yau manifolds with intersecting
  D-branes}.
\newblock JHEP 07, 026 (2002), \texttt{hep-th/0206038}.

\bibitem{SheikhJabbari:1997cv}
M.~M. Sheikh~Jabbari.
\newblock \textsl{Classification of different branes at angles}.
\newblock Phys. Lett. B420, 279--284 (1998), \texttt{hep-th/9710121}.

\bibitem{Ohta:1997fr}
  N.~Ohta and P.~K.~Townsend,
  Phys.\ Lett.\ B {\bf 418}, 77 (1998)
  [arXiv:hep-th/9710129].

\bibitem{Douglas:1999vm}
Michael~R. Douglas.
\newblock \textsl{Topics in D-geometry}.
\newblock Class. Quant. Grav. 17, 1057--1070 (2000), \texttt{hep-th/9910170}.

\bibitem{Polchinski:1998rr}
J.~Polchinski.
\newblock \textsl{String theory. Vol. 2: Superstring theory and beyond}.
\newblock Cambridge, UK: Univ. Pr. (1998) 531 p.

\bibitem{Ott2:2005}
Tassilo Ott.
\newblock \textsl{work in progress}.

\bibitem{Blumenhagen:2002gw}
Ralph Blumenhagen, Lars G{\"o}rlich, and Tassilo Ott.
\newblock \textsl{Supersymmetric intersecting branes on the type IIA
  $T^6/\mathbb{Z}_4$ orientifold}.
\newblock JHEP 01, 021 (2003), \texttt{hep-th/0211059}.

\end{thebibliography}
\end{document}